\documentclass[aps,pra,amsmath,amssymb,reprint,floatfix,superscriptaddress,showpacs]{revtex4-1}
\usepackage{graphicx}
\usepackage[bookmarks=false,colorlinks=true,linkcolor=blue,filecolor=blue,citecolor=blue,urlcolor=blue]{hyperref}
\usepackage{bm}
\usepackage{soul}

\newcommand{\tp}{\tilde{p}}
\newcommand{\tq}{\tilde{q}}

\newcommand{\tphi}{\tilde{\phi}}
\newcommand{\tpsi}{\tilde{\psi}}
\newcommand{\txi}{\tilde{\xi}}

\newcommand{\tvarphi}{\tilde{\varphi}}
\newcommand{\ttheta}{\tilde{\theta}}
\newcommand{\trho}{\tilde{\rho}}

\newcommand{\tu}{\tilde{u}}
\newcommand{\tv}{\tilde{v}}
\newcommand{\tK}{\tilde{K}}
\newcommand{\tGamma}{\tilde{\Gamma}}
\newcommand{\tbfGamma}{\tilde{{\bm \Gamma}}}

\newcommand{\brho}{\bar{\rho}}

\newcommand{\bfk}{\mathbf{k}}

\newcommand{\bfC}{\mathbf{C}}

\newcommand{\bfK}{\mathbf{K}}

\newcommand{\bft}{\mathbf{t}}

\newcommand{\bfgamma}{{\bm \gamma}}
\newcommand{\bfGamma}{{\bm \Gamma}}

\newcommand{\calL}{\mathcal{L}}

\newcommand{\sgn}{\hspace{1pt} \textrm{sgn}}

\newcommand{\ext}{\textrm{ext}}
\newcommand{\CP}{\textrm{CP}}
\newcommand{\CF}{\textrm{CF}}
\newcommand{\CQ}{\textrm{CQ}}
\newcommand{\CH}{\textrm{CH}}
\newcommand{\CB}{\textrm{CB}}
\newcommand{\VCB}{\textrm{VCB}}
\newcommand{\hole}{\textrm{hole}}
\newcommand{\VCQ}{\textrm{VCQ}}
\newcommand{\ua}{\uparrow}
\newcommand{\da}{\downarrow}

\newcommand{\red}[1]{{\textcolor{red}{#1}}}

\begin{document}

\title{Quantum Hall hierarchy from coupled wires}

\author{Yohei Fuji}
\affiliation{Condensed Matter Theory Laboratory, RIKEN CPR, Wako, Saitama 351-0198, Japan}

\author{Akira Furusaki}
\affiliation{Condensed Matter Theory Laboratory, RIKEN CPR, Wako, Saitama 351-0198, Japan}
\affiliation{RIKEN Center for Emergent Matter Science, Wako, Saitama 351-0198, Japan}

\date{\today}

\begin{abstract}
The coupled-wire construction provides a useful way to obtain microscopic Hamiltonians for various two-dimensional  topological phases, among which fractional quantum Hall states are paradigmatic examples.
Using the recently introduced flux attachment and vortex duality transformations for coupled wires, we show that this construction is remarkably versatile to encapsulate phenomenologies of hierarchical quantum Hall states: the Jain-type hierarchy states of composite fermions filling Landau levels and the Haldane-Halperin hierarchy states of quasiparticle condensation. 
The particle-hole conjugate transformation for coupled-wire models is also given as a special case of the hierarchy construction.
We also propose coupled-wire models for the composite Fermi liquid, which turn out to be compatible with a sort of the particle-hole symmetry implemented in a nonlocal way at $\nu=1/2$. 
Furthermore, our approach shows explicitly the connection between the Moore-Read Pfaffian state and a chiral $p$-wave pairing of the composite fermions.
This composite fermion picture is also generalized to a family of the Pfaffian state, including the anti-Pfaffian state and Bonderson-Slingerland hierarchy states.
\end{abstract}

\maketitle

\tableofcontents

\section{Introduction}

Fractional quantum Hall (FQH) states are quintessential examples of topological phases in which strong interaction between electrons plays a fundamental role for their stability. 
The FQH states are nontrivial many-body states that cannot be understood by perturbative approaches from a free electron gas.
After a remarkable success of the Laughlin's wave functions for incompressible states at filling fraction $\nu=1/q$ with an odd integer $q$ \cite{Laughlin83}, two physical pictures emerged that allow us to view the Laughlin states in terms of composite particles.
One is the composite boson picture, in which an odd number of flux quanta are attached to each electron to form a composite particle with bosonic statistics \cite{Girvin87}. 
In this picture, the Laughlin states can be viewed as a condensate of composite bosons \cite{Girvin87,Read89}, which are described by a Chern-Simons Ginzburg-Landau theory \cite{SCZhang89}. 
The other is the composite fermion picture, in which every electron binds an even number of flux quanta to form a composite particle with fermionic statistics \cite{Jain89,Jain90}. 
Then the Laughlin states can be seen as an integer quantum Hall state realized in a filled lowest Landau level of the composite fermions \cite{Jain89,Jain90}, and the corresponding Chern-Simons theory has also been developed \cite{Lopez91}. 

Aside from these developments, there have been earlier attempts to explain other quantum Hall plateaus that do not fall into the Laughlin states. 
Soon after the Laughlin's discovery, Haldane and Halperin independently proposed a systematic way to construct descendant FQH wave functions from the Laughlin states \cite{Haldane83,Halperin84}. 
In a heuristic point of view of their construction, a new wave function is obtained by exciting quasiparticles in a parent Laughlin state and condensing them into a bosonic Laughlin state. 
This procedure can be repeated to generate a hierarchy of FQH states.
An intriguing observation is that if we adopt the composite boson picture, quasiparticle excitations can be seen as vortex excitations with a gap above the composite boson condensate \cite{XGWen90a,DHLee89,XGWen90b}. 
Utilizing the vortex duality \cite{Peskin78,Dasgupta81,Fisher89}, these hierarchy states can also be formulated as a bosonic Chern-Simons theory through the condensation of vortices \cite{Blok90a,Blok90b,Ezawa91}. 
Several years after the proposal by Haldane and Halperin, Jain proposed different FQH wave functions at $\nu=p/(2pn \pm 1)$ with integers $p$ and $n$, which are obtained by filling $p$ Landau levels of the composite fermions \cite{Jain89,Jain90}. 
At some filling fractions both Haldane-Halperin and Jain hierarchy states can be constructed, but their wave functions have different forms. 
Despite their apparent distinction, however, these states are known to have the same quasiparticle properties and thus possess the same topological order \cite{Read90}. 
Such topological properties can be understood from an effective hydrodynamic description using the Chern-Simons theory for Abelian FQH states \cite{XGWen92a,XGWen95}.
For a more detailed account of the history and recent developments of the hierarchy states, see Ref.~\cite{Hansson17}. 

The idea of the composite fermions is also fruitful for understanding the physics of interacting electrons at the filling fraction $\nu=1/M$ with an even integer $M$, where composite fermions of electrons with $2\pi M$ flux quanta see a zero magnetic field on average. 
In this case, the composite fermions can form either a Fermi liquid or a superconductor. 
The former possibility was pointed out and examined in detail by Halperin, Lee, and Read, who proposed the composite Fermi liquid (CFL) for the compressible state of strongly interacting electrons at $\nu=1/2$ \cite{Halperin93}. 
The other possibility was discussed in depth by Read and Green \cite{Read00}. 
In particular, they argued that a spinless chiral $p$-wave pairing of composite fermions corresponds to the Moore-Read Pfaffian state \cite{Moore91}, in which quasiparticles are non-Abelian anyons. 

The composite-particle pictures are conceptually quite useful and have led to many nontrivial discoveries on the physics of FQH states and their topological order. 
However, trial wave functions must be verified by direct large-scale numerical calculations of microscopic models of finite size, while Chern-Simons approaches are effective theories that require an additional justification from numerics or experiments.

A notable progress that can bridge the gap between FQH states and microscopic Hamiltonians has been made by Kane and coworkers \cite{Kane02}, who used electron wires placed in a magnetic field as building blocks for FQH states. 
This coupled-wire approach allows us to analyze a two-dimensional (2D) array of the interacting quantum wires in a more controllable fashion with the help of bosonization technique and conformal field theory (CFT). 
It also enables us to study topological properties of the FQH states, such as edge states, quasiparticle statistics \cite{Kane02,Teo14}, and ground-state degeneracy on a torus \cite{Sagi15}.
After the original construction of Abelian FQH states \cite{Kane02}, the coupled-wire construction has been extended to non-Abelian FQH states \cite{Teo14,Fuji17,Kane17,Kane18} and surface topological orders of three-dimensional (3D) topological crystalline phases \cite{Mross15,Mross16c,Sahoo16,FLu17,SHong17,MCheng18}. 

In this paper, we show that the coupled-wire models for FQH states admit clear physical interpretations of the corresponding states in terms of the composite bosons or composite fermions. 
Key ingredients are coupled-wire versions of the flux attachment and the vortex duality that have been proposed by Mross, Alicea, and Motrunich as explicit nonlocal transformations on bosonic field variables \cite{Mross17a}. 
The coupled-wire construction turns out to be a complementary approach to the conventional wave-function-based or effective Chern-Simons approaches, giving us an insight to the free-particle pictures of various quantum Hall states from microscopic Hamiltonians beyond the Landau level physics. 
Conversely, this approach allows us to obtain a ``model'' coupled-wire Hamiltonian for the desired quantum Hall states based on their physical interpretations as the composite bosons or fermions. 
This will help us to explore microscopic realizations of the quantum Hall physics in one-dimensional (1D) or quasi-1D many-body systems on the lattice or in the continuum. 

\subsection*{Outline of the paper}

The rest of our paper proceeds as follows. 
In Sec.~\ref{sec:Dictionary}, we introduce our basic tools to construct and analyze the coupled-wire models. 
We then focus on three particular examples of quantum Hall states in the following sections: Abelian hierarchy states (Sec.~\ref{sec:AbelianHierarchy}), CFLs (Sec.~\ref{sec:CFL}), and Moore-Read Pfaffian states (Sec.~\ref{sec:Pfaffian}). 
These three sections are independent of each other to some extent, and the reader may choose to read them in any order.
We then conclude the paper in Sec.~\ref{sec:Conclusion} with several outlooks. 
We give a more detailed outline of each section below.

Section~\ref{sec:Dictionary} presents our dictionary of bosonization or fermionization in 1D and 2D, which is extensively used in the following sections. 
We start with the standard bosonization approach of 1D fermionic or bosonic systems. 
We then introduce a 2D array of coupled quantum wires as a basic setup for our construction of the FQH state. 
The flux attachment and the vortex duality transformations are explicitly defined in this 2D array as nonlocal transformations in bosonic fields \cite{Mross17a}. 
This allows us to visualize the composite particles and vortices (quasiparticles) as local objects in the coupled-wire models. 

In Sec.~\ref{sec:AbelianHierarchy}, we consider Abelian hierarchy states. 
The discussion starts with reviewing the previous construction \cite{Kane02,Teo14} of the Laughlin $\nu=1/q$ state as a prototypical example of the FQH state (Sec.~\ref{sec:Laughlin}). 
We confirm that the coupled-wire Hamiltonian admits both the composite fermion and composite boson interpretations. 
We then move to hierarchy states, mainly focusing on the $\nu=2/5$ and $\nu=2/7$ states (Sec.~\ref{sec:Hierarchy}). 
In both cases, the corresponding coupled-wire Hamiltonians yield FQH states that are regarded as integer quantum Hall (IQH) states of composite fermions in the Jain sequence and condensates of quasiparticles in the Haldane-Halperin hierarchy states. 
Generalizations to other states in the Jain sequence or the Haldane-Halperin hierarchy are also given. 
We then discuss a systematic way to obtain the particle-hole (PH) conjugates of FQH states (Sec~\ref{sec:PHConjugate}) and FQH states in higher Landau levels (Sec.~\ref{sec:NextLL}), including several examples including the Laughlin and $\nu=4/11$ states. 

In Sec.~\ref{sec:CFL}, we consider the CFL at filling fraction $\nu=1/M$, where $M$ is an even (odd) integer for fermions (bosons). 
We first construct coupled-wire models for the CFL with an open Fermi surface at general filling fraction $\nu=1/M$ (Sec~\ref{sec:GeneralCFL}) and then focus our attention to the case of $\nu=1/2$ (Sec.~\ref{sec:AntiCFL}). 
In the latter case, we can also construct the PH conjugate of the CFL at the same filling fraction, called the anti-CFL or composite hole liquid \cite{Barkeshli15,Mulligan16}, which can be distinguished from the CFL. 
It turns out, however, that the composite hole liquid is obtained from the same coupled-wire Hamiltonian as that for the CFL. 
This may indicate a possible PH symmetry in the CFL although the PH transformation is implemented in our coupled-wire models in a way that cannot be realized in a genuinely 2D lattice system.
We also discuss a similar issue for the CFL of two-component (i.e., spinful) bosons at $\nu=1/2+1/2$ and propose the PH transformations for two-component or single-component bosons (Sec.~\ref{sec:CFLTwoComp}). 

In Sec.~\ref{sec:Pfaffian}, we focus on the Pfaffian state as another candidate FQH state at $\nu=1/M$. 
After reviewing the coupled-wire construction of the Pfaffian state by Teo and Kane \cite{Teo14}, we extend the construction by applying flux-attachment transformations and discuss the Pfaffian state in terms of a chiral $p$-wave pairing state of composite fermions. 
We then propose coupled-wire models for Bonderson-Slingerland hierarchy states \cite{Bonderson08}, which are hierarchy states obtained from the Pfaffian states by condensing bound pairs of quasiparticles. 
We also construct the PH conjugate of the Pfaffian state at $\nu=1/2$, called the anti-Pfaffian state \cite{Levin07,SSLee07}, and discuss its interpretation as a composite fermion pairing (Sec.~\ref{sec:AntiPfaffian}). 
The section is closed with a brief discussion on other composite fermion pairings (Sec.~\ref{sec:OtherPairing}). 

Appendices~\ref{app:VortexAction} and \ref{app:Hole} provide the derivation of the kinetic actions for vortex and hole variables, respectively, coupled to an external electromagnetic field, whose complete forms are somewhat tedious and shortened while keeping only important terms in the main text.
Appendix~\ref{app:Pfaffian} contains a detailed discussion on the Klein factors that are introduced to ensure the anticommutation relations of fermionic fields in the coupled-wire models for the Pfaffian state.

\section{Bosonization/fermionization dictionaries}
\label{sec:Dictionary}

In this section we summarize the dictionary of bosonization/fermionization rule that we shall frequently use in this paper. 
We first present our convention of the bosonization in one-dimensional (1D) fermionic or bosonic systems and then introduce a 2D array of coupled quantum wires, each of which is described by a Tomonaga-Luttinger liquid. 
This furnishes elementary constituents of our coupled-wire models \cite{Kane02,Teo14}. 
Finally, we introduce nonlocal transformations of the bosonic fields that implement the flux attachment and vortex duality in 2D coupled wires \cite{Mross17a}.

\subsection{1D dictionary}

The building block of our analysis is a spinless (or spin polarized) electron wire placed in parallel to the $x$ axis. 
In the low-energy limit, an electron operator $\psi(x)$ is expanded around the Fermi points as 
\begin{align}
\psi(x) \sim e^{ik_F x} \psi_R(x) +e^{-ik_F x} \psi_L (x), 
\end{align}
where $\psi_{R/L}(x)$ annihilate a right-moving ($R$) or left-moving ($L$) fermion excitations near the Fermi points at the wave number $k=\pm k_F$. 
The average electron density $\brho$ is related to the Fermi momentum by $k_F = \pi \brho$. 
Linearizing the spectrum about the Fermi momenta, we obtain the effective low-energy Hamiltonian for a single wire, 
\begin{align}
H_0^\textrm{single} = \int dx \Bigl[ iv (-\psi^\dagger_R \partial_x \psi_R + \psi^\dagger_L \partial_x \psi_L) \Bigr], 
\end{align}
where $v$ is the velocity at the Fermi points $k=\pm k_F$. 
We now apply the bosonization technique \cite{Giamarchi} to write the Hamiltonian in terms of free bosons, 
\begin{align}
H_0^\textrm{single} = \frac{v}{2\pi} \int dx \Bigl[ (\partial_x \varphi)^2 +(\partial_x \theta)^2 \Bigr], 
\end{align}
where the bosonic fields $\theta(x)$ and $\varphi(x)$ satisfy the commutation relations, 
\begin{align}
\begin{split}\label{eq: original commutator}
&[\theta(x), \varphi(x')] = i\pi \Theta(x-x'), \\
&[\theta(x),\theta(x')]=[\varphi(x),\varphi(x')]=0,
\end{split}
\end{align}
with $\Theta(x)$ being the Heaviside step function. 
Equation \eqref{eq: original commutator} implies that $\partial_x \theta(x)$ and $\varphi(x)$ are canonically conjugate fields. 
With these bosonic fields, the fermionic operators are represented as 
\begin{align} \label{eq:ChiralFermion}
\psi_{R/L}(x) = \frac{1}{\sqrt{2\pi \alpha}} e^{i[\varphi(x) \pm \theta(x)]},
\end{align}
where $\alpha$ is a short-distance cutoff. 
The chiral fermion currents are then given by 
\begin{align}
\begin{split}
\rho_R &= \,:\!\psi_R^\dagger \psi_R\!: \, = \frac{1}{2\pi} (\partial_x \varphi +\partial_x \theta), \\
\rho_L &= \,:\!\psi_L^\dagger \psi_L\!:\, = -\frac{1}{2\pi} (\partial_x \varphi -\partial_x \theta),
\end{split}
\end{align}
where $:\! X \!:$ denotes normal ordering of $X$.
Thus, $\partial_x \theta$ and $\partial_x \varphi$ are related to the electron density and current,  
\begin{align} \label{eq:ElectronCurrent}
\begin{split}
j^\textrm{el}_0 &= \rho_R +\rho_L = \frac{1}{\pi} \partial_x \theta, \\
j^\textrm{el}_1 &= v(\rho_R -\rho_L) = \frac{v}{\pi} \partial_x \varphi,
\end{split}
\end{align}
respectively.
A crucial feature of this bosonic representation is that the forward scattering from density-density interactions takes a quadratic form in the bosonic field, $(\partial_x \theta)^2$.
The effect of interactions is thus incorporated in the free boson theory of the Tomonaga-Luttinger liquid Hamiltonian,
\begin{align} \label{eq:SingleLL}
H_0^\textrm{LL} = \int dx \biggl[ \frac{v}{2\pi} (\partial_x \varphi)^2 +\frac{u}{2\pi} (\partial_x \theta)^2 \biggr].
\end{align}
The ratio $u/v$ controls the forward-scattering interaction, and $u=v$ for free electrons. 

While we will mainly consider fermionic systems in this paper, many of the subsequent discussions can be similarly applied to bosonic systems. 
We therefore briefly summarize the ``bosonization'' dictionary for bosons \cite{Haldane81}. 
The effective low-energy theory of 1D interacting bosons is also commonly given by the Tomonaga-Luttinger Hamiltonian in Eq.~\eqref{eq:SingleLL}. 
The boson annihilation operator $b(x)$ and density operators $\rho(x)$ are expressed in terms of the bosonic fields as 
\begin{align} \label{eq:BosonOp}
\begin{split}
b(x) &\sim e^{i\varphi(x)}, \\
\rho(x) &\sim \brho +\frac{1}{\pi} \partial_x \theta(x) +\sum_{n \neq 0} e^{2in[\pi \brho x +\theta(x)]}, 
\end{split}
\end{align}
where $\brho$ is the average boson density. 
The last term in $\rho(x)$ represents the density fluctuations with the wave number $2\pi \brho n$, which may manifest themselves as a charge-density-wave order parameter.
We identify $\pi \brho$ with the Fermi momentum $k_F$ even for bosonic systems and use them interchangeably
in the following sections.

\subsection{Array of Luttinger liquids}

Following Refs.~\cite{Kane02,Teo14}, we consider a 2D array of electron wires in the $xy$ plane, where each wire is described by the Tomonaga-Luttinger liquid Hamiltonian in Eq.~\eqref{eq:SingleLL}. 
The wires are placed at $y=jd_0$ ($j\in\mathbb{Z}$), where $d_0$ is the spacing between adjacent wires.
We take the Landau gauge $(A_0,A_1,A_2) = (0,-By,0)$ for the magnetic field applied perpendicular to the $xy$ plane.
The electron operator on the $j$th wire is now expanded as
\begin{align} \label{eq:FermiOp}
\psi_j(x) \sim e^{i(k_F +bj)x} \psi_{R,j}(x) + e^{i(-k_F +bj)x} \psi_{L,j}(x),
\end{align}
around the Fermi points of the $j$th wire at $k=bj\pm k_F$, where $b=d_0B$ is the magnetic flux density per wire in the natural unit ($\hbar=e=c=1$).
The unit flux quantum $\phi_0$ is equal to $2\pi$.
The filling fraction $\nu$ is then given by 
\begin{align}
\nu = \frac{2\pi \brho}{b} = \frac{2k_F}{b}.
\end{align}
The right- and left-moving fermion operators are bosonized as 
\begin{align} \label{eq:ElectronOp}
\psi_{R/L,j}(x) = \frac{\kappa_j}{\sqrt{2\pi \alpha}} e^{i[\varphi_j(x) \pm \theta_j(x)]}, 
\end{align}
where the bosonic fields obey the commutation relations, 
\begin{align}
\begin{split}
[\theta_j(x), \varphi_{j'}(x')] &= i\pi \delta_{j,j'} \Theta(x-x'),\\
[\theta_j(x),\theta_{j'}(x')]&=[\varphi_j(x),\varphi_{j'}(x')]=0,
\end{split} 
\end{align}
and $\kappa_j$ is the Klein factor ensuring the anticommutation relation between fermion operators on different wires. 
We here choose $\kappa_j$ to be a Majorana fermion: $\{ \kappa_j, \kappa_{j'} \} = 2\delta_{j,j'}$. 
We take the Hamiltonian to be the sum of identical Tomonaga-Luttinger liquids over the wires, 
\begin{align}
H_0 = \int dx \sum_j \biggl[ \frac{v}{2\pi} (\partial_x \varphi_j)^2 +\frac{u}{2\pi} (\partial_x \theta_j)^2 \biggr]. 
\end{align}
We then consider an appropriate interwire interaction $H_1$ to find a desired quantum Hall state.
The interwire interactions must satisfy the charge and momentum conservations at given filling fraction $\nu$. 
Since the $\varphi_j$ and $\theta_j$ fields come with
$bjx$ and $k_Fx$, respectively, from Eqs.\ (\ref{eq:FermiOp}) and (\ref{eq:ElectronOp}), the interwire interactions of the form
\begin{align}
\exp\left[i\sum_p(n_p\varphi_{j+p}+m_p\theta_{j+p})\right]
\quad(m_p, n_p\in\mathbb{Z})
\label{eq:vertex op}
\end{align}
are allowed only if the conditions
\begin{equation}
\sum_pn_p=0,
\qquad
\nu=-\frac{2\sum_p pn_p}{\sum_p m_p},
\label{eq:conservations}
\end{equation}
are satisfied.
Furthermore, imposing that the interwire interaction in Eq.\ (\ref{eq:vertex op}) is given by a product of $\psi_{R/L,j+p}$ and $\psi_{R/L,j+p}^\dagger$ leads to another condition,
\begin{equation}
m_p=n_p \ \mathrm{mod}~2.
\end{equation}

We further introduce an external electromagnetic field $A^\ext_{\mu,j}(\tau,x)$. 
The Euclidean action for the electron wires minimally coupled to $A^\ext_{\mu,j}(\tau,x)$ is given by 
\begin{align} \label{eq:FermionPlusA}
S_0 &= \int_{\tau,x} \sum_j \Bigl\{ \psi_{R,j}^\dagger \bigl[ \partial_\tau -iA^\ext_{0,j} -iv(\partial_x -iA^\ext_{1,j}) \bigr] \psi_{R,j} \nonumber \\
&\ \ \ +\psi_{L,j}^\dagger \bigl[ \partial_\tau -iA^\ext_{0,j} +iv(\partial_x -iA^\ext_{1,j}) \bigr] \psi_{L,j} +\cdots \Bigr\},
\end{align}
where the ellipsis contains intrawire forward scattering interactions yielding $u \neq v$, and we have used the short-hand notation $\int_{\tau,x}=\int d\tau dx$. 
In terms of the bosonic fields, the action is written as 
\begin{align} \label{eq:SLLplusA}
S_0 &= \int_{\tau,x} \sum_j \biggl[ \frac{i}{\pi} \partial_x \theta_j (\partial_\tau \varphi_j -A^\ext_{0,j}) \nonumber \\
&\ \ \ +\frac{v}{2\pi} (\partial_x \varphi_j -A^\ext_{1,j})^2 +\frac{u}{2\pi} (\partial_x \theta_j)^2 \biggr].
\end{align}
We can also use the same action to describe a coupled-wire system of charged bosons with unit charge.
For simplicity, we will take the $A^\ext_{2,j}=0$ gauge in this paper. 

\subsection{2D dictionary}

We here present a coupled-wire formulation of the flux attachment and vortex duality, introduced by Mross, Alicea, and Motrunich \cite{Mross17a}, as explicit nonlocal transformations for the bosonic fields. 
This offers a bosonization/fermionization dictionary in 2D, which helps us to gain more physical insights from the coupled-wire construction of various quantum Hall states.

\subsubsection{Flux attachment}
\label{sec:FluxAttach}

Let us consider the action \eqref{eq:SLLplusA}.
We now attach the $2\pi m$ flux to electrons to obtain composite particles, which obey fermionic (bosonic) statistics when $m$ is an even (odd) integer.
The $2\pi m$ flux attachment is performed through a nonlocal transformation, 
\begin{align} \label{eq:TransFluxAttach}
\begin{split}
\Phi^\CP_j(x) &= \varphi_j(x) +m \sum_{j' \neq j} \sgn (j'-j) \theta_{j'}(x), \\
\Theta^\CP_j(x) &= \theta_j(x).
\end{split}
\end{align}
These bosonic fields satisfy the commutation relations,
\begin{align} \label{eq:CommCP}
\begin{split}
[\Phi^\CP_j(x), \Phi^\CP_{j'}(x')] &= -i\pi m \sgn (j-j'), \\
[\Theta^\CP_j(x), \Theta^\CP_{j'}(x')] &= 0, \\
[\Theta^\CP_j(x), \Phi^\CP_{j'}(x')] &= i\pi \delta_{j,j'} \Theta (x-x'). 
\end{split}
\end{align}
Substituting these expressions into Eq.~\eqref{eq:SLLplusA} yields a highly nonlocal theory, but it can be turned into a local form by introducing an auxiliary field $a_{1,j}(x)$, 
\begin{align} \label{eq:FluxAttach_a1}
a_{1,j}(x) = m \sum_{j' \neq j} \sgn (j'-j) \partial_x \Theta^\CP_{j'}(x). 
\end{align}
We implement this constraint using a Lagrange multiplier $a_{0,j+1/2}(x)$ defined between adjacent wires, 
\begin{align} \label{eq:SLLinCP}
S_0 &= \int_{\tau,x} \sum_j \biggl[ \frac{i}{\pi} \partial_x \Theta^\CP_j (\partial_\tau \Phi^\CP_j -A^\ext_{0,j}) \nonumber \\
&\ \ \ +\frac{v}{2\pi} (\partial_x \Phi^\CP_j -a_{1,j} -A^\ext_{1,j})^2 +\frac{u}{2\pi} (\partial_x \Theta^\CP_j)^2 \nonumber \\
&\ \ \ +\frac{i}{2\pi m} \Bigl( a_{1,j} -m\sum_{j' \neq j} \sgn (j'-j) \partial_x \Theta^\CP_{j'} \Bigr) \nonumber \\
&\ \ \ \times (a_{0,j+1/2}-a_{0,j-1/2}) \biggr]. 
\end{align}
This action can be rewritten as 
\begin{align} \label{eq:SLLinCP2}
S_0 &= \int_{\tau,x} \sum_j \biggl\{ \frac{i}{\pi} \partial_x \Theta^\CP_j \Bigl[ \partial_\tau \Phi^\CP_j -\frac{1}{2} (Sa_{0,j-1/2}) -A^\ext_{0,j} \Bigr] \nonumber \\
&\ \ \ +\frac{v}{2\pi} (\partial_x \Phi^\CP_j -a_{1,j} -A^\ext_{1,j})^2 +\frac{u}{2\pi} (\partial_x \Theta^\CP_j)^2 \nonumber \\
&\ \ \ +\frac{i}{2\pi m} a_{1,j} (\Delta a_{0,j-1/2}) \biggr\}, 
\end{align}
where we have introduced the shorthand notations $(S a_k) = a_{k+1} +a_k$ and $(\Delta a_k) = a_{k+1} -a_k$. 
Notice that the Lagrange multiplier terms in Eq.~\eqref{eq:SLLinCP} generated two contributions: a temporal component of the minimal coupling between the composite-particle field $\Theta_j^\CP$ and the fictitious gauge field $a_{\mu,j}$, and a discrete analog of the Chern-Simons term $(i/4\pi m) \epsilon_{\mu \nu \lambda} a_\mu \partial_\nu a_\lambda$ in the $a_2=0$ gauge. 

The action \eqref{eq:SLLinCP2} can be understood as follows.
In Eq.~\eqref{eq:SLLinCP2}, 
The constraint \eqref{eq:FluxAttach_a1} can be recast into 
\begin{align}
(\Delta a_{1,j}) = -2\pi m \times \frac{1}{2} (j^\CP_{0,j} +j^\CP_{0,j+1}), 
\end{align}
where $j^\CP_{0,j} = (1/\pi) \partial_x \Theta^\CP_j$ is the composite-particle density, which is equal to the original electron density $j^\textrm{el}_{0,j}$ [see Eqs.~\eqref{eq:ElectronCurrent} and \eqref{eq:TransFluxAttach}]. 
This precisely implements the $2\pi m$ flux attachment in our coupled-wire system, and the composite particles will see an effective magnetic flux $\propto b_\textrm{CP} = b -2\pi m \brho$. 
This can be seen as a coupled-wire version of the familiar Chern-Simons flux attachment introduced in Ref.~\cite{SCZhang89}. 
The statistics of the composite particles is encoded in the commutation relations in Eq.~\eqref{eq:CommCP}, from which we may define local composite-particle operators by vertex operators of the bosonic fields $\Phi^\CP_j$ and $\Theta^\CP_j$, which are nonlocal in the original bosonic operators. 
Detailed discussions on the construction of such operators and their statistics are given with specific examples of coupled-wire models in the subsequent sections.

\subsubsection{Vortex duality}

Suppose that we have composite bosons after performing the above $2\pi m$ flux attachment with an odd integer $m$. 
We replace the superscript ``CP'' with ``CB'' accordingly.
Following Ref.~\cite{Mross17a}, we apply the vortex duality transformation by defining vortex fields, 
\begin{align} \label{eq:VCBfromCB}
\begin{split}
\Phi^\VCB_{j+1/2}(x) &= \sum_{j'} \sgn (j'-j-1/2) \Theta^\CB_{j'}(x), \\
\Theta^\VCB_{j+1/2}(x) &= \frac{1}{2} \!\left[\Phi^\CB_{j+1}(x) -\Phi^\CB_j(x)\right]. 
\end{split}
\end{align}
The bosonic fields are defined between adjacent wires, i.e., on dual wires. 
They satisfy the commutation relations, 
\begin{align} \label{eq:CommVCB}
\begin{split}
[\Phi^\VCB_{j+1/2}(x), \Phi^\VCB_{j'+1/2}(x')] &=0, \\
[\Theta^\VCB_{j+1/2}(x), \Theta^\VCB_{j'+1/2}(x')] &= \frac{i\pi m}{4} (\delta_{j,j'-1} -\delta_{j,j'+1}), \\
[\Theta^\VCB_{j+1/2}(x), \Phi^\VCB_{j'+1/2}(x')] &= i\pi \delta_{j,j'} \bigl[ \Theta(x-x')-1 \bigr]. 
\end{split}
\end{align}
For later convenience we give the inverse transformation, 
\begin{align} \label{eq:CBfromVCB}
\begin{split}
\Phi^\CB_j(x) &= -\sum_{j'} \sgn (j'-j+1/2) \Theta^\VCB_{j'+1/2}(x), \\
\Theta^\CB_j(x) &= -\frac{1}{2} \!\left[\Phi^\VCB_{j+1/2}(x) -\Phi^\VCB_{j-1/2}(x)\right]. 
\end{split}
\end{align}
The duality transformation maps a superfluid phase of the composite bosons driven by the interaction $\cos (\Phi^\CB_j -\Phi^\CB_{j+1})$ to a Mott insulating phase of the vortices driven by $\cos (2\Theta^\VCB_{j+1/2})$ as the standard vortex duality does. 
It also similarly maps a Mott insulating phase of the composite bosons to a superfluid phase of the vortices. 

We now apply the vortex duality transformation to the action \eqref{eq:SLLinCP2}. 
Since the transformation \eqref{eq:VCBfromCB} is nonlocal, the action in terms of the vortices also becomes nonlocal. 
A cure is to introduce another auxiliary field $\alpha_\mu$ as in Ref.~\cite{Mross17a}.
For the composite bosons coupled with the ``Chern-Simons'' field $a_\mu$, we anticipate that integrating out $a_\mu$ should yield a Chern-Simons term of $\alpha_\mu$ that relates the flux of $\alpha_\mu$ with the vortex density.
We thus define the auxiliary field $\alpha_{1,j+1/2}(x)$ by
\begin{align} \label{eq:FluxAttach_alpha1}
\alpha_{1,j+1/2}(x) = -\frac{1}{m} \sum_{j' \neq j} \sgn (j'-j) \partial_x \Theta^\VCB_{j'+1/2}(x), 
\end{align}
which implies the relation between the flux of $\alpha_\mu$ (in the $\alpha_2=0$ gauge) and the vortex density $j^\VCB_{0,j+1/2}=(1/\pi)\partial_x\Theta^\VCB_{j+1/2}$,
\begin{align}
\alpha_{1,j+1/2}-\alpha_{1,j-1/2}=\frac{2\pi}{m}\!\left(j^\VCB_{0,j+1/2}+j^\VCB_{0,j-1/2}\right).
\end{align}
The constraints Eqs.~\eqref{eq:FluxAttach_a1} and \eqref{eq:FluxAttach_alpha1} are rewritten as 
\begin{align} \label{eq:GaugeCBandVCB}
\begin{split}
a_{1,j} &= \frac{m}{2} \left(\partial_x \Phi^\VCB_{j+1/2} +\partial_x \Phi^\VCB_{j-1/2}\right), \\
\alpha_{1,j+1/2} &= \frac{1}{2m} \left(\partial_x \Phi^\CB_{j+1} +\partial_x \Phi^\CB_j\right). 
\end{split}
\end{align}
Plugging these expressions into Eq.~\eqref{eq:SLLinCP2} and introducing a Lagrange multiplier $\alpha_{0,j}(x)$, we can rewrite the theory in terms of the vortex field in a local form. 
We finally obtain 
\begin{align} \label{eq:ActionVCB}
S_0 &= \int_{\tau,x} \sum_j \biggl[ \frac{i}{\pi} \partial_x \Theta^\VCB_{j+1/2} \Bigl( \partial_\tau \Phi^\VCB_{j+1/2} -\frac{1}{2} (S\alpha_{0,j}) \Bigr) \nonumber \\
&\ \ \ +\frac{m^2 v}{2\pi} (\partial_x \Phi^\VCB_{j+1/2} -\alpha_{1,j+1/2})^2 +\frac{v}{2\pi} (\partial_x \Theta^\VCB_{j+1/2})^2 \nonumber \\
&\ \ \ +\frac{u-m^2 v}{8\pi} (\Delta \partial_x \Phi^\VCB_{j-1/2})^2 -i\frac{m}{2\pi} \alpha_{1,j+1/2} (\Delta \alpha_{0,j}) \nonumber \\
&\ \ \ -\frac{v}{2\pi} \partial_x \Theta^\VCB_{j+1/2} (\Delta A^\ext_{1,j}) -\frac{i}{4\pi} (SA^\ext_{1,j}) (\Delta \alpha_{0,j}) \nonumber \\
&\ \ \ +\frac{i}{2\pi} A^\ext_{0,j} (\Delta \alpha_{1,j-1/2}) +\cdots \biggr], 
\end{align}
where the ellipsis contains terms involving the second derivative of $A^\ext_\mu$. 
The derivation is outlined in Appendix~\ref{app:VortexAction}.
In this vortex theory, the vortices are minimally coupled to the gauge field $\alpha_\mu$ through the discretized level-$m$ Chern-Simons term with $\alpha_2=0$. 
The external electromagnetic field $A^\ext_\mu$ is coupled to the $2\pi$ flux of $\alpha_\mu$ through a discrete version of the mutual Chern-Simons term $(-i/2\pi) \epsilon_{\mu \nu \lambda} A^\ext_\mu \partial_\nu \alpha_\lambda$ in the $A^\ext_2 = \alpha_2 =0$ gauge. 
On the other hand, the external magnetic field $(\Delta A^\ext_{1,j})$ is coupled to the vortex density in the form of a chemical potential. 
This physically means that the applied magnetic field dopes vortex excitations (or excites quasiparticles) on top of the condensate of composite bosons.
This will be highlighted in the Haldane-Halperin picture for hierarchy states discussed in Sec.~\ref{sec:HaldaneHalperin}. 
We summarize several field variables in Table~\ref{tab:Fields} that have been introduced in this section and will be used in the following discussion.
%%%%%%%%%%%%%%%%%%%%%%%%%%%%%%%%%%
\begin{table*}
\caption{List of field variables used in this paper. 
Their physical meanings and the equations in which they are defined are given.}
\label{tab:Fields}
\begin{ruledtabular}
\begin{tabular}{lll}
Symbol & Physical meaning & Definition \\ \hline
$\varphi_j$, $\theta_j$ & Bosonic fields for original particles & Eq.~\eqref{eq:ChiralFermion} for fermions, Eq.~\eqref{eq:BosonOp} for bosons\\
$\Phi^\CF_j$, $\Theta^\CF_j$ & Bosonic fields for composite fermions & Eq.~\eqref{eq:TransFluxAttach} with even (odd) $m$ for fermions (bosons) \\
$\psi^\CF_{R,j}$, $\psi^\CF_{L,j}$ & Composite fermion fields & Eq.~\eqref{eq:CFField} \\
$\Phi^\CB_j$, $\Theta^\CB_j$ & Bosonic fields for composite bosons & Eq.~\eqref{eq:TransFluxAttach} with odd (even) $m$ for fermions (bosons) \\
$\Phi^\VCB_{j+1/2}$, $\Theta^\VCB_{j+1/2}$ & Bosonic fields for vortices (quasiparticles) & Eq.~\eqref{eq:VCBfromCB} \\ 
$\varphi^\hole_{j+1/2}$, $\theta^\hole_{j+1/2}$ & Bosonic fields for holes & Eq.~\eqref{eq:HoleField} or \eqref{eq:HoleFromElectron} \\
\hline
$A^\ext_{0,j}$, $A^\ext_{1,j}$ & External electromagnetic gauge fields & Action in Eq.~\eqref{eq:FermionPlusA} or \eqref{eq:SLLplusA} \\
$a_{0,j+1/2}$, $a_{1,j}$ & Gauge fields coupled to composite particles & $a_{1,j}$ from Eq.~\eqref{eq:FluxAttach_a1}, $a_{0,j+1/2}$ are Lagrange multipliers \\
$\alpha_{0,j}$, $\alpha_{1,j+1/2}$ & Gauge fields coupled to vortices & $\alpha_{1,j+1/2}$ from Eq.~\eqref{eq:FluxAttach_alpha1}, $\alpha_{0,j}$ are Lagrange multipliers
\end{tabular}
\end{ruledtabular}
\end{table*}
%%%%%%%%%%%%%%%%%%%%%%%%%%%%%%%%%%%

\section{Abelian hierarchy}
\label{sec:AbelianHierarchy}

In this section, we apply the flux attachment and vortex duality transformations introduced above to the coupled-wire models of Abelian hierarchy states. 
We first review the coupled-wire model of the Laughlin states at $\nu=1/q$ \cite{Kane02,Teo14} and then show that it admits both interpretations of the Laughlin states as a filled lowest Landau level of composite fermions or a condensate of composite bosons. 
We extend the discussion to hierarchy states obtained from a parent Laughlin state, whose coupled-wire models are shown to admit the Jain and/or Haldane-Halperin interpretations. 
At the end of this section we introduce a PH conjugate state as a special case of the Haldane-Halperin hierarchy state, and propose a simple method to construct coupled-wire models for FQH states at higher Landau levels.

\subsection{Laughlin states} \label{sec:Laughlin}

\begin{figure}
\includegraphics[clip,width=0.45\textwidth]{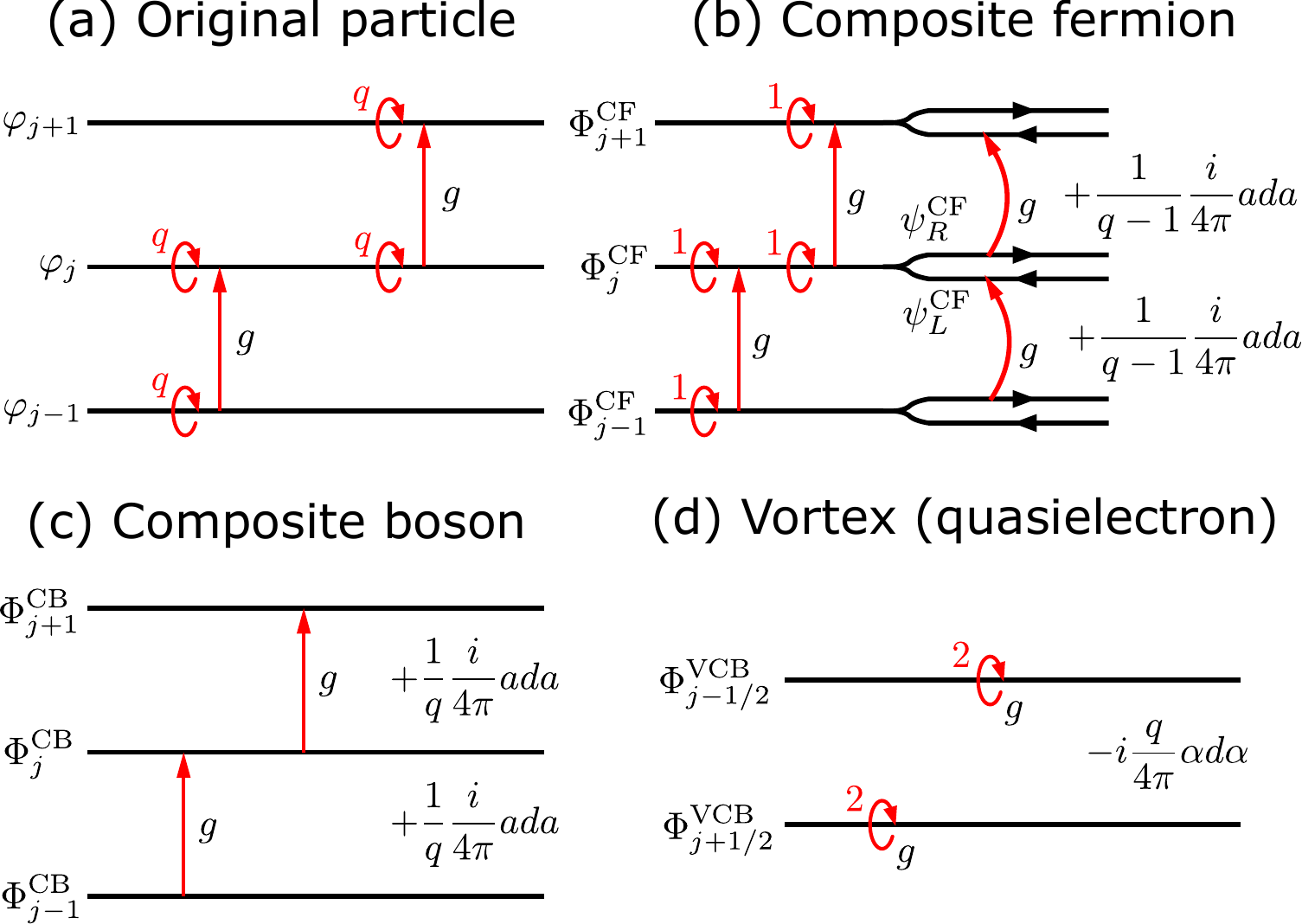}
\caption{Tunneling Hamiltonian for the $\nu=1/q$ Laughlin state in terms of (a) the original particles, (b) composite fermions, (c) composite bosons, and (d) vortices from the composite bosons. 
The horizontal lines represent wires where the bosonic fields are defined. 
The red straight arrows represent single-particle hopping involving the $\varphi_j$ field. 
The red curly arrows represents charge density fluctuations involving $\theta_j$ with the wave number $k_F$ multiplied by the factor indicated in red.
For the composite fermions, the black horizontal arrows and the red curved arrows, respectively, represent chiral fermion modes $\psi^\textrm{CF}_{R/L}$ and single-fermion hopping terms.
The associated Chern-Simons couplings are also shown.}
\label{fig:Laughlin}
\end{figure}

At filling fraction $\nu=1/q$ with odd (even) $q$, fermions (bosons) can form the celebrated Laughlin state. 
We review the coupled-wire construction of the Laughlin state, which has been proposed in Ref.~\cite{Kane02} and extensively analyzed in Ref.~\cite{Teo14}. 
Suppose that the kinetic action is given in the form of Eq.~\eqref{eq:SLLplusA} for each wire. 
The interwire tunneling interaction for the $\nu=1/q$ Laughlin state is given by 
\begin{align} \label{eq:TunnelingLaughlin}
H_1 = g \int_x \sum_j \kappa_j \kappa_{j+1} e^{i(\varphi_j +q\theta_j -\varphi_{j+1} +q\theta_{j+1})} +\textrm{H.c.} 
\end{align}
To treat the bosonic and fermionic Laughlin states on equal footing, we choose the factor $\kappa_j$ to be a Majorana fermion for odd $q$ and to be the identity operator for even $q$. 
It is important to note that the interwire interaction \eqref{eq:TunnelingLaughlin} is built out of local electron operators \eqref{eq:FermiOp} for odd $q$ or local boson operators \eqref{eq:BosonOp} for even $q$. 
The tunneling Hamiltonian is schematically shown in Fig.~\ref{fig:Laughlin}(a).
%%%%%%%%%%%%%%%%%%%%%%%%%%%%%%%%%%%%%
%%%%%%%%%%%%%%%%%%%%%%%%%%%%%%%%%%%%%
This tunneling process picks up oscillation factors $e^{-ibx}$ from the hopping of a particle in the applied magnetic field and $e^{i2qk_Fx}$ from density fluctuations, which are precisely canceled at $\nu=1/q$.

The chiral bosonic fields defined by 
\begin{align}
\begin{split}
\tphi_{R,j} &= \varphi_j +q\theta_j, \\
\tphi_{L,j} &= \varphi_j -q\theta_j,
\end{split}
\end{align}
satisfy the commutation relations, 
\begin{align} \label{eq:LaughlinComm}
[\partial_x \tphi_{r,j}(x), \tphi_{r',j'}(x')] = 2i\pi rq \delta_{r,r'} \delta_{j,j'} \delta (x-x'), 
\end{align}
where $r=R/L=+/-$. 
The tunneling term is then written as 
\begin{align}
H_1 = g \int_x \sum_j \kappa_j \kappa_{j+1} e^{i(\tphi_{R,j} -\tphi_{L,j+1})} +\textrm{H.c.} 
\end{align}
Assuming that the coupling constant $g$ flows to the strong-coupling limit, we see that the tunneling term opens a gap in the bulk and leaves unpaired gapless chiral modes at the outermost wires, which correspond to the boundaries of the FQH system.
The commutation relations of the edge modes are given by Eq.~\eqref{eq:LaughlinComm} and consistent with those obtained from the Chern-Simons theory \cite{XGWen95}, 
\begin{align}
\calL = -i\frac{q}{4\pi} \epsilon_{\mu \nu \lambda} \alpha_\mu \partial_\nu \alpha_\lambda.
\end{align}
We show in Sec.~\ref{sec:LaughlinCB} that a discrete analog of this Chern-Simons theory is obtained from the condensation of composite bosons.

We remark that the tunneling term \eqref{eq:TunnelingLaughlin} is actually irrelevant in the renormalization group (RG) sense at the fixed point of decoupled Luttinger liquids \eqref{eq:SLLplusA}, because the tunneling $g$ has scaling dimension $1/(2K)+q^2K/2 \geq q$, where $K$ is the Luttinger parameter defined by $K=\sqrt{v/u}$,
However, the tunneling becomes relevant in the presence of additional interwire forward scattering interactions. 
To see this, let us define fields on dual wires by 
\begin{align}
\begin{split}
\tvarphi_{j+1/2} &= \frac{1}{2}(\tphi_{R,j}+\tphi_{L,j+1}), \\
\ttheta_{j+1/2} &= \frac{1}{2}(\tphi_{R,j} -\tphi_{L,j+1}), 
\end{split}
\end{align}
which satisfy the commutation relations,
\begin{align}
[\partial_x \ttheta_{j+1/2}(x), \tvarphi_{j'+1/2}(x')] = i\pi q \delta_{j,j'} \delta (x-x').
\end{align} 
The particle density on the dual wire at $y=j+1/2$ can be defined by $\trho_{j+1/2}=(1/\pi q) \partial_x \ttheta_{j+1/2}$, as the unit-charge particle operators $e^{i\tphi_{R,j}(x)}$ or $e^{i\tphi_{L,j+1}(x)}$ create a kink of the height $-\pi q$ in $\tilde\theta_{j+1/2}$.
If the coupled-wire Hamiltonian of each wire has the form 
\begin{align}
\tilde{H}_0 = \frac{\tv}{2\pi} \int_x \sum_j \Bigl[ \tK (\partial_x \tvarphi_{j+1/2})^2 +\frac{1}{\tK} (\partial_x \ttheta_{j+1/2})^2 \Bigr], 
\end{align}
then the tunneling term $g \cos (2\ttheta_{j+1/2})$ has the scaling dimension $q \tK$ and becomes relevant for $\tK<2/q$ \cite{Teo14}.  
Thus, we expect that the Laughlin state should be stabilized by adding to the action \eqref{eq:SLLplusA} the interwire forward scattering interactions, 
\begin{align} \label{eq:InterForward}
& H^\textrm{inter-forward}_0 = \frac{w}{2\pi} \int dx \sum_j (\partial_x \ttheta_{j+1/2})^2 \nonumber \\
&= \frac{w}{8\pi} \int_x \sum_j \left[\partial_x (\varphi_j +q \theta_j - \varphi_{j+1} +q \theta_{j+1})\right]^2, 
\end{align}
with $w>0$.
In the following analysis, we assume the sliding Luttinger liquid action $S_0 +\int_\tau H^\textrm{inter-forward}_0$.

A pair of a quasiparticle and its antiparticle is created by the bare $2k_F$ backscattering operator \cite{Teo14}, 
\begin{align} \label{eq:2kFBackscattering}
e^{i2\theta_j} = e^{i(\tphi_{R,j} -\tphi_{L,j})/q}. 
\end{align}
As this operator creates $\mp\pi$-kinks in $\ttheta_{j \pm 1/2}$, it hops a quasiparticle with charge $-1/q$ from the dual wire $j+1/2$ to $j-1/2$.
The quasiparticle excitations are deconfined not only along the wires (as typical in 1D systems) but also across the wires. 
This can be understood by considering a string of the backscattering operators $e^{i2\theta_j} e^{i2\theta_{j+1}} \cdots e^{i2\theta_{j+k}}$ with length $k$. 
The bosonic fields inside the string acquire a finite expectation value $e^{i(2/q) \langle \ttheta_{j+1/2} \rangle}$ when acting on the ground state and thus are replaced by a constant. 
As a result the string of the backscattering operators leaves quasiparticle excitations with charge $\pm 1/q$ at the ends of the string, which are separated by $k$ wires away from each other. 
Furthermore, one can transfer a quasiparticle along the dual wire $j+1/2$ from $x_1$ to $x_2$ using a string operator $\exp(\frac{i}{q} \int^{x_2}_{x_1} dx \partial_x \tphi_{R,j})$.
With these string operators, we can create quasiparticle excitations deconfined in the full 2D space. 
We can then compute the braiding statistics of quasiparticles \cite{Teo14} or the ground-state degeneracy on a torus \cite{Sagi15} using the string operators.

\subsubsection{Composite fermion picture}
\label{sec:LaughlinJain}

We here perform the $2\pi (q-1)$ flux attachment to bosons (fermions) for even (odd) $q$, and convert them to composite fermions. 
As proposed by Jain \cite{Jain89,Jain90}, the Laughlin $\nu=1/q$ state is understood as the state corresponding to the filled lowest Landau level of the composite fermions at the effective filling fraction $\nu_\CF=1$. 
We now implement the flux attachment, as discussed in Sec.~\ref{sec:FluxAttach}, through the transformation \eqref{eq:TransFluxAttach}, 
\begin{align} \label{eq:FluxAttach(q-1)}
\begin{split}
\Phi^\CF_j = \varphi_j +(q-1) \sum_{j' \neq j} \sgn (j'-j) \theta_{j'}, \ \ \ 
\Theta^\CF_j = \theta_j. 
\end{split}
\end{align}
In terms of these bosonic fields, the kinetic action \eqref{eq:SLLplusA} with the forward scattering \eqref{eq:InterForward} is given by 
\begin{align} \label{eq:SLLActionCF}
S_0 &= \int_{\tau,x} \sum_j \biggl[ \frac{i}{\pi} \partial_x \Theta^\CF_j \Bigl( \partial_\tau \Phi^\CF_j -\frac{1}{2}(Sa_{0,j-1/2}) -A^\ext_{0,j} \Bigr) \nonumber \\
&\ \ \ +\frac{v}{2\pi} (\partial_x \Phi^\CF_j -a_{1,j} -A^\ext_{1,j})^2 +\frac{u}{2\pi} (\partial_x \Theta^\CF_j)^2 \nonumber \\
&\ \ \ +\frac{w}{8\pi} (\partial_x \Phi^\CF_j +\partial_x \Theta^\CF_j -\partial_x \Phi^\CF_{j+1} +\partial_x \Theta^\CF_{j+1})^2 \nonumber \\
&\ \ \ +\frac{i}{2\pi(q-1)} a_{1,j} (\Delta a_{0,j-1/2}) \biggr], 
\end{align}
which is written in the local form by introducing the gauge field $a_\mu$. 
The tunneling term \eqref{eq:TunnelingLaughlin} takes the form 
\begin{align}
H_1 = g \int_x \sum_j \kappa_j \kappa_{j+1} e^{i(\Phi^\CF_j +\Theta^\CF_j -\Phi^\CF_{j+1} +\Theta^\CF_{j+1})} +\textrm{H.c.}, 
\end{align}
which is schematically shown in Fig.~\ref{fig:Laughlin}~(b). 
We then define the composite fermion fields by 
\begin{align} \label{eq:CFField}
\psi^\CF_{R/L,j}(x) \sim \frac{\kappa_j}{\sqrt{2\pi \alpha}} e^{i\Phi^\CF_j(x) \pm i\Theta^\CF_j(x)}. 
\end{align}
Recall that $\kappa_j$ is a Majorana fermion for odd $q$, while $\kappa_j=1$ for even $q$. 
This ensures the anticommuting property of $\psi^\CF_{R/L,j}$ from the commutation relations \eqref{eq:CommCP} with $m=q-1$. 
With these fermionic fields, the action can be written as 
\begin{align}
\label{eq:ActionLaughlinCF}
S_0 &= \int_{\tau,x} \sum_j \biggl[ \sum_{r=\pm} \psi^{\CF \dagger}_{r,j} \Bigl( \partial_\tau -\frac{i}{2} (Sa_{0,j-1/2}) -iA^\ext_{0,j} \Bigr) \psi^\CF_{r,j} \nonumber \\
&\ \ \ -\sum_{r=\pm} irv \psi^{\CF \dagger}_{r,j} (\partial_x -a_{1,j} -A^\ext_{1,j}) \psi^\CF_{r,j} \nonumber \\
&\ \ \ +\frac{i}{2\pi (q-1)} a_{1,j} (\Delta a_{0,j-1/2}) +\cdots \biggr], \\
\label{eq:TunnelingLaughlinCF}
H_1 &= 2\pi \alpha g \int dx \sum_j e^{i\pi(q-1)/2} \psi^{\CF \dagger}_{R,j} \psi^\CF_{L,j+1} +\textrm{H.c.}
\end{align}
Here, the ellipsis in $S_0$ contains four-fermion forward scattering terms of the composite fermions and the subscript $r=R/L=+/-$. 
This theory may be seen as a discrete version of the fermion Chern-Simons theory \cite{Lopez91}. 
We note that at the naive mean-field level with $a_\mu = \langle a_\mu \rangle$, the tunneling term \eqref{eq:TunnelingLaughlinCF} becomes a mass term of the composite fermions and appears to be relevant in the RG sense. 
However, this is not quite correct as discussed above, and the actual RG flow is controlled by the forward scattering interactions.
If the coupling constant $g$ flows to the strong-coupling fixed point, the bulk is gapped while a single gapless chiral fermion mode remains at the boundary. 
Thus, the composite fermions form the IQH state with Chern number $C=1$, which may be thought of as the filled lowest Landau level of the composite fermions. 
This illustrates the Jain picture of the Laughlin state in the coupled-wire approach. 

\subsubsection{Composite boson picture}
\label{sec:LaughlinCB}

We now attach $2\pi q$ flux to bosons (fermions) for even (odd) $q$ and convert their statistics to bosonic. 
In this composite boson picture, the Laughlin $\nu=1/q$ state is interpreted as a condensate of the composite bosons since they now see a zero magnetic field on average, as described by the Ginzburg-Landau theory for the FQH states \cite{Girvin87,Read89,SCZhang89}.
The flux attachment is again achieved by the nonlocal transformation \eqref{eq:TransFluxAttach},
\begin{align}
\Phi^\CB_j = \varphi_j +q\sum_{j' \neq j} \sgn(j'-j) \theta_{j'}, \ \ \ \Theta^\CB_j = \theta_j. 
\end{align}
We then find the local kinetic action after introduction of the Chern-Simons gauge field $a_\mu$, 
\begin{align} \label{eq:SLLActionCB}
S_0 &= \int_{\tau,x} \sum_j \biggl[ \frac{i}{\pi} \partial_x \Theta^\CB_j \Bigl( \partial_\tau \Phi^\CB_j -\frac{1}{2}(Sa_{0,j-1/2}) -A^\ext_{0,j} \Bigr) \nonumber \\
&\ \ \ +\frac{v}{2\pi} (\partial_x \Phi^\CB_j -a_{1,j} -A^\ext_{1,j})^2 +\frac{u}{2\pi} (\partial_x \Theta^\CB_j)^2 \nonumber \\
&\ \ \ +\frac{w}{8\pi} (\Delta \partial_x \Phi^\CB_j)^2 +\frac{i}{2\pi q} a_{1,j} (\Delta a_{0,j-1/2}) \biggr].
\end{align}
The tunneling term \eqref{eq:TunnelingLaughlin} becomes 
\begin{align} \label{eq:TunnelingLaughlinCB}
H_1 = g \int dx \sum_j \kappa_j \kappa_{j+1} e^{i(\Phi^\CB_j -\Phi^\CB_{j+1})} +\textrm{H.c.} 
\end{align}
The commutation relations in Eq.~\eqref{eq:CommCP} with $m=q$ ensure that the operators $\kappa_j e^{i\Phi^\CB_j}$ and $e^{i2\Theta^\CB_j}$ are bosonic. 
Thus the tunneling term is a hopping of the composite bosons between neighboring wires [Fig.~\ref{fig:Laughlin}~(c)].
When the coupling constant $g$ flows to the strong-coupling limit, the interaction \eqref{eq:TunnelingLaughlinCB} leads to condensation of the composite bosons. 
As the composite bosons are not local objects in terms of microscopic variables, this boson condensate does not host gapless Goldstone modes.
Instead, they are Higgsed by the Chern-Simons gauge field $a_\mu$ and become massive excitations. 

We now switch to the dual vortex picture \cite{XGWen90a,DHLee89,XGWen90b}, which enables us to deduce an effective hydrodynamic description of the Laughlin state in terms of the Chern-Simons theory. 
The vortex duality transformation in the coupled wires is performed via the nonlocal transformation in Eq.~\eqref{eq:VCBfromCB}. 
In terms of the vortices, the kinetic action \eqref{eq:SLLActionCB} is given by
\begin{align} \label{eq:ActionLaughlinVCB}
S_0 &= \int_{\tau,x} \sum_j \biggl[ \frac{i}{\pi} \partial_x \Theta^\VCB_{j+1/2} \Bigl( \partial_\tau \Phi^\VCB_{j+1/2} -\frac{1}{2} (S\alpha_{0,j}) \Bigr) \nonumber \\
&\ \ \ +\frac{q^2 v}{2\pi} (\partial_x \Phi^\VCB_{j+1/2} -\alpha_{1,j+1/2})^2 +\frac{v+w}{2\pi} (\partial_x \Theta^\VCB_{j+1/2})^2 \nonumber \\
&\ \ \ +\frac{u-q^2 v}{8\pi} (\Delta \partial_x \Phi^\VCB_{j-1/2})^2 -i\frac{q}{2\pi} \alpha_{1,j+1/2} (\Delta \alpha_{0,j}) \nonumber \\
&\ \ \ -\frac{v}{2\pi} \partial_x \Theta^\VCB_{j+1/2} (\Delta A^\ext_{1,j}) -\frac{i}{4\pi} (SA^\ext_{1,j}) (\Delta \alpha_{0,j}) \nonumber \\
&\ \ \ +\frac{i}{2\pi} A^\ext_{0,j} (\Delta \alpha_{1,j-1/2}) +\cdots \biggr].
\end{align}
The tunneling term \eqref{eq:TunnelingLaughlinCB} is written as 
\begin{align}
H_1 = g \int dx \sum_j \kappa_j \kappa_{j+1} e^{-i2\Theta^\VCB_{j+1/2}} +\textrm{H.c.},
\end{align}
which pins $\Theta^\VCB$ when this term is relevant.
The condensate of the composite bosons is now seen as a Mott insulator of the vortices coupled to the gauge field $\alpha_\mu$ with the level-$q$ Chern-Simons term [Fig.~\ref{fig:Laughlin}~(d)]. 
The interwire forward scattering interaction \eqref{eq:InterForward} gives rise to a repulsive interaction between the vortices and therefore enhances an instability towards a Mott insulator of the vortices. 
According to the commutation relations in Eq.~\eqref{eq:CommVCB} with $m=q$, the operators $e^{i\Phi^\VCB_{j+1/2}}$ and $\kappa_j \kappa_{j+1} e^{i2\Theta^\VCB_{j+1/2}}$ behave as bosonic operators. 

A physical meaning of the vortex operator $e^{i\Phi^\VCB_{j+1/2}}$ is to create a \emph{single} gapped quasiparticle excitation on the dual wire $j+1/2$. 
This can be seen by writing it in terms of the bosonic fields on dual wires (links), 
\begin{align}
e^{i\Phi^\VCB_{j+1/2}(x)} &\propto \cdots e^{-i\theta_{j-1}(x)} e^{-i\theta_j(x)} e^{i\theta_{j+1}(x)} e^{i\theta_{j+2}(x)} \cdots \nonumber \\
&= \cdots e^{-\frac{i}{q} \ttheta_{j-1/2}(x)} e^{-\frac{i}{q} \tvarphi_{j+1/2}(x)} e^{\frac{i}{q} \ttheta_{j+3/2}(x)} \cdots.
\end{align} 
The operator $e^{-\frac{i}{q} \tvarphi_{j+1/2}(x)}$ creates a $\pi$-kink in $\ttheta_{j+1/2}(x)$, while the string of $e^{\pm \frac{i}{q} \ttheta_{j+1/2}(x)}$ trivially acts on the ground state where $\tilde\theta_{j+1/2}(x)$ are pinned.
Since $\partial_x \Phi^\VCB_{j+1/2} -\alpha_{1,j+1/2} = -(1/q) \partial_x \tvarphi_{j+1/2}$, the above vortex operator can be rewritten as
\begin{align}
e^{i\Phi^\VCB_{j+1/2}(x)} \propto \exp\!\left(-\frac{i}{q} \tvarphi_{j+1/2}(x) +i\int^x_{-\infty}\! dx' \alpha_{1,j+1/2}(x')\right) \! ,
\end{align}
which shows that the vortex operator $e^{i\Phi^\VCB_{j+1/2}}$ is the $\pi$-kink operator $e^{-\frac{i}{q} \tvarphi_{j+1/2}}$ with a Dirac string of the gauge field $\alpha_\mu$ inserted from an infinitely distant point.
The vortex operator $e^{i\Phi^\VCB_{j+1/2}(x)}$ thus creates a single \emph{quasielectron} with charge $-1/q$ at $x$ on the dual wire $j+1/2$.
In a similar way, the antivortex operator $e^{-i\Phi^\VCB_{j+1/2}}$ may be seen as an operator creating a single \emph{quasihole} with charge $1/q$. 
Such quasiparticle operators are by no means local operators in terms of the original particles, since any local operator must create quasiparticles in pairs as in Eq.~\eqref{eq:2kFBackscattering}. 
Finally, neglecting the second and higher derivative terms, we can regard the action \eqref{eq:ActionLaughlinVCB} as a discrete analog of the effective Chern-Simons theory \cite{XGWen95},
\begin{align}
\calL = -i\frac{q}{4\pi} \epsilon_{\mu \nu \lambda} \alpha_\mu \partial_\nu \alpha_\lambda -\frac{i}{2\pi} \epsilon_{\mu \nu \lambda} A^\ext_\mu \partial_\nu \alpha_\lambda +ij^\textrm{QP}_\mu \alpha_\mu, 
\end{align}
in the $\alpha_2 = A^\ext_2 =0$ gauge, where the quasiparticle current is given by $j^\textrm{QP}_{0,j+1/2} = -(1/\pi) \partial_x \Theta^\VCB_{j+1/2}$ and $j^\textrm{QP}_{1,j+1/2} = -(q^2 v/\pi) \partial_x \Phi^\VCB_{j+1/2}$ for the fundamental (smallest charge) quasielectron. 
We note that there have been several attempts to obtain the effective Chern-Simons theory from coupled wires from different perspectives \cite{RASantos15,Imamura16}.

\subsection{Hierarchy states}
\label{sec:Hierarchy}

After warming up with the Laughlin states, we are now ready to consider hierarchy states. 
We focus, in particular, on the hierarchy states at $\nu=2/5$ and $\nu=2/7$, which are in the Haldane-Halperin hierarchy obtained by condensation of quasielectrons or quasiholes of the Laughlin $\nu=1/3$ state, respectively \cite{Haldane83,Halperin84}. 
They also appear in the Jain hierarchy as the $\nu=2$ IQH states of composite fermions \cite{Jain89,Jain90}. 
It has been shown that these apparently different approaches lead to FQH states that belong to the same universality class \cite{Read90,Blok90a,Blok90b}, i.e., they are described by the same Chern-Simons theory \cite{XGWen92a,XGWen95}. 
We here show that the coupled-wire approach is also capable of unifying the Jain and Haldane-Halperin hierarchies at the corresponding filling fractions in terms of the coupled-wire Hamiltonian.

Let us first briefly review the construction of the first-level hierarchy states proposed in Ref.~\cite{Teo14}. 
We suppose that the action of decoupled wires takes the form of Eq.~\eqref{eq:SLLplusA}.
The coupled-wire Hamiltonian for the first-level hierarchy states involves tunneling of particles between second-neighbor wires. 
To obtain the hierarchy state at the filling fraction $\nu=2n/(m_0+m_1)$, Teo and Kane proposed the tunneling Hamiltonian \cite{Teo14},
\begin{align} \label{eq:Tunneling_qqp}
H_1 &= g \int_x \sum_j \kappa^n_j \kappa^n_{j+2}
% \nonumber \\
%&\ \ \ \times
\exp\!\left[i(n\varphi_j +m_0 \theta_j +2m_1\theta_{j+1}
\right.\nonumber\\
&\left.\hspace*{35mm} -n\varphi_{j+2} +m_0 \theta_{j+2})\right] +\textrm{H.c.} 
\end{align}
For electronic (bosonic) FQH hierarchy states, this interaction is built from local electron operators with Majorana fermions $\kappa_j$ and an even integer $n+m_1$ (from local boson operators with $\kappa_j=1$ and an even integer $m_1$). 
In order to see that this tunneling term produces the correct edge physics, we group every two successive wires and  define the bosonic fields, 
\begin{align} \label{eq:ChiralField_qqp}
\begin{split}
\tphi^1_{R,l} &= n\varphi_{2l} +m_0 \theta_{2l} +2m_1 \theta_{2l+1}, \\
\tphi^2_{R,l} &= n\varphi_{2l+1} +m_0 \theta_{2l+1}, \\
\tphi^1_{L,l} &= n\varphi_{2l} -m_0 \theta_{2l}, \\
\tphi^2_{L,l} &= n\varphi_{2l+1} -m_0 \theta_{2l+1} -2m_1 \theta_{2l},
\end{split}
\end{align}red{which} satisfy the commutation relations, 
\begin{align} \label{eq:Comm_qqp}
[\partial_x \tphi^I_{r,l}(x), \tphi^J_{r',l'}(x')] = 2i\pi r K_{IJ} \delta_{r,r'} \delta_{l,l'} \delta (x-x'), 
\end{align}
with the $K$ matrix, 
\begin{align} \label{eq:Kmat_qqp}
\bfK = n\begin{pmatrix} m_0 & m_1 \\ m_1 & m_0 \end{pmatrix}. 
\end{align}
The tunneling Hamiltonian \eqref{eq:Tunneling_qqp} is then written as  
\begin{align}
H_1 &= g \int_x \sum_l \Bigl[ \kappa^n_{2l} \kappa^n_{2l+2} e^{i(\tphi^1_{R,l} -\tphi^1_{L,l+1})} \nonumber \\
&\qquad\qquad + \kappa^n_{2l+1} \kappa^n_{2l+3} e^{i(\tphi^2_{R,l} -\tphi^2_{L,l+1})} +\textrm{H.c.} \Bigr]. 
\end{align}
When the coupling constant $g$ flows to the strong-coupling limit, it opens a bulk gap while there remain two uncoupled gapless modes at the boundaries. 
These gapless modes satisfy the commutation relations \eqref{eq:Comm_qqp}, which are exactly the same as those derived from the two-component Chern-Simons theory \cite{XGWen95},
\begin{align} \label{eq:MultiCompCSAction}
\calL = -\frac{i}{4\pi} \sum_{I,J} K_{IJ} \epsilon_{\mu \nu \lambda} \alpha^I_\mu \partial_\nu \alpha^J_\lambda -\frac{i}{2\pi} \sum_I t_I \epsilon_{\mu \nu \lambda} A^\ext_\mu \partial_\nu \alpha^I_\lambda, 
\end{align}
with the $K$ matrix given in Eq.~\eqref{eq:Kmat_qqp} in the basis of charge vector $t=(n,n)$. 
Similarly to the case of the Laughlin states discussed above, we should add an interwire forward scattering interaction,
\begin{align} \label{eq:InterForward_qqp}
H^\textrm{inter-forward}_0 = \frac{w}{8\pi} \int_x \sum_l \sum_{I=1,2} (\partial_x \tphi^I_{R,l} -\partial_x \tphi^I_{L,l+1})^2,
\end{align}
to make the coupling constant $g$ relevant. 
In the following discussion, this term is assumed to be added to the decoupled-wire action \eqref{eq:SLLplusA}. 

The $\nu=2/5$ state corresponds to the $K$ matrix \eqref{eq:Kmat_qqp} with $(n,m_0,m_1)=(1,3,2)$, while the $\nu=2/7$ state corresponds to $(n,m_0,m_1)=(1,3,4)$. 
The above $K$ matrix \eqref{eq:Kmat_qqp} is given in the multilayer basis with $t=(n,n)$ as the bosonic fields \eqref{eq:ChiralField_qqp} carry charge $n$.
On the other hand, the hierarchical construction naturally gives the Chern-Simons theory in the hierarchical basis with charge vector $t=(1,0)$ \cite{XGWen95}. 
The two bases can be transformed to each other by a $GL(2,\mathbb{Z})$ transformation with the determinant $\pm n$. 
In the hierarchical basis, the $K$ matrix for the $\nu=2/5$ state is given by 
\begin{align} \label{eq:Kmat332h}
\bfK = \begin{pmatrix} 3 & -1 \\ -1 & 2 \end{pmatrix}, 
\end{align}
while the one for the $\nu=2/7$ state is 
\begin{align} \label{eq:Kmat334h}
\bfK = \begin{pmatrix} 3 & 1 \\ 1 & -2 \end{pmatrix}.
\end{align}
The corresponding Chern-Simons theories can be obtained from the composite boson approach to the coupled-wire model, as we will demonstrate below. 

\subsubsection{Composite fermion: Jain hierarchy}
\label{sec:JainPicture}

\begin{figure}
\includegraphics[clip,width=0.45\textwidth]{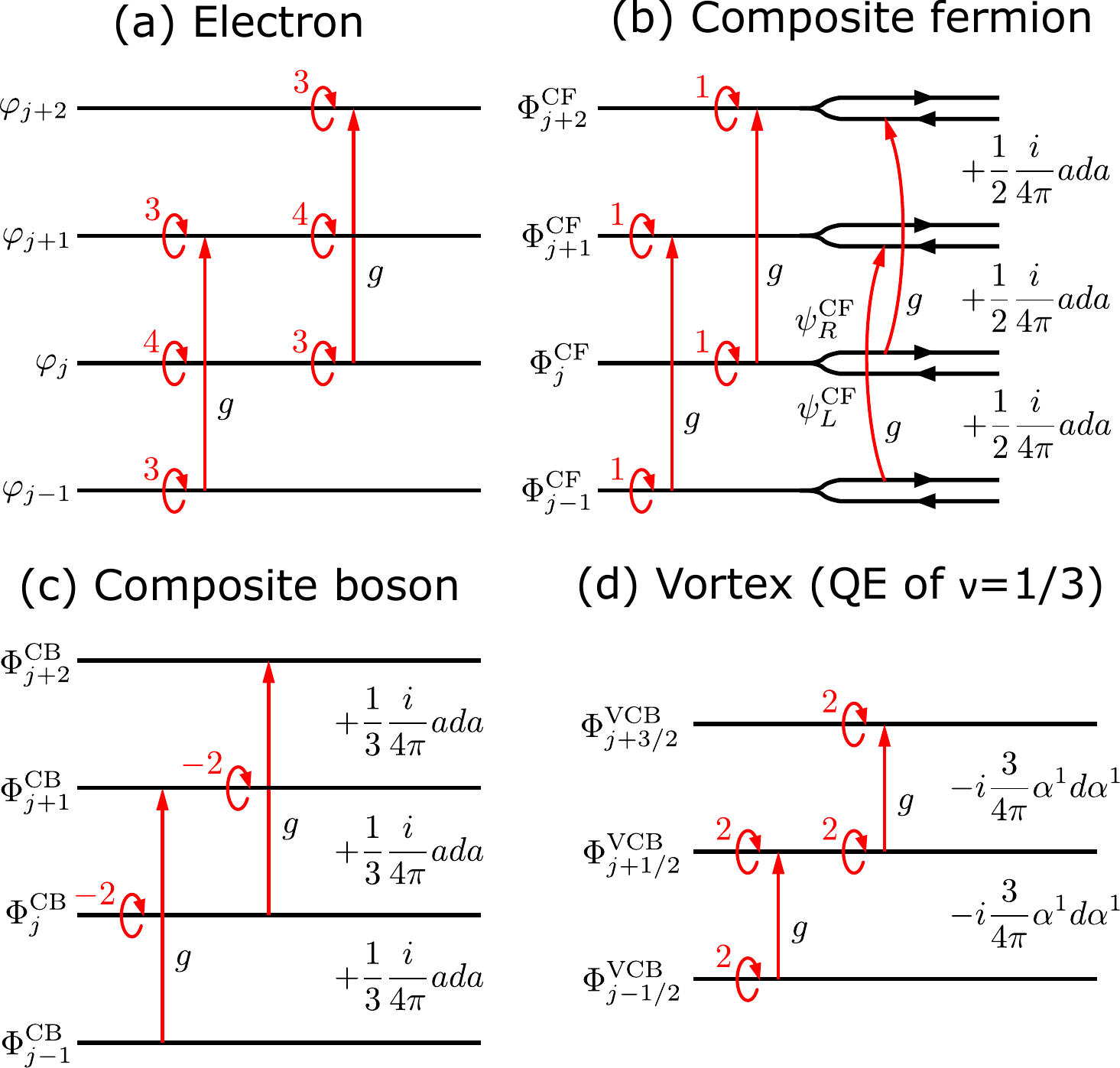}
\caption{Tunneling Hamiltonian for the $\nu=2/5$ state in terms of (a) the electrons, (b) composite fermions, (c) composite bosons, and (d) vortices of the composite bosons. 
The same notation as in Fig.~\ref{fig:Laughlin} is used.}
\label{fig:Hierarchy}
\end{figure}

Let us first consider the $\nu=2/5$ state, for which interwire tunneling is given by 
\begin{align} \label{eq:Tunneling332}
H_1 = g\int_x \sum_j \kappa_j \kappa_{j+2} e^{i(\varphi_j +3\theta_j +4\theta_{j+1} -\varphi_{j+2} +3\theta_{j+2})} +\textrm{H.c.},
\end{align}
which is schematically shown in Fig.~\ref{fig:Hierarchy}~(a).
%%%%%%%%%%%%%%%%%%%%%%%%%%%%%%%%%%%%%
%%%%%%%%%%%%%%%%%%%%%%%%%%%%%%%%%%%%%
We now attach $4\pi$ flux to electrons using the nonlocal transformation \eqref{eq:TransFluxAttach} with $m=2$, 
\begin{align} \label{eq:4piFluxAttach}
\Phi^\CF_j = \varphi_j +2\sum_{j' \neq j} \sgn(j'-j) \theta_{j'}, \ \ \ \Theta^\CF_j = \theta_j.
\end{align}
The kinetic action is then written as 
\begin{align}
S_0 &= \int_{\tau,x} \sum_j \biggl[ \frac{i}{\pi} \partial_x \Theta^\CF_j \Bigl( \partial_\tau \Phi^\CF_j -\frac{1}{2} (Sa_{0,j-1/2}) -A^\ext_{0,j} \Bigr) \nonumber \\
&\qquad\qquad +\frac{v}{2\pi} (\partial_x \Phi^\CF_j -a_{1,j} -A^\ext_{1,j})^2 +\frac{u}{2\pi} (\partial_x \Theta^\CF_j)^2 \nonumber \\
&\qquad\qquad +\frac{w}{8\pi} ( \partial_x \Phi^\CF_j +\partial_x \Theta^\CF_j -\partial_x \Phi^\CF_{j+2} +\partial_x \Theta^\CF_{j+2})^2 \nonumber \\
&\qquad\qquad +\frac{1}{2} \frac{i}{2\pi} a_{1,j} (\Delta a_{0,j-1/2}) \biggr], 
\end{align}
and the tunneling Hamiltonian \eqref{eq:Tunneling332} is written as 
\begin{align}
H_1 = g \int_x \sum_j \kappa_j \kappa_{j+2} e^{i(\Phi^\CF_j +\Theta^\CF_j -\Phi^\CF_{j+2} +\Theta^\CF_{j+2})} +\textrm{H.c.}
\end{align}
We find that the interwire tunneling for the $\nu=2/5$ state is a second-neighbor hopping of the composite fermions [see also Fig.~\ref{fig:Hierarchy}~(b)], 
\begin{align}
H_1 \sim -2\pi \alpha g \int dx \sum_j \psi^\CF_{R,j} \psi^{\CF \dagger}_{L,j+2} +\textrm{H.c.},
\end{align}
where $\psi^\CF$ is the composite fermion field defined in \eqref{eq:CFField}.
When $g$ flows to the strong-coupling limit, the tunneling Hamiltonian $H_1$ leaves two chiral Dirac modes propagating in the same direction at each boundary and thus gives an IQH state of the composite fermions with Chern number $C=2$. 
Thus the $\nu=2/5$ state can be understood as the composite fermions filling the two lowest Landau levels in the coupled wire model.

The $\nu=2/7$ state is obtained by the following second-neighbor tunneling Hamiltonian, 
\begin{align} \label{eq:Tunneling334}
H_1 = g \int_x \sum_j \kappa_j \kappa_{j+2} e^{i(\varphi_j +3\theta_j +8\theta_{j+1} -\varphi_{j+2} +3\theta_{j+2})} +\textrm{H.c.}
\end{align}
Applying the $8\pi$ flux attachment transformation, 
\begin{align}
\Phi^\CF_j = \varphi_j +4\sum_{j' \neq j} \sgn(j'-j) \theta_{j'}, \ \ \ \Theta^\CF_j = \theta_j, 
\end{align}
we obtain the tunneling Hamiltonian, 
\begin{align}
H_1 = g \int_x \sum_j \kappa_j \kappa_{j+2} e^{i(\Phi^\CF_j -\Theta^\CF_j -\Phi^\CF_{j+2} -\Theta^\CF_{j+2})} +\textrm{H.c.}
\end{align}
In terms of the composite fermion fields, we have 
\begin{align}
H_1 = 2\pi \alpha g \int dx \sum_j \psi^\CF_{L,j} \psi^{\CF \dagger}_{R,j+2} +\textrm{H.c.},
\end{align}
which, in the strong-coupling limit, opens a bulk gap and leaves two chiral fermion modes at the boundaries with the opposite chirality to the $\nu=2/5$ state, yielding the IQH state with Chern number $C=-2$. 
We conclude that the $\nu=2/7$ state is understood as the composite fermions filling two ``negative'' Landau levels.

One can readily generalize the construction of the $\nu=2/5$ state to the Jain hierarchy states at filling fractions $\nu=p/[p(q-1)+1]$ with integers $p$ and $q$ \cite{Jain89,Jain90}.
These states are obtained by attaching the $2\pi (q-1)$ flux to electrons (bosons) for odd (even) $q$ and filling $p$ Landau levels of the composite fermions. 
The corresponding interwire tunneling is given by 
\begin{align} \label{eq:TunnelingJain}
H_1 &= g \int_x \sum_j \kappa_j \kappa_{j+p} \exp \biggl\{i \Bigl[ \varphi_j +q\theta_j
-\varphi_{j+p} +q\theta_{j+p}
\nonumber \\
&\hspace*{35mm} +2(q-1) \sum_{k=1}^{p-1} \theta_{j+k} 
 \Bigr]\biggr\} +\textrm{H.c.},
\end{align}
which involves a $p$th neighbor hopping of electrons or bosons. 
In this hopping process, a particle feels the magnetic flux $pb$ that must be canceled by the density fluctuations of $[2(p-1)(q-1)+2q]k_F$, which is the case at the filling fraction of our interest. 
After the $2\pi (q-1)$ flux attachment \eqref{eq:TransFluxAttach}, we have 
\begin{align}
H_1 &= g \int_x \sum_j \kappa_j \kappa_{j+p} e^{i(\Phi^\CF_j +\Theta^\CF_j -\Phi^\CF_{j+p} +\Theta^\CF_{j+p})} +\textrm{H.c.}
\end{align}
The composite fermions see an effective magnetic flux $pb_\CF$ that must be canceled by $2k_F$, resulting in the integer filling of composite fermion $\nu_\CF=2k_F/b_\CF=p$.
This interaction is written in terms of the composite fermion fields \eqref{eq:CFField} as 
\begin{align}
H_1 = 2\pi \alpha g \int dx \sum_j e^{i\pi (q-1)/2} \psi^\CF_{R,j} \psi^{\CF \dagger}_{L,j+p} +\textrm{H.c.}, 
\end{align}
i.e., a $p$th neighbor hopping of composite fermions.
This interaction leaves $p$ decoupled chiral composite-fermion modes at the boundaries, which is consistent with the picture of $p$ filled Landau levels of the composite fermions. 
From the tunneling Hamiltonian in terms of the original bosonic fields in Eq.~\eqref{eq:TunnelingJain}, we can read off the corresponding $K$ matrix by examining the commutation relations of the edge states by grouping $p$ wires. 
We then find 
\begin{align}
K_{IJ} = \delta_{I,J} +(q-1) C_{IJ}, 
\end{align}
where $\bfC$ is the $p \times p$ pseudo identity matrix whose every entry is one. 
This $K$ matrix agrees with the one obtained from the Chern-Simons approach in the multilayer basis \cite{XGWen95}.

We can similarly obtain the negative Jain hierarchy states at the filling fractions $\nu=p/[(q-1)p-1]$ including $\nu=2/7$ ($p=2$ and $q=5$).
These states are obtained by attaching $2\pi (q-1)$ flux to electrons (bosons) and filling $p$ Landau levels of the composite fermions in a magnetic field antiparallel to the originally applied one. 
The corresponding tunneling Hamiltonian is given by 
\begin{align} \label{eq:TunnelingNegativeJain}
H_1 &= g \int_x \sum_j \kappa_j \kappa_{j+p} \exp \biggl\{i\Bigl[ \varphi_j +(q-2) \theta_j \nonumber \\
&\ \ \ +2(q-1) \sum_{k=1}^{p-1} \theta_{j+k} -\varphi_{j+p} +(q-2) \theta_{j+p} \Bigr]\biggr\} +\textrm{H.c.}
\end{align}
After the $2\pi (q-1)$ flux attachment defined in Eq.~\eqref{eq:TransFluxAttach}, this interaction is written as 
\begin{align}
H_1 = g \int_x \sum_j \kappa_j \kappa_{j+p} e^{i(\Phi^\CF_j -\Theta^\CF_j -\Phi^\CF_{j+p} -\Theta^\CF_{j+p})} +\textrm{H.c.},
\end{align}
and is given in terms of the composite fermion fields by 
\begin{align}
H_1 = 2\pi \alpha g \int dx \sum_j e^{i\pi(q-1)/2} \psi^\CF_{L,j} \psi^{\CF \dagger}_{R,j+p} +\textrm{H.c.}
\end{align}
Thus $p$ chiral fermion modes, with the opposite chirality to the positive Jain hierarchy states, remain gapless at the boundaries and give the Chern number $C=-p$. 
The corresponding $K$ matrix can be read off as 
\begin{align}
K_{IJ} = -\delta_{I,J} +(q-1) C_{IJ}.
\end{align}

\subsubsection{Composite boson: Haldane-Halperin hierarchy}
\label{sec:HaldaneHalperin}

The $\nu=2/5$ state is obtained in the Haldane-Halperin hierarchy construction by exciting quasielectrons on top of the parent Laughlin $\nu=1/3$ state and condensing them into the Laughlin $\nu=1/2$ state \cite{Haldane83,Halperin84}. 
In a field theoretical description, this picture is accommodated in the Ginzburg-Landau theory for the FQH states or in the composite boson formulation via the $6\pi$ flux attachment \cite{SCZhang89,Read89}. 
Let us emulate the hierarchy construction of the $\nu=2/5$ state in the coupled-wire approach. 
Applying the $6\pi$ flux attachment transformation defined in Eq.~\eqref{eq:TransFluxAttach} with $m=3$, 
\begin{align} \label{eq:CB332}
\Phi^\CB_j = \varphi_j +3\sum_{j' \neq j} \sgn (j'-j) \theta_{j'}, \ \ \ \Theta^\CB_j = \theta_j, 
\end{align}
the interwire tunneling \eqref{eq:Tunneling332} is written as [Fig.~\ref{fig:Hierarchy}~(c)]
\begin{align} \label{eq:Tunneling332CB}
H_1 = g \int_x \sum_j \kappa_{j-1} \kappa_{j+1} e^{i(\Phi^\CB_{j-1} -2\Theta^\CB_j -\Phi^\CB_{j+1})} +\textrm{H.c.}, 
\end{align}
where we have shifted the wire label as $j \to j-1$ to simplify the presentation.
The kinetic action \eqref{eq:SLLplusA} with the interwire forward scattering interaction \eqref{eq:InterForward_qqp} is now given by 
\begin{align} \label{eq:Action332CB}
S_0 &= \int_{\tau,x} \sum_j \biggl[ \frac{i}{\pi} \partial_x \Theta^\CB_j \Bigl( \partial_\tau \Phi^\CB_j -\frac{1}{2} (S a^1_{0,j-1/2}) -A^\ext_{0,j} \Bigr) \nonumber \\
&\qquad\qquad +\frac{v}{2\pi} (\partial_x \Phi^\CB_j -a^1_{1,j} -A^\ext_{1,j})^2 +\frac{u}{2\pi} (\partial_x \Theta^\CB_j)^2 \nonumber \\
&\qquad\qquad +\frac{w}{8\pi} (\partial_x \Phi^\CB_{j-1} -2\partial_x \Theta^\CB_j -\partial_x \Phi^\CB_{j+1})^2 \nonumber \\
&\qquad\qquad +\frac{1}{3} \frac{i}{2\pi} a^1_{1,j} (\Delta a^1_{0,j-1/2}) \biggr], 
\end{align}
where we have introduced the Chern-Simons gauge fields $a^1_{0,j+1/2}$ and $a^1_{1,j}$ to make the theory local.
The tunneling term \eqref{eq:Tunneling332CB} would simply result in the condensation of the composite bosons if the backscattering operator $e^{-i2\Theta^\CB_j}$ from the middle wire $j$ were not involved in $H_1$.
As explained in Sec.~\ref{sec:Laughlin}, the $2k_F$ backscattering operator $e^{i2\Theta^\CB_j} = e^{i2\theta_j}$ creates a pair of quasiparticle excitations of the Laughlin $\nu=1/3$ state. 
The operator $e^{-i2\Theta^\CB_j}$ hops a quasielectron with charge $-1/q$ from the dual wire $j-1/2$ to $j+1/2$. 
Thus the tunneling term \eqref{eq:Tunneling332CB} can be seen to create quasielectrons hopping between adjacent (dual) wires on top of the condensate of the composite bosons. 

We now move to the dual picture in terms of vortices \cite{DHLee89,Blok90a,Blok90b,Ezawa91}. 
As discussed in Sec.~\ref{sec:LaughlinCB}, vortices represent point-like single-quasielectron excitations in this picture, which sharpens our view of hierarchy states as a condensate of quasiparticles.
Applying the vortex duality transformation \eqref{eq:VCBfromCB}, we rewrite the tunneling term \eqref{eq:Tunneling332CB} in terms of the vortex fields [Fig.~\ref{fig:Hierarchy}~(d)], 
\begin{align} \label{eq:Tunneling332VCB}
H_1 &= g \int dx \sum_j \kappa_{j-1} \kappa_{j+1} \nonumber \\
&\ \ \ \times e^{-i(\Phi^\VCB_{j-1/2} +2\Theta^\VCB_{j-1/2} -\Phi^\VCB_{j+1/2} +2\Theta^\VCB_{j+1/2})} +\textrm{H.c.} 
\end{align}
The kinetic action \eqref{eq:Action332CB} becomes 
\begin{align} \label{eq:Action332VCB}
S_0 &= \int_{\tau,x} \sum_j \biggl[ \frac{i}{\pi} \partial_x \Theta^\VCB_{j+1/2} \Bigl( \partial_\tau \Phi^\VCB_{j+1/2} -\frac{1}{2} (S\alpha^1_{0,j}) \Bigr) \nonumber \\
&\qquad\qquad +\frac{9v}{2\pi} (\partial_x \Phi^\VCB_{j+1/2} -\alpha^1_{1,j+1/2})^2 + \! \frac{v}{2\pi} (\partial_x \Theta^\VCB_{j+1/2})^2 \nonumber \\
&\qquad\qquad +\frac{u-9v}{8\pi} (\Delta \partial_x \Phi^\VCB_{j-1/2})^2 \! - \!\frac{v}{2\pi} \partial_x \Theta^\VCB_{j+1/2} (\Delta A^\ext_{1,j}) \nonumber \\
&\qquad\qquad +\frac{w}{8\pi} \Bigl( (\Delta \partial_x \Phi^\VCB_{j-1/2}) -2(S \partial_x \Theta^\VCB_{j-1/2}) \Bigr)^2 \nonumber \\
&\qquad\qquad -i\frac{3}{2\pi} \alpha^1_{1,j+1/2} (\Delta \alpha^1_{0,j}) -\frac{i}{4\pi} (SA^\ext_{1,j}) (\Delta \alpha^1_{0,j}) \nonumber \\
&\qquad\qquad +\frac{i}{2\pi} A^\ext_{0,j} (\Delta \alpha^1_{1,j-1/2}) +\cdots \biggr], 
\end{align}
where we have dropped terms with higher-order derivatives of the gauge fields that can make only quantitative changes in the low-energy dynamics. 
The vortex operators $e^{i\Phi^\VCB_{j+1/2}}$ satisfy the bosonic statistics [see Eq.~\eqref{eq:CommVCB}] and create a quasielectron on the dual wire $j+1/2$. 
In the presence of the composite-boson condensate, quasielectrons see an effective magnetic flux $b_\VCB$ that is produced by the original electrons, since the vortex field $\Phi^\VCB$ couples to the gauge field $\alpha^1_\mu$ whose flux is the original electron density. 
The composite boson hopping $e^{-i(\Phi^\CB_{j-1}-\Phi^\CB_{j+1})}$ gives rise to a density-density interaction between vortices, $e^{i2(\Theta^\VCB_{j-1/2}+\Theta^\VCB_{j+1/2})}$, with the wave number $4\pi \brho_\VCB$.
The effective magnetic flux is canceled when $b_\VCB = 4\pi \brho_\VCB$, giving an effective filling fraction $\nu_\VCB =2\pi\brho_\VCB/b_\VCB =1/2$ for the vortices. 
Indeed, the interwire tunneling \eqref{eq:Tunneling332VCB} has exactly the same form as that for the $\nu=1/2$ Laughlin state in Eq.~\eqref{eq:TunnelingLaughlin}.
Hence, the quasielectrons form the bosonic Laughlin $\nu=1/2$ state in the strong-coupling limit of $g$. 
Therefore, the very notion of quasiparticle condensation in the Haldane-Halperin hierarchy naturally comes out in the coupled-wire construction.

In order to obtain the effective Chern-Simons theory, we repeat what we have done for the Laughlin states in Sec.~\ref{sec:LaughlinCB}.
Thus we attach $4\pi$ flux to the bosonic quasielectrons (vortices),
\begin{align}
\begin{split}
\Phi^\CQ_{j+1/2} &= \Phi^\VCB_{j+1/2} +2\sum_{j' \neq j} \sgn (j'-j) \Theta^\VCB_{j'+1/2}, \\
\Theta^\CQ_{j+1/2} &= \Theta^\VCB_{j+1/2},
\end{split}
\end{align}
to define bosonic composite quasiparticle fields $\Phi^\CQ$ and $\Theta^\CQ$.
The tunneling Hamiltonian \eqref{eq:Tunneling332VCB} then becomes
\begin{align} \label{eq:Tunneling332CQ}
H_1 = g \int_x \sum_j \kappa_{j-1} \kappa_{j+1} e^{-i(\Phi^\CQ_{j-1/2} -\Phi^\CQ_{j+1/2})} +\textrm{H.c.}
\end{align}
The kinetic action \eqref{eq:Action332VCB} is kept in a local form by introducing a new Chern-Simons gauge field. 
The composite quasiparticle operators $\kappa_j \kappa_{j+1} e^{i\Phi^\CQ_{j+1/2}}$ also obey the bosonic statistics and can be condensed. 
As a final step, we introduce the second vortex fields $\Phi^\VCQ$ and $\Theta^\VCQ$,
\begin{align}
\begin{split}
\Phi^\VCQ_j &= \sum_{j'} \sgn (j'-j+1/2) \Theta^\CQ_{j'+1/2}, \\
\Theta^\VCQ_j &= \frac{1}{2} (\Phi^\CQ_{j+1/2} -\Phi^\CQ_{j-1/2}),
\end{split}
\end{align}
which represent point-like quasiparticle excitations of the daughter $\nu=1/2$ state.
In the strong-coupling limit the tunneling term \eqref{eq:Tunneling332CQ} pins the $\Theta^\VCQ$ field and turns the system into a Mott insulator of these vortices.
Finally, the kinetic action \eqref{eq:Action332VCB} is written as 
\begin{align} \label{eq:Action332VCQ}
S_0 &= \int_{\tau,x} \sum_j \biggl[ \frac{i}{\pi} \partial_x \Theta^\VCQ_j \Bigl( \partial_\tau \Phi^\VCQ_j -\frac{1}{2} (S\alpha^2_{0,j-1/2}) \Bigr) \nonumber \\
&\ \ \ +\frac{18v}{\pi} (\partial_x \Phi^\VCQ_j -\alpha^2_{1,j})^2 +\frac{w-9v}{2\pi} (\partial_x \Theta^\VCQ_j)^2 \nonumber \\
&\ \ \ -\frac{9v}{2\pi} \partial_x \Theta^\VCQ_j (\Delta \alpha^1_{1,j-1/2}) -i\frac{3}{2\pi} \alpha^1_{1,j+1/2} (\Delta \alpha^1_{0,j}) \nonumber \\
&\ \ \ -i\frac{2}{2\pi} \alpha^2_{1,j} (\Delta \alpha^2_{0,j-1/2}) -\frac{i}{4\pi} (S\alpha^1_{1,j-1/2}) (\Delta \alpha^2_{0,j-1/2}) \nonumber \\
&\ \ \ +\frac{i}{4\pi} (S \alpha^1_{0,j}) (\Delta \alpha^2_{1,j}) -\frac{i}{4\pi} (SA^\ext_{1,j}) (\Delta \alpha^1_{0,j}) \nonumber \\
&\ \ \ +\frac{i}{2\pi} A^\ext_{0,j} (\Delta \alpha^1_{1,j-1/2}) +\cdots \biggr],
\end{align}
where we have dropped higher derivative terms for brevity. 
The gauge fields $\alpha^1_\mu$ and $\alpha^2_\mu$ constitute a discrete version of the Chern-Simons action \eqref{eq:MultiCompCSAction} in the hierarchical basis with the $K$ matrix \eqref{eq:Kmat332h} via a redefinition of the fields $\Phi^\VCQ_j \to -\Phi^\VCQ_j$, $\Theta^\VCQ_j \to -\Theta^\VCQ_j$, and $\alpha^2_\mu \to -\alpha^2_\mu$, appropriately reflecting the sign of the quasielectron current of the parent $\nu=1/3$ state. 
The full kinetic action for the first-level hierarchy state is given in Appendix~\ref{app:VortexAction}. 

The $\nu=2/7$ state can be understood in a way parallel to the $\nu=2/5$ state. 
The corresponding tunneling Hamiltonian \eqref{eq:Tunneling334} is written in terms of the composite bosons via the $6\pi$ flux attachment transformation \eqref{eq:CB332} as 
\begin{align}
H_1 = g \int_x \sum_j \kappa_{j-1} \kappa_{j+1} e^{i(\Phi^\CB_{j-1} +2\Theta^\CB_j -\Phi^\CB_{j+1})} +\textrm{H.c.}
\end{align}
Compared with Eq.~\eqref{eq:Tunneling332CB} for the $\nu=2/5$ state, this tunneling Hamiltonian can be seen to excite quasiholes, instead of quasielectrons, on top of the composite boson condensate, as the operator $e^{i2\Theta^\CB_j}$ hops a quasihole from the dual wire $j-1/2$ to $j+1/2$.
In the vortex picture, the tunneling Hamiltonian becomes 
\begin{align}
H_1 &= g\!\int \! dx \sum_j \kappa_{j-1} \kappa_{j+1} \nonumber \\
&\ \ \ \times e^{i(\Phi^\VCB_{j-1/2} -2\Theta^\VCB_{j-1/2} -\Phi^\VCB_{j+1/2} -2\Theta^\VCB_{j+1/2})} +\textrm{H.c.}
\end{align}
This leads to the Laughlin $\nu=1/2$ state of quasiholes. 
Hence, the vortices are condensed by attaching $-4\pi$ flux. 
The subsequent vortex duality transformation yields a kinetic action similar to Eq.~\eqref{eq:Action332VCQ} but with the Chern-Simons term associated with the $K$ matrix \eqref{eq:Kmat334h} in the hierarchy basis.

\begin{figure}
\includegraphics[clip,width=0.47\textwidth]{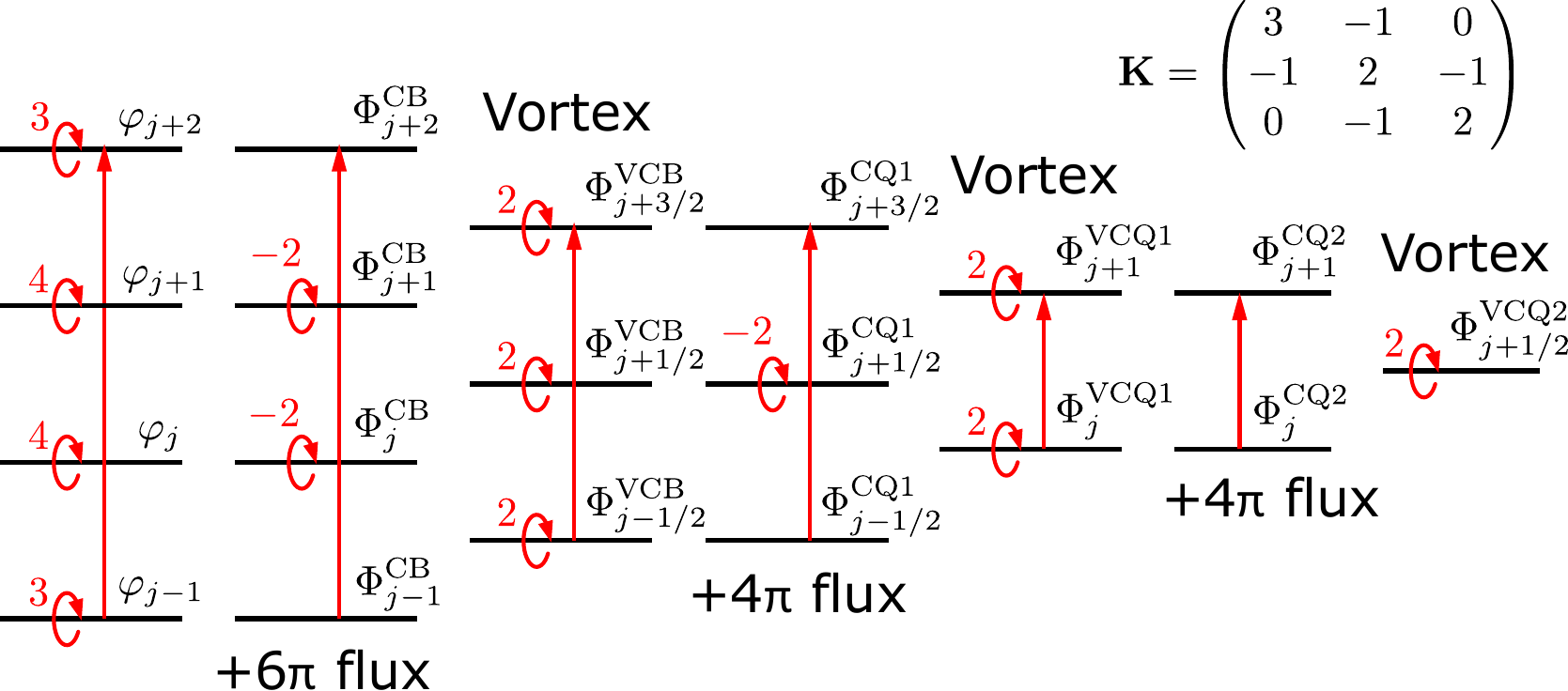}
\caption{Tunneling Hamiltonian for a second-level hierarchy state at $\nu=3/7$. 
Successive applications of the flux attachment and vortex duality transformations finally yield the Hamiltonian describing the Mott insulator of vortices. 
The resulting $K$ matrix is also given above.}
\label{fig:Hierarchy3rd}
\end{figure}

The construction is easily generalized to the hierarchy states at the filling fractions \cite{Haldane83,Halperin84},
\begin{align}
\nu = \cfrac{1}{q-\cfrac{1}{2p_1-\cfrac{1}{2p_2 -\cdots}}}, 
\end{align}
where $q$ is an odd (even) integer for fermions (bosons), and $p_1, p_2, \cdots$ are arbitrary integers.
The corresponding $K$ matrix in the hierarchy basis is given by \cite{XGWen95}
\begin{align}
\bfK = \begin{pmatrix} q & -\sgn(p_1) & 0 & & \\ -\sgn(p_1) & 2p_1 & -\sgn(p_2) & 0 \\ 0 & -\sgn(p_2) & 2p_2 & \ddots \\ & 0 & \ddots & \ddots \end{pmatrix}.
\end{align}
The tunneling Hamiltonian for the hierarchy state with this $K$ matrix is obtained by reverse engineering of the above procedure in such a way that quasiparticles of the parent $\nu=1/q$ state are condensed into the daughter $\nu=1/2p_1$ state, quasiparticles of the daughter $\nu=1/2p_1$ state are condensed into the granddaughter $\nu=1/2p_2$ state, and so on. 
An example of the $\nu=3/7$ state is illustrated in Fig.~\ref{fig:Hierarchy3rd}. 
%%%%%%%%%%%%%%%%%%%%%%%%%%%%%%%%%%%%%
%%%%%%%%%%%%%%%%%%%%%%%%%%%%%%%%%%%%%
We then find the tunneling Hamiltonian for the first-level hierarchy states, 
\begin{align} \label{eq:TunnelingFirstHierarchy}
H_1 &= g \int_{\tau,x} \sum_j f(\{ \kappa_j \}) \exp i\bigl[ p_1 \varphi_j +qp_1 \theta_j \nonumber \\
&\ \ \ +2(qp_1-1) \theta_{j+1} -p_1 \varphi_{j+2} +qp_1 \theta_{j+2} \bigr] +\textrm{H.c.}, 
\end{align}
and for the second-level hierarchy states,
\begin{align}
H_1 &= g \int_{\tau,x} \sum_j f(\{ \kappa_j \}) \exp i\bigl[ p_1 p_2 \varphi_j +qp_1 p_2 \theta_j \nonumber \\
&\ \ \ +(p_1 p_2-1) \varphi_{j+1} +(3qp_1 p_2 -q -2p_2) \theta_{j+1} \nonumber \\
&\ \ \ -(p_1 p_2 -1) \varphi_{j+2} +(3q p_1 p_2 -q -2p_2) \theta_{j+2} \nonumber \\
&\ \ \ -p_1 p_2 \varphi_{j+3} +q p_1 p_2 \theta_{j+3} \bigr] +\textrm{H.c.},
\end{align}
where the product of Klein factors $f(\{ \kappa_j \})$ should be appropriately chosen, depending on how the interaction is microscopically built out of fermion operators. 
Note that the tunneling Hamiltonian \eqref{eq:TunnelingFirstHierarchy} coincides with the one that Teo and Kane proposed for the first-level hierarchy states \cite{Teo14}, which is given in Eq.~\eqref{eq:Tunneling_qqp} and identified with Eq.~\eqref{eq:TunnelingFirstHierarchy} by choosing $(n,m_0,m_1)=(p_1,qp_1,qp_1-1)$. 
The general higher-level hierarchy states require complicated coupled-wire Hamiltonians with multiparticle hopping processes. 
However, for the special case $p_1=p_2=\cdots=1$, the tunneling Hamiltonian reduces to that for the positive Jain hierarchy states in Eq.~\eqref{eq:TunnelingJain}, which involves only a single-particle hopping process. 
Indeed, as illustrated for the $\nu=2/5$ and $\nu=2/7$ states, the coupled-wire construction yields the same Hamiltonian for both Jain and Haldane-Halperin hierarchy states where their filling fractions match.

\subsection{Particle-hole conjugate}
\label{sec:PHConjugate}

We can also obtain the coupled-wire models for the PH conjugates of fermionic FQH states realized at filling fraction $1-\nu$ \cite{Girvin84}. 
Following the strategy in Refs.~\cite{SSLee07,Barkeshli15}, we first attach $2\pi$ flux to electrons for converting them to composite bosons and then apply the vortex duality to the composite bosons. 
In this case, the vortex action \eqref{eq:ActionVCB} has the level-1 Chern-Simons term with the opposite sign to that for the composite bosons \eqref{eq:SLLinCP2}. 
Hence, each vortex is attached $-2\pi$ flux and converted to a fermion. 
In this way we obtain the bosonic fields for holes, 
\begin{align} \label{eq:HoleField}
\begin{split}
\varphi^\hole_{j+1/2} &= \Phi^\VCB_{j+1/2} +\sum_{j' \neq j} \sgn (j'-j) \Theta^\VCB_{j'+1/2}, \\
\theta^\hole_{j+1/2} &= \Theta^\VCB_{j+1/2}, 
\end{split}
\end{align}
through the flux attachment to the vortex fields \eqref{eq:VCBfromCB} using the transformation \eqref{eq:TransFluxAttach} with $m=1$. 
Let us call these fields as the hole fields. 

Integrating out $\alpha_\mu$ in Eq.~\eqref{eq:ActionVCB} yields the theory of holes. 
The vortex action \eqref{eq:ActionVCB} is then written as 
\begin{align} \label{eq:ActionHole}
S_0 &= \int_{\tau,x} \sum_j \biggl[ \frac{i}{\pi} \partial_x \theta^\hole_{j+1/2} \Bigl( \partial_\tau \varphi^\hole_{j+1/2} +\frac{1}{2} (SA^\ext_{0,j}) \Bigr) \nonumber \\
&\qquad\qquad
+\frac{v}{2\pi} \Bigl( \partial_x \varphi^\hole_{j+1/2} +\frac{1}{2} (SA^\ext_{1,j}) -\frac{i}{2v} (\Delta A^\ext_{0,j}) \Bigr)^2 \nonumber \\
&\qquad\qquad
+\frac{v}{2\pi} (\partial_x \theta^\hole_{j+1/2})^2 -\frac{v}{2\pi} \partial_x \theta^\hole_{j+1/2} (\Delta A^\ext_{1,j}) \nonumber \\
&\qquad\qquad
+\frac{u-v}{8\pi} \Bigl( (\Delta \partial_x \varphi^\hole_{j-1/2}) +(S\partial_x \theta^\hole_{j-1/2}) \Bigr)^2 \nonumber \\
&\qquad\qquad
+\frac{i}{4\pi} (SA^\ext_{1,j}) (\Delta A^\ext_{0,j}) +\cdots \biggr],
\end{align}
where we have omitted higher derivative terms of the electromagnetic field $A^\ext_\mu$. 
The derivation of the full action is given in Appendix~\ref{app:HoleTheory}. 
Since this is the theory of holes, the corresponding bosonic fields carry electric charge with the opposite sign to electrons. 
There is also a discrete analog of the Chern-Simons term, $(i/4\pi) \epsilon_{\mu \nu \lambda} A^\ext_\mu \partial_\nu A^\ext_\lambda$, producing the Hall response of the filled lowest Landau level. 
The action is free from any fluctuating gauge field and should be identified with the original electron action \eqref{eq:SLLplusA}. 

The hole fields defined in Eq.~\eqref{eq:HoleField} are related to the original bosonic fields by 
\begin{align} \label{eq:HoleFromElectron}
\begin{split}
\varphi^\hole_{j+1/2} &= -\frac{1}{2} (\varphi_j +\theta_j +\varphi_{j+1} -\theta_{j+1}), \\
\theta^\hole_{j+1/2} &= -\frac{1}{2} (\varphi_j +\theta_j -\varphi_{j+1} +\theta_{j+1}). 
\end{split}
\end{align}
This is a local redefinition of the original bosonic fields and similar to the relation between a composite fermion and a fermionized vortex, each of which is obtained by attaching $\pm 2\pi$ flux to a boson or a vortex, respectively \cite{Mross17a}. 
From Eqs.~\eqref{eq:ElectronOp} and \eqref{eq:HoleFromElectron},
the electron operators are written as 
\begin{align}
\begin{split}
\psi_{R,j} &= \frac{\kappa_j}{\sqrt{2\pi \alpha}} e^{-i(\varphi^\hole_{j+1/2} +\theta^\hole_{j+1/2})}, \\
\psi_{L,j} &= \frac{\kappa_j}{\sqrt{2\pi \alpha}} e^{-i(\varphi^\hole_{j-1/2} -\theta^\hole_{j-1/2})}.
\end{split}
\end{align}
The hole fields \eqref{eq:HoleFromElectron} can be used to systematically generate coupled-wire Hamiltonians for the PH conjugates of various FQH states in the lowest Landau level. 
We define the PH conjugate transformation by the combination of the replacement,
\begin{subequations} 
\label{eq:PHTrans}
\begin{align}
\varphi_j \to -\varphi^\hole_{j+1/2}, \ \ \ \theta_j \to \theta^\hole_{j+1/2},
\end{align}
and complex conjugation
\begin{equation}
i \to -i.
\end{equation}
\end{subequations}
The electron operators on the $j$th wire \eqref{eq:ElectronOp} are transformed by the PH transformation as
\begin{align} \label{eq:PHTransFermi}
\psi_{L,j} \to \psi^\dagger_{R,j}, \ \ \ \psi_{R,j} \to \psi^\dagger_{L,j+1}.
\end{align}
Equation \eqref{eq:PHTrans} defines the coupled-wire version of the PH transformation. 
This transformation is essentially equivalent to the PH transformation defined for a Dirac theory in Refs.~\cite{Mross16a,Mross17a}, as we will discuss in Sec.~\ref{sec:AntiCFL} in more detail. 
The PH conjugation with ``shifted wires'' is also anticipated in Ref.~\cite{Kane17}. 

As a sanity check, let us apply the PH transformation \eqref{eq:PHTrans} to the filled lowest Landau level of electrons, i.e., the $\nu=1$ IQH state. 
Its tunneling Hamiltonian may be given by 
\begin{align}
H_1 \sim g \int dx \sum_j e^{i(\varphi_j +\theta_j -\varphi_{j+1} +\theta_{j+1})} +\textrm{H.c.}, 
\end{align}
which leaves a single chiral fermion at the boundaries in the strong-coupling limit. 
Here we have dropped the Klein factors, which will be appropriately supplemented after the PH transformation. 
We apply Eq.~\eqref{eq:PHTrans} to replace the bosonic fields by the hole fields, 
\begin{align}
H_1 \sim g \int dx \sum_j e^{i(\varphi^\hole_{j-1/2} -\theta^\hole_{j-1/2} -\varphi^\hole_{j+1/2} -\theta^\hole_{j+1/2})} +\textrm{H.c.}
\end{align}
Using Eq.~\eqref{eq:HoleFromElectron}, we obtain the Hamiltonian in terms of the original bosonic fields, 
\begin{align}
H_1 \sim g \int dx \sum_j e^{i2\theta_j} +\textrm{H.c.}
\end{align}
This is a backscattering operator with the wave number $2k_F$ and leads to a trivial band insulator, which may be thought of as an empty state of electrons with $k_F=0$. 
Thus the PH transformation interchanges the filled and empty Landau levels as desired. 

\begin{figure}
\includegraphics[clip,width=0.47\textwidth]{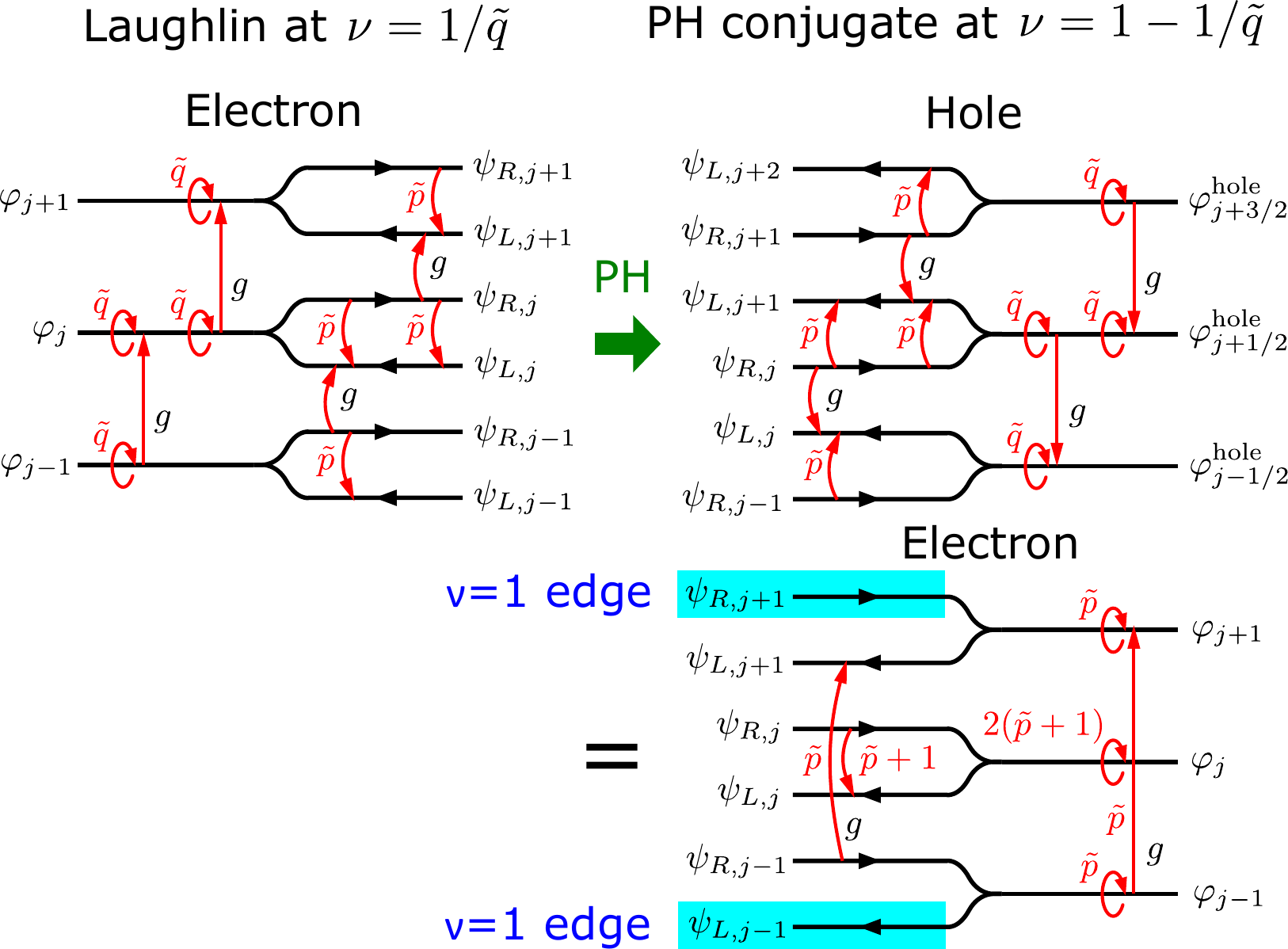}
\caption{PH transformation to the Laughlin $\nu=1/\tilde{q} = 1/(2\tilde{p}+1)$ state. 
The PH conjugate of the Laughlin state involves $\tilde{p}$ electron hopping and obviously leaves the chiral edge states corresponding to the $\nu=1$ IQH state.}
\label{fig:PHLaughlin}
\end{figure}

As a next example, we apply the PH transformation \eqref{eq:PHTrans} to the $\nu=1/\tq$ Laughlin state of electrons, where $\tq$ is an odd integer. 
This is illustrated in Fig.~\ref{fig:PHLaughlin}. 
%%%%%%%%%%%%%%%%%%%%%%%%%%%%%%%%%%%%%
%%%%%%%%%%%%%%%%%%%%%%%%%%%%%%%%%%%%%
Applying the PH transformation to Eq.~\eqref{eq:TunnelingLaughlin} yields 
\begin{align}
H_1 \sim g\int dx \sum_j e^{i(\varphi^\hole_{j-1/2} -\tq \theta^\hole_{j-1/2} -\varphi^\hole_{j+1/2} -\tq \theta^\hole_{j+1/2})} +\textrm{H.c.}
\end{align}
Setting $\tq=2\tp+1$, we find the tunneling Hamiltonian in terms of the original bosonic fields, 
\begin{align} \label{eq:TunnelingPHLaughlin}
H_1 &\sim g \int dx \sum_j \kappa_{j-1}^{\tp} \kappa_{j+1}^{\tp} \nonumber \\
&\ \ \ \times e^{i(\tp \varphi_{j-1} +\tp \theta_{j-1} +2(\tp +1) \theta_j -\tp \varphi_{j+1} +\tp \theta_{j+1})} +\textrm{H.c.}
\end{align}
This interaction is allowed at filling fraction,
\begin{align}
\nu=\frac{2\tp}{2\tp+1}=1-\frac{1}{\tq}.
\end{align}
When $\tq=3$, the tunneling Hamiltonian in Eq.~\eqref{eq:TunnelingPHLaughlin} agrees with the one proposed in Ref.~\cite{Kane02} for the $\nu=2/3$ FQH state that has counter-propagating edge modes. 
This Hamiltonian corresponds to the tunneling Hamiltonian \eqref{eq:TunnelingFirstHierarchy} for the first-level hierarchy state with $q=1$ and $p_1=-\tp$, in which hole excitations with charge $+1$ are condensed into the Laughlin $\nu=1/2\tp$ state. 
In the basis of charge vector $\bft=(1,-1)$ for the Chern-Simons theory \eqref{eq:MultiCompCSAction}, the corresponding $K$ matrix takes a diagonal form, 
\begin{align}
\bfK = \begin{pmatrix} 1 & 0 \\ 0 & -\tq \end{pmatrix}.
\end{align}
Thus this state can be viewed as the stacking of the $\nu=1$ IQH state of electrons and the $\nu=1/\tq$ Laughlin state of holes and precisely interpreted as the PH conjugate of the Laughlin state at $\nu=1-1/\tq$. 

Another application is that the PH transformation \eqref{eq:PHTrans} interchanges the coupled-wire Hamiltonian for the positive Jain state at $\nu=p/(2p+1)$ in Eq.~\eqref{eq:TunnelingJain} with that for the negative Jain state at $\nu=(p+1)/[2(p+1)-1]$ in Eq.~\eqref{eq:TunnelingNegativeJain}. 
In the following sections, we apply this transformation to the coupled-wire Hamiltonians for the CFL and the Moore-Read Pfaffian state.

\subsection{FQH sates in higher Landau levels}
\label{sec:NextLL}

Pursuing the above idea of defining the hole fields, we can also discuss bosonic fields for electrons in the $(n+1)$th Landau level in the presence of $n$ filled Landau levels ($n\in\mathbb{N}$).
First let us define the bosonic fields for electrons added on top of the filled lowest Landau level, 
\begin{align} \label{eq:FieldOn_n1}
\begin{split}
\varphi^{(n=1)}_{j+1/2} &= \frac{1}{2} (\varphi_j +\theta_j +\varphi_{j+1} -\theta_{j+1}), \\
\theta^{(n=1)}_{j+1/2} &= \frac{1}{2} (\varphi_j +\theta_j -\varphi_{j+1} +\theta_{j+1}), 
\end{split}
\end{align}
which are just a redefinition of the hole fields \eqref{eq:HoleFromElectron} such that they carry charge $-1$. 
We then recursively define the bosonic fields for electrons on top of $n+1$ filled Landau levels, 
\begin{align} \label{eq:FieldFilledLL}
\begin{split}
\varphi^{(n+1)}_{j+\frac{n+1}{2}} &= \frac{1}{2} (\varphi^{(n)}_{j+n/2} +\theta^{(n)}_{j+n/2} +\varphi^{(n)}_{j+n/2+1} -\theta^{(n)}_{j+n/2+1}), \\
\theta^{(n+1)}_{j+\frac{n+1}{2}} &= \frac{1}{2} (\varphi^{(n)}_{j+n/2} +\theta^{(n)}_{j+n/2} -\varphi^{(n)}_{j+n/2+1} +\theta^{(n)}_{j+n/2+1}).
\end{split}
\end{align}
This transformation is designed in such a way that an empty state plus $n$ filled Landau levels corresponds to the $\nu=n$ IQH state of electrons,
\begin{align}
H_1 &\sim g \int_x \sum_j e^{i2\theta^{(n)}_{j+n/2}} +\textrm{H.c.} \nonumber \\
&= g \int_x \sum_j e^{i(\varphi_j +\theta_j -\varphi_{j+n} +\theta_{j+n})} +\textrm{H.c.}
\end{align}
Accordingly, the kinetic action \eqref{eq:SLLplusA} written in terms of $\varphi^{(n)}_j$ and $\theta^{(n)}_j$ produces a discrete analog of the Chern-Simons term $(in/4\pi) \epsilon_{\mu \nu \lambda} A^\ext_\mu \partial_\nu A^\ext_\lambda$. 

The coupled-wire Hamiltonian for FQH states in the $(n+1)$th Landau level is obtained by writing the corresponding Hamiltonian for the desired FQH state in terms of the bosonic fields $\varphi^{(n+1)}_{j+\frac12(n+1)}$ and $\theta^{(n+1)}_{j+\frac12(n+1)}$ in Eq.~\eqref{eq:FieldFilledLL}. 
For example, the $\nu=1+1/3$ state will be given by the tunneling Hamiltonian, 
\begin{align} \label{eq:Tunneling1+1/3}
H_1 \sim g \int dx \sum_j e^{i(\varphi^{(n=1)}_{j-1/2} +3 \theta^{(n=1)}_{j-1/2} -\varphi^{(n=1)}_{j+1/2} +3 \theta^{(n=1)}_{j+1/2})} +\textrm{H.c.},
\end{align}
which is written, in terms of the original bosonic fields, as
\begin{align}
H_1 \sim g \int_x \sum_j e^{i(2\varphi_j +2\theta_j +2\theta_{j+1} -2\varphi_{j+2} +2\theta_{j+2})} +\textrm{H.c.}
\end{align}
Next, we consider the $\nu=4/11$ state, which is an enigmatic state observed in experiments \cite{Pan03} whose physical interpretation remains unsettled \cite{Mukherjee14}. 
This filling fraction actually admits the first-level Haldane-Halperin hierarchy state with $q=3$ and $p_1=2$, whose coupled-wire Hamiltonian is given by [see Eq.~\eqref{eq:TunnelingFirstHierarchy}]
\begin{align}
H_1 = g \int_x \sum_j e^{i(2\varphi_j +6\theta_j +10\theta_{j+1} -2\varphi_{j+2} +6\theta_{j+2})} +\textrm{H.c.},
\end{align}
which can be written, in terms of the composite fermions defined through the $4\pi$ flux attachment \eqref{eq:4piFluxAttach}, as
\begin{align}
H_1 = g \int_x \sum_j e^{i(2\Phi^\CF_j +2\Theta^\CF_j +2\Theta^\CF_{j+1} -2\Phi^\CF_{j+2} +2\Theta^\CF_{j+2})} +\textrm{H.c.}
\end{align}
This takes the same form as Eq.~\eqref{eq:Tunneling1+1/3} and thus may be seen as the composite fermions forming the $\nu=1+1/3$ state as proposed in Refs.~\cite{Goerbig04,CCChang04}. 

\section{Composite Fermi liquid}
\label{sec:CFL}

In this section we construct the coupled-wire Hamiltonian for the CFL \cite{Halperin93}. 
This is a compressible liquid state of electrons at filling fraction $\nu=1/M$ with $M$ even, where the composite fermions see a zero magnetic field on average and thus may form a Fermi liquid \cite{Kalmeyer92,Halperin93}. 
We begin with the coupled-wire construction for general $M$ and then specialize our attention to the filling fraction $\nu=1/2$ where electrons in the lowest Landau level are expected to have the PH symmetry in the limit of large Landau level spacing. 
The issue of the PH symmetry for the CFL at $\nu=1/2$ has been discussed \cite{Girvin84,Kivelson97,DHLee98,Rezayi00,Barkeshli15} and recently reexamined by replacing the nonrelativistic CFL with a Dirac theory \cite{Son15,ChongWang16a,ChongWang16b,Metlitski16,Mross16a,Geraedts16,Mulligan16,Murthy16,Balram16,Potter16,Levin17,ChongWang17a}. 
We show that our coupled-wire Hamiltonian for the CFL at $\nu=1/2$ is invariant under the PH transformation proposed in Sec.~\ref{sec:PHConjugate}, although the PH symmetry for coupled wires involves a translation and therefore is not a symmetry that is realized in an original microscopic Hamiltonian.
We also discuss the CFL of two-component bosons at $\nu=1/2+1/2$ \cite{ChongWang16b,Mross16a,Mross16b,Geraedts17}. 

\subsection{General construction at $\nu=1/M$}
\label{sec:GeneralCFL}

\begin{figure}
\includegraphics[clip,width=0.47\textwidth]{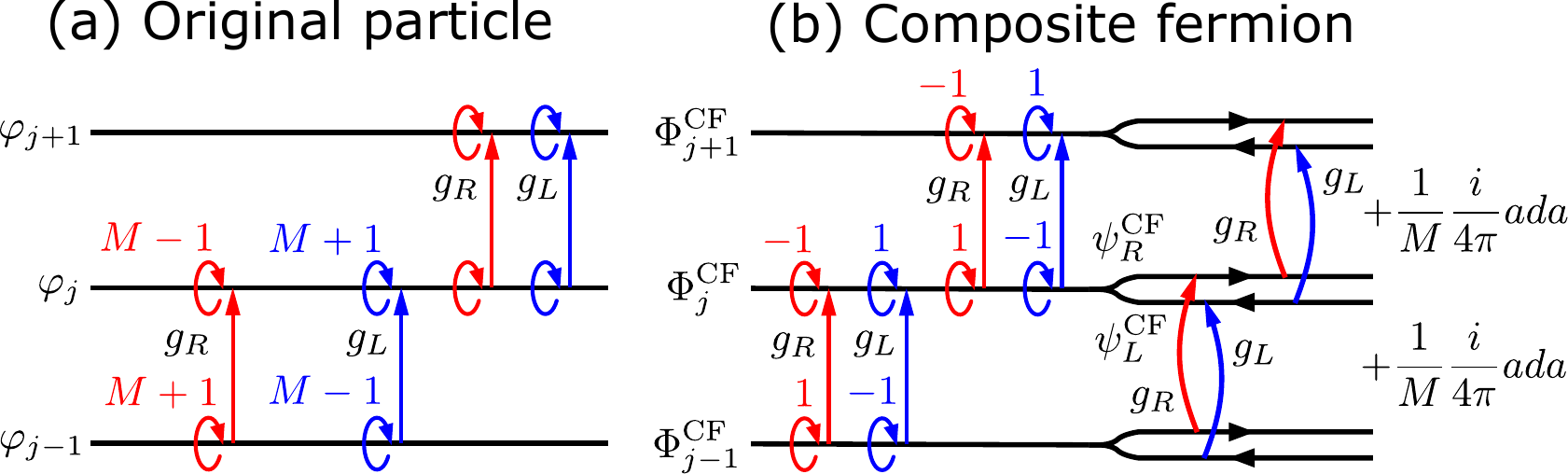}
\caption{Tunneling Hamiltonian for the CFL at $\nu=1/M$ in terms of (a) the original particles and (b) composite fermions.}
\label{fig:CFL}
\end{figure}

The composite fermions obtained through the $2\pi M$ flux attachment to fermions (bosons) with an even (odd) integer $M$ realize the Jain sequence $\nu=p/(pM+1)$ when they fill $p$ Landau levels. 
The tunneling Hamiltonian for the Jain sequence proposed in Sec.~\ref{sec:JainPicture} involves $p$-th neighbor hopping of particles. 
In the limit $p \to \infty$ where the filling fraction approaches $\nu=1/M$, the tunneling Hamiltonian becomes long ranged. 
Instead, we propose a simpler nearest-neighbor tunneling Hamiltonian \footnote{The same tunneling Hamiltonian for $M=2$ is considered in Ref.~\cite{Haller18} for a two-leg fermionic ladder at $\nu=1/2$.},
\begin{align} \label{eq:CFLTunneling}
H_1 &= \int_x \sum_j \Bigl[ g_R \kappa_j \kappa_{j+1} e^{i[ \varphi_j +(M+1) \theta_j -\varphi_{j+1} +(M-1) \theta_{j+1}]} \nonumber \\
&\ \ \ +g_L \kappa_j \kappa_{j+1} e^{i[ \varphi_j +(M-1) \theta_j -\varphi_{j+1} +(M+1) \theta_{j+1}]} +\textrm{H.c.} \Bigr], 
\end{align}
where we assign $\kappa_j$ to be a Majorana fermion for even $M$ while $\kappa_j=1$ for odd $M$. 
This tunneling Hamiltonian is schematically shown in Fig.~\ref{fig:CFL}. 
%%%%%%%%%%%%%%%%%%%%%%%%%%%%%%%%%%%%%
%%%%%%%%%%%%%%%%%%%%%%%%%%%%%%%%%%%%%
The operators in the tunneling Hamiltonian are chiral operators with a nonzero conformal spin and cannot open a gap even in the strong-coupling limit of $g_{R/L}$. 
Thus the resulting state is expected to be gapless.
Indeed, applying the $2\pi M$ flux attachment transformation \eqref{eq:TransFluxAttach} with $m=M$, we obtain 
\begin{align}
H_1 &= \int_x \sum_j \Bigl[ g_R \kappa_j \kappa_{j+1} e^{i(\Phi^\CF_j +\Theta^\CF_j -\Phi^\CF_{j+1} -\Theta^\CF_{j+1})} \nonumber \\
&\qquad\qquad  +g_L \kappa_j \kappa_{j+1} e^{i(\Phi^\CF_j -\Theta^\CF_j -\Phi^\CF_{j+1} +\Theta^\CF_{j+1})} +\textrm{H.c.} \Bigr],
\end{align}
which can be written in terms of the composite fermion fields \eqref{eq:CFField} as 
\begin{align} \label{eq:CFLTunnelingCF}
H_1 &= 2\pi \alpha e^{i\pi M/2} \int_x \sum_j \Bigl[ g_R \psi^\CF_{R,j} \psi^{\CF \dagger}_{R,j+1} \nonumber \\
&\hspace*{30mm} +g_L \psi^\CF_{L,j} \psi^{\CF \dagger}_{L,j+1} +\textrm{H.c.} \Bigr]. 
\end{align}
This Hamiltonian gives a simple nearest-neighbor hopping of the composite fermions within the same branch.
With the kinetic action given in Eq.~\eqref{eq:ActionLaughlinCF}, 
\begin{align} \label{eq:CFLActionCF}
S_0 &= \int_{\tau,x} \sum_j \biggl[ \sum_{r=\pm} \psi^{\CF \dagger}_{r,j} \Bigl( \partial_\tau -\frac{i}{2} (Sa_{0,j-1/2}) -iA^\ext_{0,j} \Bigr) \psi^\CF_{r,j} \nonumber \\
&\qquad\qquad -\sum_{r=\pm} r iv \psi^{\CF \dagger}_{r,j} (\partial_x -a_{1,j} -A^\ext_{1,j}) \psi^\CF_{r,j} \nonumber \\
&\qquad\qquad +\frac{i}{2\pi M} a_{1,j} (\Delta a_{0,j-1/2}) +\cdots \biggr],
\end{align}
the coupled-wire model may be seen as a discrete version of the Chern-Simons CFL theory proposed by Halperin, Lee, and Read \cite{Halperin93} in the $a_2=A^\ext_2=0$ gauge. 
Similarly to the hierarchy states discussed so far, there is a caveat that the tunneling term \eqref{eq:CFLTunnelingCF}, consisting of bilinears of the composite fermion fields, are not relevant in the RG sense in the limit of decoupled wires. 
Hence the ellipsis in Eq.~\eqref{eq:CFLActionCF} is understood to contain some interwire forward scattering interactions of original particles that make the coupling constants $g_{R/L}$ relevant.

\begin{figure}
\includegraphics[clip,width=0.4\textwidth]{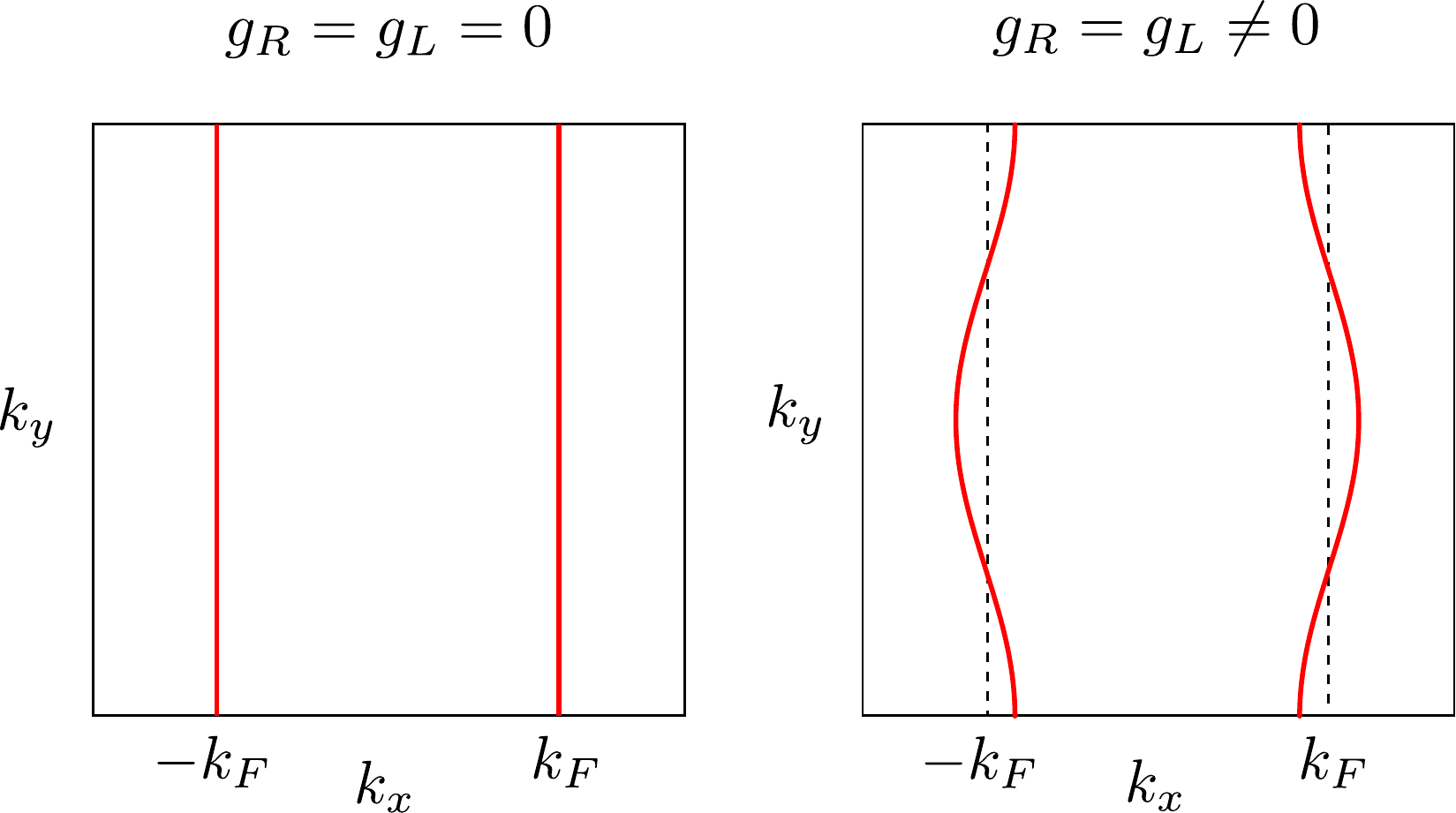}
\caption{The Fermi surface of the composite fermion from Eq.~\eqref{eq:BandCF}. 
For $g_R=g_L=0$, the composite fermion has linear dispersions at $k_x=\pm k_F$ while a flat dispersion along $k_y$. 
Nonzero $g_R = g_L$ develop an open Fermi surface with the shape of cosine.}
\label{fig:CFSurface}
\end{figure}

Applying a mean-field approximation to the gauge field $a_\mu = \langle a_\mu \rangle$ and neglecting forward scattering interactions, we can examine the band structure of the composite fermions. 
We here set $\langle a_\mu \rangle =0$ as a nonvanishing average merely shifts the origin of momentum space.
The mean-field Hamiltonian is given by
\begin{align}
H_\textrm{MF} &= \int \frac{dk_x}{2\pi} \sum_{k_y} \left[ \psi^{\CF \dagger}_{R,\bfk} \mathcal{H}_R(\bfk) \psi^\CF_{R,\bfk} +\psi^{\CF \dagger}_{L,\bfk} \mathcal{H}_L(\bfk) \psi^\CF_{L,\bfk} \right]
\end{align}
with
\begin{align} \label{eq:BandCF}
\begin{split}
\mathcal{H}_R(\bfk) &= v(k_x -k_F) -\tilde{g}_R \cos (k_y -\pi M/2), \\
\mathcal{H}_L(\bfk) &= -v(k_x +k_F) -\tilde{g}_L \cos (k_y -\pi M/2), 
\end{split}
\end{align}
where $\tilde{g}_{R/L} = 4\pi \alpha g_{R/L}$ and the chemical potential for the composite fermions is set to be at zero energy.
The composite-fermion's Fermi surface is schematically shown in Fig.~\ref{fig:CFSurface}. 
%%%%%%%%%%%%%%%%%%%%%%%%%%%%%%%%%%%%%
%%%%%%%%%%%%%%%%%%%%%%%%%%%%%%%%%%%%%
The interaction $g_{R/L}$ cannot exceed an energy cutoff \red{$\Lambda \sim v/\alpha$} below which the linearized approximation for the dispersion in individual quantum wires is justified. 
This imposes a restriction that one can only obtain the CFL with an \emph{open} Fermi surface from the coupled-wire construction. 
The dispersion of the composite fermions can be quadratic in the $y$ direction while it remains linear in the $x$ direction.

\subsection{Fermion at $\nu=1/2$}
\label{sec:AntiCFL}

We here focus on the CFL at $\nu=1/2$. 
In the limit of vanishing Landau level mixing, a half-filled Landau level at $\nu=1/2$ possesses an exact PH symmetry. 
However, the Halperin-Lee-Read theory for the CFL at $\nu=1/2$ \cite{Halperin93} is not explicitly PH symmetric, 
and the PH conjugate of the CFL, called the anti-CFL or the composite hole liquid, has also been discussed \cite{Barkeshli15,Mulligan16}. 
Furthermore, it has been argued that the CFL can have an emergent PH symmetry at low energies \cite{ChongWang17a,Kumar18}. 

We have defined the PH transformation for coupled-wire models in Eq.~\eqref{eq:PHTrans}.
We note that the PH transformation does not represent a microscopic symmetry; in other words there is no way to regularize the transformation \eqref{eq:PHTrans} in a purely 2D lattice system with short-range interactions. 
A simple way to see this is to examine how the PH transformation acts on the electron operators. 
Equation \eqref{eq:PHTransFermi} tells us that a left-moving fermion $\psi_{L,j}$ is transformed to $\psi^\dagger_{R,j}$ in the same wire, while a right-moving fermion $\psi_{R,j}$ is transformed to $\psi^\dagger_{L,j+1}$ in a neighboring wire. 
Such a PH transformation cannot be implemented for a local fermionic operator $\psi_j \sim e^{ik_F x} \psi_{R,j} +e^{-ik_F x} \psi_{L,j}$. 
Nevertheless, the PH transformation \eqref{eq:PHTrans} can be used to derive the PH conjugates of FQH states with proper topological properties as discussed in Sec.~\ref{sec:PHConjugate}. 
It is thus interesting to examine how the PH transformation acts on our coupled-wire model for the CFL at $\nu=1/2$. 

\begin{figure}
\includegraphics[clip,width=0.47\textwidth]{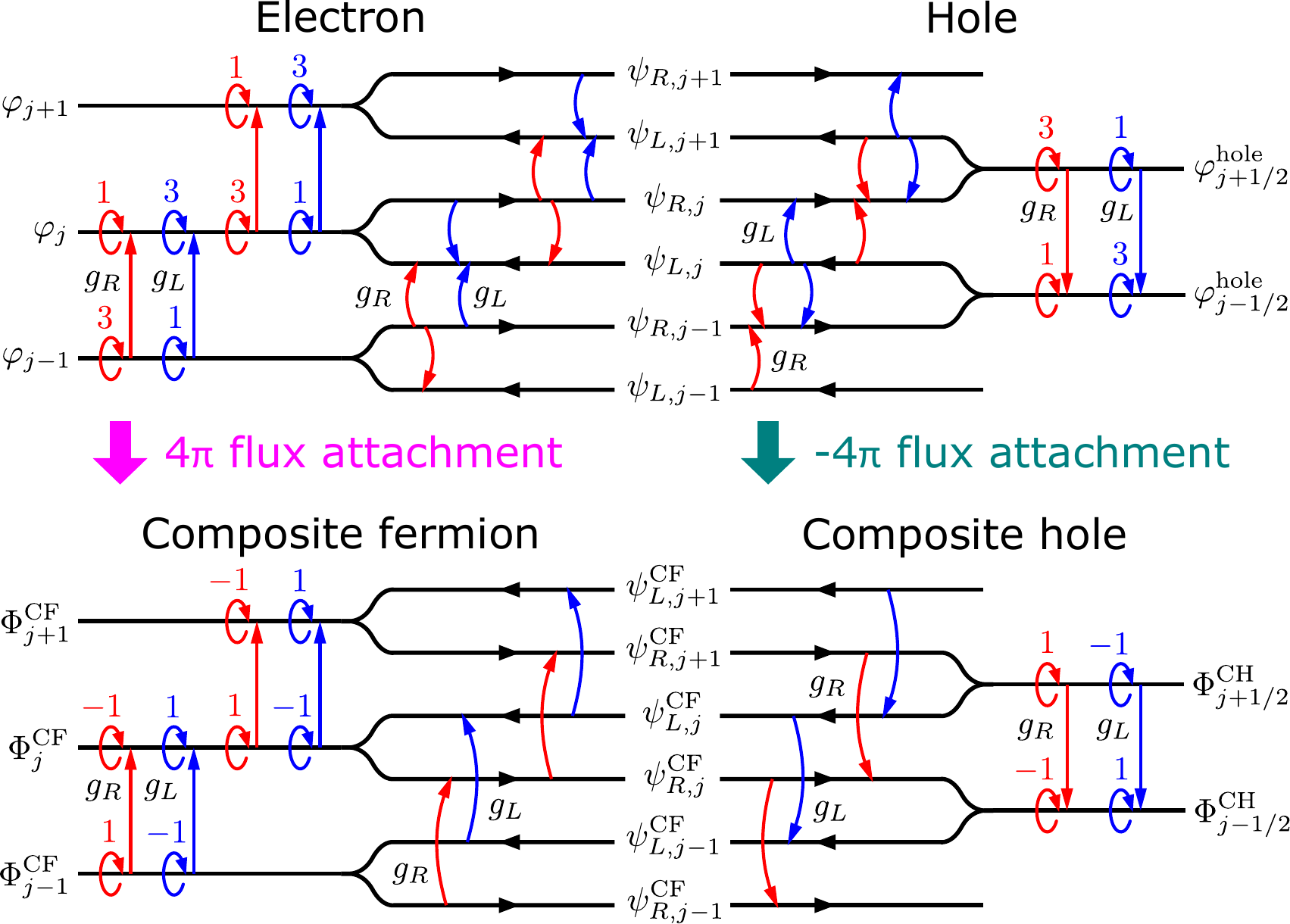}
\caption{Tunneling Hamiltonian for the CFL at $\nu=1/2$. 
After $4\pi$ flux attachment to electrons, the Hamiltonian is written in terms of composite fermions. 
Starting from the hole picture, we can also write the Hamiltonian in terms of composite holes by attaching $-4\pi$ flux to the holes.}
\label{fig:CFL_CHL}
\end{figure}

Our coupled-wire model for the CFL at $\nu=1/2$ has the kinetic action,
\begin{align} \label{eq:ActionHFLL}
S_0 &= \int_{\tau,x} \sum_j \biggl\{ \frac{i}{\pi} \partial_x \theta_j (\partial_\tau \varphi_j -A^\ext_{0,j}) +\frac{v}{2\pi} (\partial_x \varphi_j -A^\ext_{1,j})^2 \nonumber \\
&\ \ \ +\frac{u}{2\pi} (\partial_x \theta_j)^2 +\frac{\tu-v}{8\pi} \left[ (\Delta \partial_x \varphi_j) -(S \partial_x \theta_j) \right]^2 \biggr\},
\end{align}
and tunneling Hamiltonian, 
\begin{align} \label{eq:TunnelingHFLL}
H_1 &= \int_x \sum_j \Bigl[ g_R \kappa_j \kappa_{j+1} e^{i(\varphi_j +3\theta_j -\varphi_{j+1} +\theta_{j+1})} \nonumber \\
&\qquad\qquad +g_L \kappa_j \kappa_{j+1} e^{i(\varphi_j +\theta_j -\varphi_{j+1} +3\theta_{j+1})} +\textrm{H.c.} \Bigr],
\end{align}
which is depicted in Fig.~\ref{fig:CFL_CHL}. 
%%%%%%%%%%%%%%%%%%%%%%%%%%%%%%%%%%%%%
%%%%%%%%%%%%%%%%%%%%%%%%%%%%%%%%%%%%%
Here we have added the interwire forward scattering term with the coupling $\tu-v$ as a tuning parameter for the kinetic action. 
For simplicity, we have omitted other forward scattering interactions that would be required to make the tunneling terms relevant in the RG sense, and we here concentrate on the above simple form of the kinetic action.
We then rewrite this theory in terms of the hole fields defined in Eq.~\eqref{eq:HoleFromElectron}. 
The kinetic action \eqref{eq:ActionHFLL} becomes 
\begin{align} \label{eq:ActionHFLLHole}
S_0 &= \int_{\tau,x} \sum_j \biggl[ \frac{i}{\pi} \partial_x \theta^\hole_{j+1/2} \Bigl( \partial_\tau \varphi^\hole_{j+1/2} +\frac{1}{2} (SA^\ext_{0,j}) \Bigr) \nonumber \\
&\ \ \ +\frac{v}{2\pi} \Bigl( \partial_x \varphi^\hole_{j+1/2} +\frac{1}{2} (SA^\ext_{1,j}) ) \Bigr)^2 +\frac{\tu}{2\pi} (\partial_x \theta^\hole_{j+1/2})^2 \nonumber \\
&\ \ \ +\frac{u-v}{8\pi} \Bigl( (\Delta \partial_x \varphi^\hole_{j-1/2}) +(S\partial_x \theta^\hole_{j-1/2}) \Bigr)^2 \nonumber \\
&\ \ \ +\frac{i}{4\pi} (SA^\ext_{1,j}) (\Delta A^\ext_{0,j}) +\cdots \biggr],
\end{align}
where we have omitted higher-order derivative terms containing $A^\ext_\mu$.
The tunneling Hamiltonian \eqref{eq:TunnelingHFLL} is now rewritten as 
\begin{align} \label{eq:TunnelingHFLLHole}
H_1 &= \int_x \sum_j \Bigl[ g_R \kappa_j \kappa_{j+1} e^{i(\varphi^\hole_{j-1/2} -\theta^\hole_{j-1/2} -\varphi^\hole_{j+1/2} -3\theta^\hole_{j+1/2})} \nonumber \\
&\ \ \ +g_L \kappa_j \kappa_{j+1} e^{i(\varphi^\hole_{j-1/2} -3\theta^\hole_{j-1/2} -\varphi^\hole_{j+1/2} -\theta^\hole_{j+1/2})} +\textrm{H.c.} \Bigr]. 
\end{align}
We thus find that the CFL action defined by Eqs.~\eqref{eq:ActionHFLL} and \eqref{eq:TunnelingHFLL} is \emph{symmetric} under the PH transformation \eqref{eq:PHTrans} when $u=\tu$ and $g_R=g_L$ and the Klein factors are appropriately assigned. 
In deriving the CFL action in terms of the hole fields in Eqs.~\eqref{eq:ActionHFLLHole} and \eqref{eq:TunnelingHFLLHole}, we have employed the $2\pi$ flux attachment to electrons and the vortex duality transformation so that we can identify the vortices attached by $-2\pi$ flux with holes, as argued in Sec.~\ref{sec:PHConjugate}. 
In this intermediate step, we obtain the CFL action in terms of the composite bosons and its vortices. 
Actually, these bosonic formulations of the CFL action turn out to be self dual and give another hallmark of the PH symmetry in the fermionic theory \cite{Mross17a}. 
The detailed discussion is provided in Appendix~\ref{app:HoleTheory}.

We can thus write the CFL action in terms of either the composite fermions with $M=2$ discussed above, 
\begin{align}
\Phi^\CF_j &= \varphi_j +2\sum_{j' \neq j} \sgn (j'-j) \theta_{j'}, \ \ \ \Theta^\CF_j = \theta_j, 
\end{align}
or the composite holes, which can be obtained by attaching the $-4\pi$ flux to the hole fields, 
\begin{align} \label{eq:CompositeHole}
\begin{split}
\Phi^\CH_{j+1/2} &= \varphi^\hole_{j+1/2} -2\sum_{j' \neq j} \sgn (j'-j) \theta^\hole_{j'+1/2}, \\
\Theta^\CH_{j+1/2} &= \theta^\hole_{j+1/2}.
\end{split}
\end{align}
The chiral bosonic fields corresponding to the composite fermions and the composite holes are related to each other in a local manner, 
\begin{align}
\begin{split}
\Phi^\CH_{j+1/2} +\Theta^\CH_{j+1/2} &= \Phi^\CF_{j+1} +\Theta^\CF_{j+1}, \\
\Phi^\CH_{j+1/2} -\Theta^\CH_{j+1/2} &= \Phi^\CF_j -\Theta^\CF_j.
\end{split}
\end{align}
As a result, the open Fermi surface of the composite holes has the same shape as that of the composite fermions discussed above. 
However, the physical origins of the Chern-Simons gauge fields are different between two formulations, as they have Chern-Simons terms with opposite signs. 
We conclude that, within the coupled-wire approach, the CFL and the composite hole liquid belong to the same phase when the PH symmetry in the sense of Eq.~\eqref{eq:PHTransFermi} exists, since both theories can be obtained from the same microscopic Hamiltonian. 
However, in the presence of boundaries, the CFL action violates the PH symmetry, and there may be chiral fermion edge modes from a filled Landau level for the composite hole liquid as seen from Fig.~\ref{fig:CFL_CHL}. 

We now make a comparison between our model and the coupled-wire model with a single Dirac cone at $\nu=1/2$ discussed in Refs.~\cite{Mross16a,Mross17a}. 
In fact, the PH transformation \eqref{eq:PHTransFermi} is essentially the same as what is defined in Refs.~\cite{Mross16a,Mross17a}, and the flux attachment transformation \eqref{eq:TransFluxAttach} with $m=2$ is essentially the duality transformation defined in Refs.~\cite{Mross16a,Mross17a}. 
The apparent distinction just stems from where the fermionic fields $\psi_{R/L}$ are defined; in our model each wire has both a right-going fermion mode and a left-going fermion mode, while the chiral fermion modes are defined separately on neighboring wires in an alternating manner in Refs.~\cite{Mross16a,Mross17a}.
When a gauge field is introduced on each wire to make the theory local after the flux attachment or duality transformation, a Chern-Simons term remains in our model while it does not in the model in Refs.~\cite{Mross16a,Mross17a}. 
In the latter model where each wire has only a chiral fermion mode, the simplest Hamiltonian with fermion hoppings between neighboring wires yields a single Dirac cone. 
Such a system is not regularized on a lattice but it gives an effective description of the surface of a certain topological crystalline insulator or the Son's theory \cite{Son15} for the half-filled Landau level \cite{Mross16a,Mross17a}. 
On the other hand, we have restricted ourselves to considering coupled-wire models that can be realized in a strictly 2D lattice system such that each wire must consist of right- and left-going fermions. 
This naturally led to the CFL with an open Fermi surface at $\nu=1/2$ under the assumption of the uniform flux configuration.

At this stage, it is not clear what one can say from our coupled-wire analysis of the CFL about the PH symmetry in the actual half-filled Landau level. 
As mentioned above, our PH transformation is implemented in a nonlocal way involving a ``half'' translation of wires. 
Therefore, even after taking the continuum limit with respect to discrete wire variables, our model does not necessarily describe the same physics as in the Landau level where the PH symmetry acts locally in Landau level variables (while it still acts nonlocally in microscopic variables). 
A similar subtlety has been pointed out in a coupled-wire model for the surface topological order of interacting 3D topological superconductors, where the 32-fold classification has been obtained for the antiferromagnetic time-reversal symmetry while the classification is known to be 16-fold for the usual time-reversal symmetry \cite{Sahoo16}.
Another issue is the shape of the Fermi surface. 
In our approach, we can only deal with an open Fermi surface with a linear dispersion in one direction and a quadratic dispersion in the other direction, which is topologically different from a closed Fermi surface.

\subsection{Two-component boson at $\nu=1/2+1/2$}
\label{sec:CFLTwoComp}

In analogy with fermions where the PH conjugate is taken with respect to a filled Landau level (IQH state), we may also define the PH conjugate for bosons, which is now taken with respect to a bosonic IQH state \cite{Senthil13} whose smallest filling is $\nu=2$.
One can then expect to apply a similar argument for the PH symmetry to bosons at $\nu=1$. 
Although the PH symmetry is not an exact symmetry for bosons in the Landau level, there can be an emergent PH symmetry at $\nu=1$ in the long wave-length limit as it is expected in the Read's theory for the lowest Landau level \cite{Read98} and has been discussed recently in Refs.~\cite{Murthy16,ChongWang16b,Mross16b}. 
The case for two-component bosons at $\nu=1/2+1/2$ is of particular interest, since a two-flavor Dirac theory, which is a natural extension of the Son's Dirac theory for fermions at $\nu=1/2$ \cite{Son15}, becomes a good candidate for an incompressible state that manifests a PH symmetry \cite{ChongWang16b,Mross16a,Mross16b}. 
We here discuss a coupled-wire model of the CFL for bosons at $\nu=1/2+1/2$ with a kind of PH symmetry that cannot be realized in a 2D lattice system, in a similar spirit to fermions at $\nu=1/2$ discussed above. 

We consider two species of bosonic fields $\varphi^\sigma_j(x)$ and $\theta^\sigma_j(x)$, which are labeled by up or down spin $\sigma=\ua,\da$. 
The CFL at $\nu=1/2+1/2$ may be described by the action with the kinetic terms, 
\begin{align} \label{eq:ActionTwoComp}
S_0 &= \int_{\tau,x} \sum_j \sum_{\sigma=\ua,\da} \biggl[ \frac{i}{\pi} \partial_x \theta^\sigma_j (\partial_\tau \varphi^\sigma_j -A^\sigma_{0,j}) \nonumber \\
&\ \ \ +\frac{v}{2\pi} (\partial_x \varphi^\sigma_j -A^\sigma_{1,j})^2 +\frac{u}{2\pi} (\partial_x \theta^\sigma_j)^2 \nonumber \\
&\ \ \ +\frac{u -v}{8\pi} \Bigl( (\Delta \partial_x \varphi^\sigma_j) -(S\partial_x \theta^{-\sigma}_j) \Bigr)^2 \biggr], 
\end{align}
and the tunneling terms, 
\begin{align} \label{eq:TunnelingTwoComp}
H_1 &= g\int_x \sum_j \sum_{\sigma=\ua,\da} \Bigl[ e^{i(\varphi^\sigma_j +2\theta^\sigma_j +\theta^{-\sigma}_j -\varphi^\sigma_{j+1} +\theta^{-\sigma}_{j+1})} \nonumber \\
&\ \ \ +e^{i(\varphi^\sigma_j +\theta^{-\sigma}_j -\varphi^\sigma_{j+1} +2\theta^\sigma_{j+1} +\theta^{-\sigma}_{j+1})} +\textrm{H.c.} \Bigr],
\end{align}
where $-\sigma$ stands for $\da$($\ua$) for $\sigma= \ua$($\da$) and we have coupled the bosonic fields with external gauge fields $A^\sigma_\mu$ for each species. 
We have assumed that the action is symmetric under the exchange of two species, while a more general form of the action is considered in Appendix~\ref{app:HoleTheoryBoson}, where the detailed derivation of the hole theory is provided. 
We here simply state the strategy of the derivation and consequences. 
This action can be regarded as the CFL with two open Fermi surfaces, each of which carries different spins, coupled to a single Chern-Simons gauge field by applying the $2\pi$ flux attachment to both species of bosons. 
The hole description of this CFL action is obtained by applying the mutual $2\pi$ flux attachment and the subsequent vortex duality to each species of mutual composite bosons. 
The resulting vortex theory has a mutual Chern-Simons term with the opposite sign to that for the mutual composite bosons. 
Integrating out the mutual Chern-Simons gauge fields in the vortex theory yields the desired hole theory. 
The kinetic action \eqref{eq:ActionTwoComp} is then written as
\begin{align}
S_0 &= \int_{\tau,x} \sum_{j,\sigma} \biggl[ \frac{i}{\pi} \partial_x \theta^{\hole,\sigma}_{j+1/2} \Bigl( \partial_\tau \varphi^{\hole,\sigma}_{j+1/2} +\frac{1}{2} (SA^\sigma_{0,j}) \Bigr) \nonumber \\
&\ \ \ +\frac{v}{2\pi} \Bigl( \partial_x \varphi^{\hole,\sigma}_{j+1/2} +\frac{1}{2} (SA^\sigma_{1,j}) \Bigr)^2 +\frac{u}{2\pi} (\partial_x \theta^{\hole ,\sigma}_{j+1/2})^2 \nonumber \\
&\ \ \ +\frac{u-v}{8\pi} \Bigl( (\Delta \partial_x \varphi^{\hole,\sigma}_{j-1/2}) +(S \partial_x \theta^{\hole,\sigma}_{j-1/2}) \Bigr)^2 \nonumber \\
&\ \ \ +\frac{i}{4\pi} (SA^\sigma_{1,j}) (\Delta A^{-\sigma}_{0,j}) +\cdots \biggr],
\label{S_0 of holes at 1/2+1/2}
\end{align}
where we have dropped higher-order derivative terms involving the external gauge fields. 
The tunneling Hamiltonian \eqref{eq:TunnelingTwoComp} reads as
\begin{align}
H_1 &= g\int_x \sum_{j,\sigma} \Bigl[ 
e^{i\left(\varphi^{\textrm{hole} ,\sigma}_{j-1/2} -2\theta^{\textrm{hole} ,\sigma}_{j-1/2} -\theta^{\textrm{hole} ,-\sigma}_{j-1/2} -\varphi^{\textrm{hole} ,\sigma}_{j+1/2} -\theta^{\textrm{hole} ,-\sigma}_{j+1/2}\right)} \nonumber \\
&\ \ \ +e^{i\left(\varphi^{\textrm{hole},\sigma}_{j-1/2} -\theta^{\textrm{hole},-\sigma}_{j-1/2} -\varphi^{\textrm{hole},\sigma}_{j+1/2} -2\theta^{\textrm{hole},\sigma}_{j+1/2} -\theta^{\textrm{hole},-\sigma}_{j+1/2}\right)} +\textrm{H.c.}
 \Bigr].
\label{H_1 of holes at 1/2+1/2}
\end{align}
In the kinetic action, we find that the bosonic fields for each species carry the opposite charge compared with the original bosons, and there exists a discrete analog of the mutual Chern-Simons term $(i/2\pi) \epsilon_{\mu \nu \lambda} A^\ua_\mu \partial_\nu A^\da_\lambda$ in the $A^\sigma_2=0$ gauge, which produces the Hall response of the bosonic IQH state \cite{Senthil13}. 
As there is no dynamical gauge fields in Eqs.~\eqref{S_0 of holes at 1/2+1/2} and \eqref{H_1 of holes at 1/2+1/2}, this action must be related to the original action by a local transformation of the bosonic fields, which is given by 
\begin{align}
\begin{split}
\varphi^{\hole,\sigma}_{j+1/2} &= -\frac{1}{2} (\varphi^\sigma_j +\theta^{-\sigma}_j +\varphi^\sigma_{j+1} -\theta^{-\sigma}_{j+1}), \\
\theta^{\hole,\sigma}_{j+1/2} &= -\frac{1}{2} (\varphi^{-\sigma}_j +\theta^\sigma_j -\varphi^{-\sigma}_{j+1} +\theta^\sigma_{j+1}).
\end{split}
\end{align}
We then find that the CFL action is symmetric under the transformation, 
\begin{align} \label{eq:PHTransTwoComp}
\varphi^\sigma_j \to -\varphi^{\hole,\sigma}_{j+1/2}, \qquad \theta^\sigma_j \to \theta^{\hole,\sigma}_{j+1/2}, 
\end{align}
with complex conjugation. 
This transformation may be regarded as a coupled-wire version of the antiunitary PH transformation for two-component bosons in the following way. 
Let us introduce new bosonic fields by $\phi^\sigma_j = \varphi^\sigma_j +\theta^{-\sigma}_j$ and $\tilde{\phi}^\sigma_j = \varphi^\sigma_j -\theta^{-\sigma}_j$, which satisfy the commutation relations $[\partial_x \phi^\ua_j(x), \phi^\da_{j'}(x')] = -[\partial_x \tilde{\phi}^\ua_j(x), \tilde{\phi}^\da_{j'}(x')] =2i\pi \delta_{j,j'} \delta (x-x')$ while the other commutators vanish. 
These bosonic fields actually correspond to gapless edge modes of the bosonic IQH state at $\nu=1+1$, and $\phi^\sigma_j$ and $\tilde{\phi}^\sigma_j$ have the opposite chirality to each other (see also Ref.~\cite{Fuji16}). 
If we define bosonic operators by $b_{\sigma,j}=e^{i\phi^\sigma_j}$ and $\tilde{b}_{\sigma,j}=e^{i\tilde{\phi}^\sigma_j}$, the transformation \eqref{eq:PHTransTwoComp} acts on these bosonic operators as $b_{\sigma,j} \to \tilde{b}^\dagger_{\sigma,j+1}$ and $\tilde{b}_{\sigma,j} \to b_{\sigma,j}^\dagger$. 
Thus it can be seen as a PH transformation \cite{Mross16a}. 
However, due to a reason similar to the one for the PH transformation in the fermionic case, such a transformation cannot be properly defined in purely 2D lattice systems. 

When the numbers of each species of bosons are not separately conserved, i.e., for the case of single-component bosons, the above derivation of PH conjugate states through the mutual $2\pi$ flux attachment and vortex duality is not appropriate. 
Nevertheless, we may still define a PH transformation by looking at the bosonic fields corresponding to edge modes of the bosonic IQH state. 
For example, for the single-component case, the bosonic IQH state in fact belongs to the same universality class as the bosonic negative Jain hierarchy state at $\nu=2$, whose tunneling Hamiltonian is given in Eq.~\eqref{eq:TunnelingNegativeJain} with $p=2$ and $q=2$.
Its edge states are given by $\phi^1_l = \varphi_{2l} +2\theta_{2l+1}$, $\phi^2_l = \varphi_{2l+1}$, $\tilde{\phi}^1_{l} = \varphi_{2l}$, and $\tilde{\phi}^2_l = \varphi_{2l+1} -2\theta_{2l}$. 
We then define the PH transformation by $\phi^I_l \to \tilde{\phi}^I_{l+1}$ and $\tilde{\phi}^I_l \to \phi^I_l$ with complex conjugation. 
The CFL Hamiltonian with a single Fermi surface, Eq.~\eqref{eq:CFLTunneling} with $M=1$, does not have the PH symmetry in this sense, but one can see, by extending the construction of the CFL in Sec.~\ref{sec:GeneralCFL}, that a Hamiltonian with two Fermi surfaces does. 
This transformation can also be used to obtain the PH conjugates of several other bosonic FQH states. 
When the above PH transformation is applied to the tunneling Hamiltonian for the bosonic Laughlin $\nu=1/2$ state in Eq.~\eqref{eq:TunnelingLaughlin}, the transformed Hamiltonian turns out to describe the same topological order as a negative Jain state at $\nu=3/2$. 
This PH transformation may be used to obtain the bosonic anti-Pfaffian state from the coupled-wire Hamiltonian for the Pfaffian state discussed below.

\section{Pfaffian states}
\label{sec:Pfaffian}

As discussed in the previous section, the composite fermions obtained via the $2\pi M$ flux attachment at filling fraction $\nu=1/M$ see a vanishing magnetic field on average. 
Aside from forming a Fermi liquid, the composite fermions have another option of forming a superconducting state.
Read and Green \cite{Read00} have argued that a spinless chiral $p$-wave superconductor of the composite fermions with orbital angular momentum $\ell=-1$ is the Moore-Read Pfaffian state, which is known to harbor non-Abelian anyons as its quasiparticles \cite{Moore91}. 
In this section, we first confirm that the coupled-wire model proposed by Teo and Kane \cite{Teo14} for the Pfaffian state is indeed consistent with this picture in terms of the composite fermions.
We note that more phenomenological coupled-wire models for pairing states have been recently proposed in Ref.~\cite{Kane17}.
We then apply the idea of hierarchy construction in Sec.~\ref{sec:HaldaneHalperin} to the Pfaffian state to obtain coupled-wire models for the so-called Bonderson-Slingerland hierarchy \cite{Bonderson08}. 
We also construct a coupled-wire model for the anti-Pfaffian state at $\nu=1/2$, which is the PH conjugate of the Pfaffian state \cite{Levin07,SSLee07}, and discuss a possible way to obtain other pairing states of the composite fermions.

\subsection{Pfaffian state}
\label{sec:GeneralPfaffian}

\begin{figure}
\includegraphics[clip,width=0.42\textwidth]{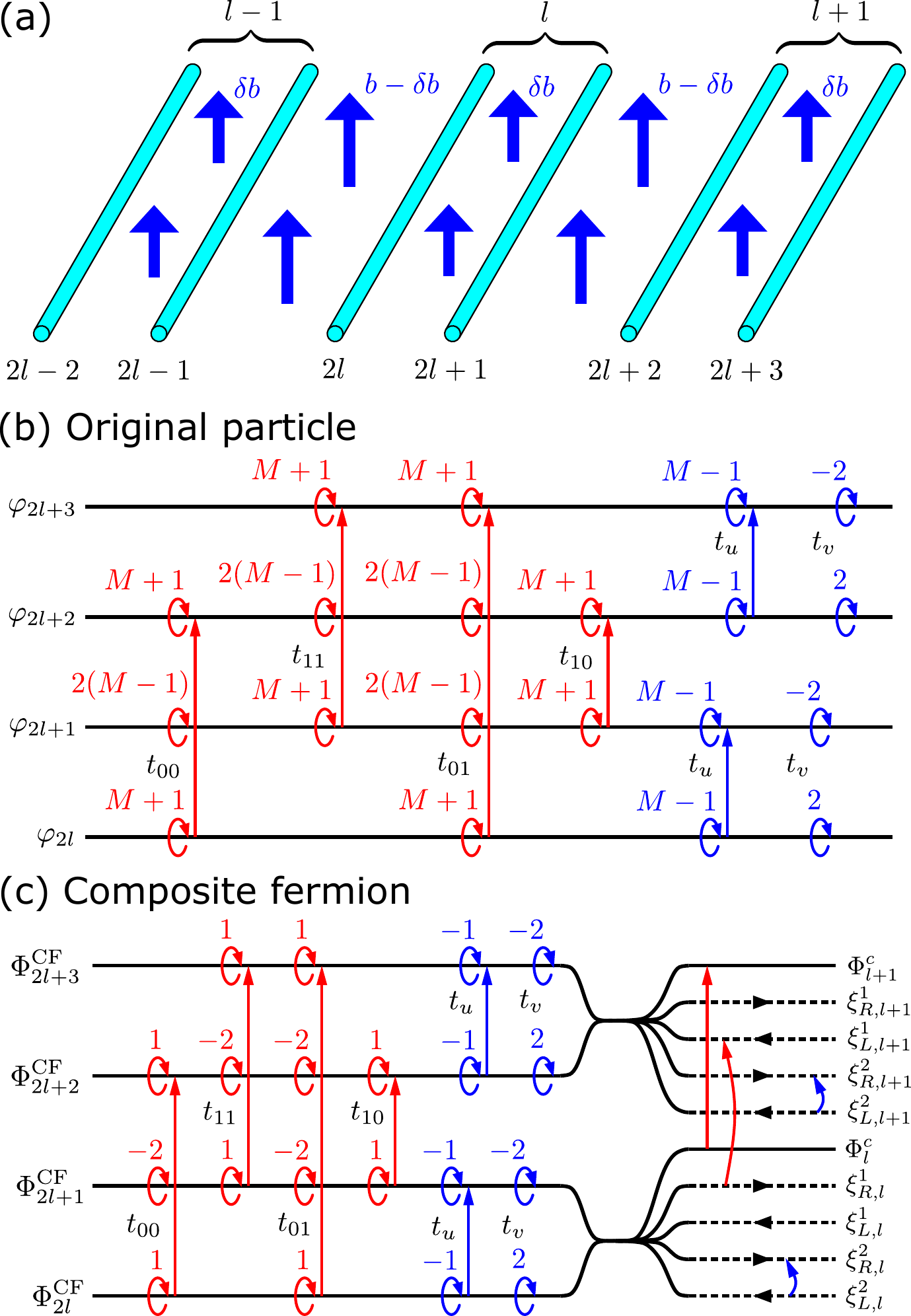}
\caption{Tunneling Hamiltonian for the Pfaffian state at $\nu=1/M$. 
(a) Quantum wires are placed in a magnetic field of the average flux $b=2Mk_F$ with modulation $\delta b = 2k_F$. 
The Hamiltonian is given in terms of the bosonic fields corresponding to (a) the original particles and (b) composite fermions. 
In the latter case, two adjacent wires can be decomposed into a bosonic charge mode $\Phi^c_l$ (solid line) and neutral chiral Majorana modes $\xi^{1,2}_{p,l}$ (dashed arrow).}
\label{fig:Pfaffian}
\end{figure}

We first review the construction of the Moore-Read Pfaffian states from coupled wires by Teo and Kane \cite{Teo14}. 
We consider fermions at $\nu=1/M$ for an even integer $M$ (bosons for an odd integer $M$). 
In this case, we have to start with an array of wires equally spaced in a spatially modulated magnetic field or of wires unequally spaced in a uniform magnetic field as schematically shown in Fig.~\ref{fig:Pfaffian}~(a).
%%%%%%%%%%%%%%%%%%%%%%%%%%%%%%%%%%%%%
%%%%%%%%%%%%%%%%%%%%%%%%%%%%%%%%%%%%%
The tunneling Hamiltonian is given by 
\begin{align} \label{eq:TunnelingPfaffian}
H_1 &= \int_x \sum_l \Biggl\{ \sum_{a=0}^1 \sum_{b=0}^1 t_{ab} \kappa_{2l+a} \kappa_{2l+b+2} \nonumber \\
&\qquad\quad \times \!\exp \!\left[i\!\left(\! \varphi_{2l+a} \! +\sum_{c=0}^3 \Gamma_{ab}^c(M) \theta_{2l+c} -\varphi_{2l+b+2}\!\right)\! \right] \nonumber \\
&\ \ \ +t_u \kappa_{2l} \kappa_{2l+1} e^{i[\varphi_{2l} +(M-1) \theta_{2l} -\varphi_{2l+1} +(M-1) \theta_{2l+1}]} \nonumber \\
&\ \ \ +t_v e^{i(2\theta_{2l} -2\theta_{2l+1})} +\textrm{H.c.} \Biggr\}, 
\end{align}
where $\bfGamma_{ab}(M)$ are integer vectors given by 
\begin{align} \label{eq:VectorGammaPfaffian}
\begin{pmatrix} \bfGamma_{00} \\ \bfGamma_{11} \\ \bfGamma_{01} \\ \bfGamma_{10} \end{pmatrix} 
= \begin{pmatrix} M+1 & 2(M-1) & M+1 & 0 \\ 0 & M+1 & 2(M-1) & M+1 \\ M+1 & 2(M-1) & 2(M-1) & M+1 \\ 0 & M+1 & M+1 & 0 \end{pmatrix}.
\end{align}
Here we have assumed that the coupling constants $t_{ab}$, $t_u$, and $t_v$ are complex numbers.
This Hamiltonian is pictorially given in Fig.~\ref{fig:Pfaffian}~(b). 
Again, the Klein factors $\kappa_j$ are chosen to be Majorana fermions obeying $\{ \kappa_j, \kappa_{j'} \} = \delta_{jj'}$ for the fermionic case while $\kappa_j=1$ for the bosonic case. 
As reviewed in Appendix~\ref{app:PfaffianTeoKane}, when the coupling constants are fine tuned, Teo and Kane showed that this tunneling Hamiltonian leaves a chiral bosonic field carrying charge and a neutral Majorana fermion field propagating in the same direction at the boundaries \cite{Teo14}. 
They also argued that the bare $2k_F$ backscattering operator $e^{i2\theta_j}$ creates a pair of quasiparticles with charge $\pm 1/2M$ and its neutral sector corresponds to spin fields of the Ising CFT, while the spin field does not admit an explicit bosonic (vertex) representation due to its non-Abelian nature.
The $M=0$ case corresponds to the chiral $p$-wave superconductor in which a single chiral Majorana fermion mode is left at the boundary while there exist bulk collective excitations (Goldstone modes) from the condensate of charge-2 bosons \cite{Teo14}.

We now perform the $2\pi M$ flux attachment transformation given in Eq.~\eqref{eq:TransFluxAttach} with $m=M$ to find the tunneling Hamiltonian in terms of the composite fermions, 
\begin{align} \label{eq:TunnelingPfaffianCF}
H_1 &= \int_x \sum_l \Biggl[ \sum_{a,b} t_{ab} \kappa_{2l+a} \kappa_{2l+b+2} \nonumber \\
&\ \ \ \times \exp i\Bigl[ \Phi^\CF_{2l+a} +\sum_{c=0}^3 \Gamma^c_{ab}(0) \Theta^\CF_{2l+c} -\Phi^\CF_{2l+b+2} \Bigr] \nonumber \\
&\ \ \ +t_u \kappa_{2l} \kappa_{2l+1} e^{i(\Phi^\CF_{2l} -\Theta^\CF_{2l} -\Phi^\CF_{2l+1} -\Theta^\CF_{2l+1})} \nonumber \\
&\ \ \ +t_v e^{i(2\Theta^\CF_{2l} -2\Theta^\CF_{2l+1})} +\textrm{H.c.} \Biggr], 
\end{align}
which is depicted in Fig.~\ref{fig:Pfaffian}~(c). 
Here the vectors $\bfGamma_{ab}(0)$ are those given in Eq.~\eqref{eq:VectorGammaPfaffian} with $M=0$.
The interaction thus takes the same form as the original interaction \eqref{eq:TunnelingPfaffian} with $M=0$. 
Hence, this tunneling Hamiltonian should be interpreted as a chiral $p$-wave superconductor of the composite fermions. 
Following the prescription of Ref.~\cite{Teo14}, we define the charge and neutral bosonic fields by grouping two neighboring wires, 
\begin{align} \label{eq:FieldChargeNeutral}
\begin{split}
\Phi^c_l &= \frac{1}{2} (\Phi^\CF_{2l} +\Theta^\CF_{2l} +\Phi^\CF_{2l+1} -\Theta^\CF_{2l+1}), \\
\Theta^c_l &= \Theta^\CF_{2l} +\Theta^\CF_{2l+1}, \\
\phi^n_{R,l} &= \frac{1}{2} (\Phi^\CF_{2l} +\Theta^\CF_{2l} -\Phi^\CF_{2l+1} -3\Theta^\CF_{2l+1}), \\
\phi^n_{L,l} &= \frac{1}{2} (\Phi^\CF_{2l} -3\Theta^\CF_{2l} -\Phi^\CF_{2l+1} +\Theta^\CF_{2l+1}), 
\end{split}
\end{align}
which satisfy the commutation relations,
\begin{align} \label{eq:CommChargeNeutral}
\begin{split}
[\Phi^c_l(x), \Phi^c_{l'}(x')] &= -i\pi M \sgn (l-l'), \\
[\Theta^c_l(x), \Phi^c_{l'}(x')] &= i\pi \delta_{l,l'} \Theta(x-x'), \\
[\phi^n_{r,l}(x), \phi^n_{r',l'}(x')] &= i\pi \delta_{l,l'} [r \delta_{r,r'} \sgn (x-x') - \varepsilon_{r,r'}],\\
[\Phi^c_l(x), \phi^n_{r',l'}(x')] &= -\frac{i\pi}{2} (M-1) \delta_{l,l'}, 
\end{split}
\end{align}
while the other commutators vanish. 
Here we have used the notation $\varepsilon_{R,L}=-\varepsilon_{L,R}=1$ and $\varepsilon_{R,R}=\varepsilon_{L,L}=0$.
We here defined the charged bosonic fields labeled by $c$ to be nonchiral, whereas the neutral bosonic fields labeled by $n$ to be chiral.
The tunneling Hamiltonian \eqref{eq:TunnelingPfaffianCF} is then written as 
\begin{align} \label{eq:TunnelingPfaffianCF2}
H_1 &= \int_x \sum_l \Bigl[ t_{00} \kappa_{2l} \kappa_{2l+2} e^{i(\Phi^c_l -\Phi^c_{l+1} +\phi^n_{R,l} -\phi^n_{L,l+1})} \nonumber \\
&\ \ \ +t_{11} \kappa_{2l+1} \kappa_{2l+3} e^{i(\Phi^c_l -\Phi^c_{l+1} -\phi^n_{R,l} +\phi^n_{L,l+1})} \nonumber \\
&\ \ \ +t_{01} \kappa_{2l} \kappa_{2l+3} e^{i(\Phi^c_l -\Phi^c_{l+1} +\phi^n_{R,l} +\phi^n_{L,l+1})} \nonumber \\
&\ \ \ +t_{10} \kappa_{2l+1} \kappa_{2l+2} e^{i(\Phi^c_l -\Phi^c_{l+1} -\phi^n_{R,l} -\phi^n_{L,l+1})} \nonumber \\
&\ \ \ +t_u \kappa_{2l} \kappa_{2l+1} e^{i(\phi^n_{R,l} +\phi^n_{L,l})} +t_v e^{i(\phi^n_{R,l} -\phi^n_{L,l})} +\textrm{H.c.} \Bigr].
\end{align}
When forward scattering interactions are appropriately incorporated and tuned in such a way that the operators $e^{i\phi^n_{r,l}}$ have conformal weight $1/2$, we can define neutral Dirac fermion operators by 
\begin{align} \label{eq:NeutralFermi}
\psi^n_{r,l}(x) = \frac{\eta_l}{\sqrt{2\pi \alpha}} e^{i\phi^n_{r,l}(x)}. 
\end{align}
These operators are ensured to satisfy fermionic anticommutation relations by the commutation relation \eqref{eq:CommChargeNeutral} and the Klein factor $\eta_l$. 
Specifically, the Klein factors are chosen to be $\eta_l = \kappa_{2l}$ for even $M$ while they are defined to be new Majorana operators obeying $\{ \eta_l, \eta_{l'} \} = 2\delta_{l,l'}$ for odd $M$. 
As different treatments for the Klein factor are required for the bosonic and fermionic cases, we need to treat them separately. 
We consider the fermionic (even $M$) case below. 
The detailed discussion including the bosonic case can be found in Appendix~\ref{app:KleinPfaffian}. 
After appropriately scaling the coupling constants, we find the tunneling Hamiltonian \eqref{eq:TunnelingPfaffianCF2} in terms of the neutral Dirac fermion fields \eqref{eq:NeutralFermi}, 
\begin{align} \label{eq:TunnelingPfaffianCF3}
H_1 &= \int_x \sum_l \Bigl[ e^{i(\Phi^c_l -\Phi^c_{l+1})} \bigl( g_{00} \psi^n_{R,l} \psi^{n\dagger}_{L,l+1} +g_{11} \psi^{n\dagger}_{R,l} \psi^n_{L,l+1} \nonumber \\
&\qquad\qquad +g_{01} \psi^n_{R,l} \psi^n_{L,l+1} +g_{10} \psi^{n\dagger}_{R,l} \psi^{n\dagger}_{L,l+1} \bigr) \nonumber \\
&\qquad\qquad -ig_u \psi^n_{R,l} \psi^n_{L,l} +ig_v \psi^n_{R,l} \psi^{n\dagger}_{L,l} +\textrm{H.c.} \Bigr], 
\end{align}
where $g$'s are taken to be real ($g_{ab}=|t_{ab}|$, $g_u=|t_u|$, and $g_v=|t_v|$). 

We first consider the case where only the coupling constants $g_{00}$ and $g_{11}$ are nonvanishing. 
In this case, the numbers of the composite fermions in two layers, consisting of wires labeled by even ($2l$) or odd ($2l+1$) integers, are separately conserved, as seen from Fig.~\ref{fig:Pfaffian}~(b).
When $g_{00}$ and $g_{11}$ flow to the strong-coupling limit, we obtain the Halperin $(M+1,M+1,M-1)$ state \cite{Halperin83}, e.g., the 331 state for the case of $\nu=1/2$. 
This can be checked by examining the $K$ matrix for the edge states in terms of the original bosonic fields, which is given by Eq.~\eqref{eq:Kmat_qqp} with $(n,m_0,m_1)=(1,M+1,M-1)$. 
It can also be shown that this state is a generalized hierarchy state obtained from the Laughlin $\nu=1/(M+1)$ state by condensing charge-$[2/(M+1)]$ quasielectrons into the Laughlin $\nu=1/4$ state in the way discussed in Sec.~\ref{sec:HaldaneHalperin}.
In the tunneling Hamiltonian \eqref{eq:TunnelingPfaffianCF3}, the neutral Dirac fermions will be gapped in the bulk while leaving an unpaired gapless Dirac fermion mode at the boundary. 
Once the neutral fermions are gapped in the bulk, a charge-$2$ boson tunneling $e^{i(2\Phi^c_l -2\Phi^c_{l+1})}$ will be generated and induce condensation of the charge-2 Cooper pairs of the composite fermions as in superconductors. 
However, concomitant Goldstone modes are Higgsed by the Chern-Simons gauge field and the bulk is entirely gapped while a gapless chiral charge mode is left at the edge. 
As there is a chiral neutral Dirac fermion at the edge (in addition to the charge mode), this may be interpreted as the weak-pairing phase of a chiral triplet $p$-wave superconductor of the composite fermion if we regard the layer degrees of freedom as spin \cite{Read00,TLHo95,Milovanovic96}. 
The coupling $g_v$ gives an interaction between the two layers of composite fermions and induces a local mass term for the neutral Dirac fermions. 
The system will undergo a transition by increasing $g_v$ to a phase in which the neutral fermions are gapped out without leaving any gapless edge mode.
This can be understood as the strong-pairing phase of composite fermions \cite{Read00} and corresponds to the Laughlin $\nu=1/4M$ state of tightly bound charge-2 bosons. 

The interlayer tunneling terms $g_{01}$, $g_{10}$, and $g_u$ in Eq.~\eqref{eq:TunnelingPfaffianCF3} violates the particle number conservation of the neutral fermions and induces pairing terms. 
In this case, it is more natural to split the neutral Dirac (complex) fermions \eqref{eq:NeutralFermi} into two Majorana (real) fermions, 
\begin{align} \label{eq:DiracToMajorana}
\psi^n_{r,l}(x) = \frac{1}{\sqrt{2}} \bigl[ \xi^1_{r,l}(x) +i \xi^2_{r,l}(x) \bigr], 
\end{align}
each of which is associated with the Ising CFT. 
As argued in Ref.~\cite{Read00}, the interlayer tunneling terms give rise to an intermediate phase between the weak- and strong-pairing Abelian phases. 
In this intermediate phase, only a single chiral Majorana mode survives at the edge, implying a spinless chiral $p$-wave pairing.
Thus, there will be a transition from the Halperin $(M+1,M+1,M-1)$ state corresponding to the $\ell=-2$ pairing to the Moore-Read state corresponding to the $\ell=-1$ pairing by tuning the onsite term $g_v$.
In particular, when the coupling constants are fine tuned such that $g_{ab} \equiv g/4$, we can further rewrite the tunneling Hamiltonian \eqref{eq:TunnelingPfaffianCF3} in terms of the Majorana fermions \eqref{eq:DiracToMajorana}, 
\begin{align} \label{eq:TunnelingPfaffianCF4}
H_1 &= \int_x \sum_l \Bigl[ g \cos (\Phi^c_l -\Phi^c_{l+1}) \xi^1_{R,l} \xi^1_{L,l+1} \nonumber \\
&\qquad\qquad +\frac{g_v-g_u}{2} i \xi^1_{R,l} \xi^1_{L,l} +\frac{g_v +g_u}{2} i\xi^2_{R,l} \xi^2_{L,l} \Bigr]. 
\end{align}
For the bosonic case, we need to multiply the first term by $\eta_l \eta_{l+1}$ (see Appendix~\ref{app:KleinPfaffian}). 
Let us assume $g_u=g_v$ for simplicity.
The tunneling Hamiltonian \eqref{eq:TunnelingPfaffianCF4} opens a gap in the $\xi^2_{r,l}$ Majorana fermions in the neutral sector within each $l$.
By contrast, the other Majorana fermions $\xi^1_{r,l}$ are gapped by forming pairs between neighboring $l$'s through the $g$ term, leaving an unpaired gapless chiral Majorana mode at each boundary. 
This is schematically depicted in Fig.~\ref{fig:Pfaffian}~(c). 
Again, once these neutral Majorana fermions are gapped in the bulk and integrated out, a charge-$2$ condensate of the bosonic modes is induced, whose Goldstone mode is Higgsed (gapped) by Chern-Simons gauge fields.
Therefore, we conclude that the tunneling Hamiltonian \eqref{eq:TunnelingPfaffianCF4} produces a spinless chiral $p$-wave superconductor of the composite fermions with Majorana Chern number $C_M=1$ or orbital angular momentum $\ell=-1$, which is indeed the non-Abelian Pfaffian state as discussed in Ref.~\cite{Read00}. 
The chiral central charge of this state is given by $c=1+1/2=3/2$, where $1$ comes from the bosonic charge mode and $1/2$ is from the neutral Majorana mode. 
It is also argued in Ref.~\cite{Kane17} that, when $g_u$ and $g_v$ flow to the strong-coupling limit, the tunneling Hamiltonian can also give rise to an anisotropic quantum Hall state aside from the strong-pairing Abelian phase, depending on the signs of the coupling constants.

From the commutation relations \eqref{eq:CommChargeNeutral}, the vertex operators of the charged bosonic fields $e^{i\Phi^c_l}$ ($\eta_l e^{i\Phi^c_l}$) for even (odd) $M$ obey bosonic statistics.
One may then apply the vortex duality transformation \eqref{eq:VCBfromCB} to the charged bosonic fields $\Phi^c_l$ and obtain the effective Chern-Simons theory for the charge sector $-(iM/4\pi) \epsilon_{\mu \nu \lambda} \alpha_\mu \partial_\nu \alpha_\lambda$. 
This is nothing but the Chern-Simons theory for the $\nu=1/M$ Laughlin state. 
The corresponding electron operator of the Laughlin state is bosonic (fermionic) for even (odd) $M$ and combined with the Majorana fermion $\xi^1$ to form the electron operator of the Pfaffian state with fermionic (bosonic) statistics. 
With quasiparticles of the Laughlin $\nu=1/M$ state and the Majorana fermion $\xi^1$, one can also generate the spectrum of \emph{Abelian} quasiparticles for the $\nu=1/M$ Pfaffian state. 
Chern-Simons theories of this sense also appear in the following discussion of the Bonderson-Slingerland hierarchy.

\subsection{Bonderson-Slingerland hierarchy}

The idea of the Haldane-Halperin hierarchy for Abelian FQH states can also be generalized to a family of non-Abelian states \cite{Bonderson08,TLan17}. 
We here employ the hierarchy construction built on the Pfaffian state proposed by Bonderson and Slingerland \cite{Bonderson08}. 
Their idea is to excite bound pairs of fundamental quasiparticles, which have only a single fusion channel and thus are Abelian quasiparticles, and to condense them into Laughlin states as in the standard hierarchy construction. 
The neutral sector of fundamental quasiparticles of the Pfaffian state, the spin field $\sigma$ with conformal weight $1/16$ in the Ising CFT, has two fusion channels $\sigma \times \sigma = \mathbf{1} + \xi$, where $\mathbf{1}$ and $\xi$ represent the trivial and Majorana fields with conformal weight $0$ and $1/2$ in the Ising CFT, respectively. 
Depending on the energetics of microscopic models, one of the channels will be chosen and result in new incompressible states at different filling fractions. 
As discussed above and also in Appendix~\ref{app:KleinPfaffian}, the spin field $\sigma$ cannot be expressed in terms of the bosonic fields, but their bound pair can.
The most natural choice of quasiparticle operators that create such a bound pair and can be easily built on coupled wires is the $4k_F$ backscattering operator $e^{i(2\theta_{2l} +2\theta_{2l+1})}$. 
In terms of the chiral charged bosonic fields defined in Eq.~\eqref{eq:ChargeNeutralField}, this backscattering operator is written as $e^{i(\tphi^c_{R,l} -\tphi^c_{L,l})/M}$. 
It does not involve the neutral fields and is thus identified with a creation operator of the bound pair of quasiparticles in the trivial fusion channel $\mathbf{1}$ with charge $\pm 1/M$. 
In the following, we construct hierarchy states of the Pfaffian state obtained by condensing such charge $\pm 1/M$ quasiparticles. 

\begin{figure}
\includegraphics[clip,width=0.42\textwidth]{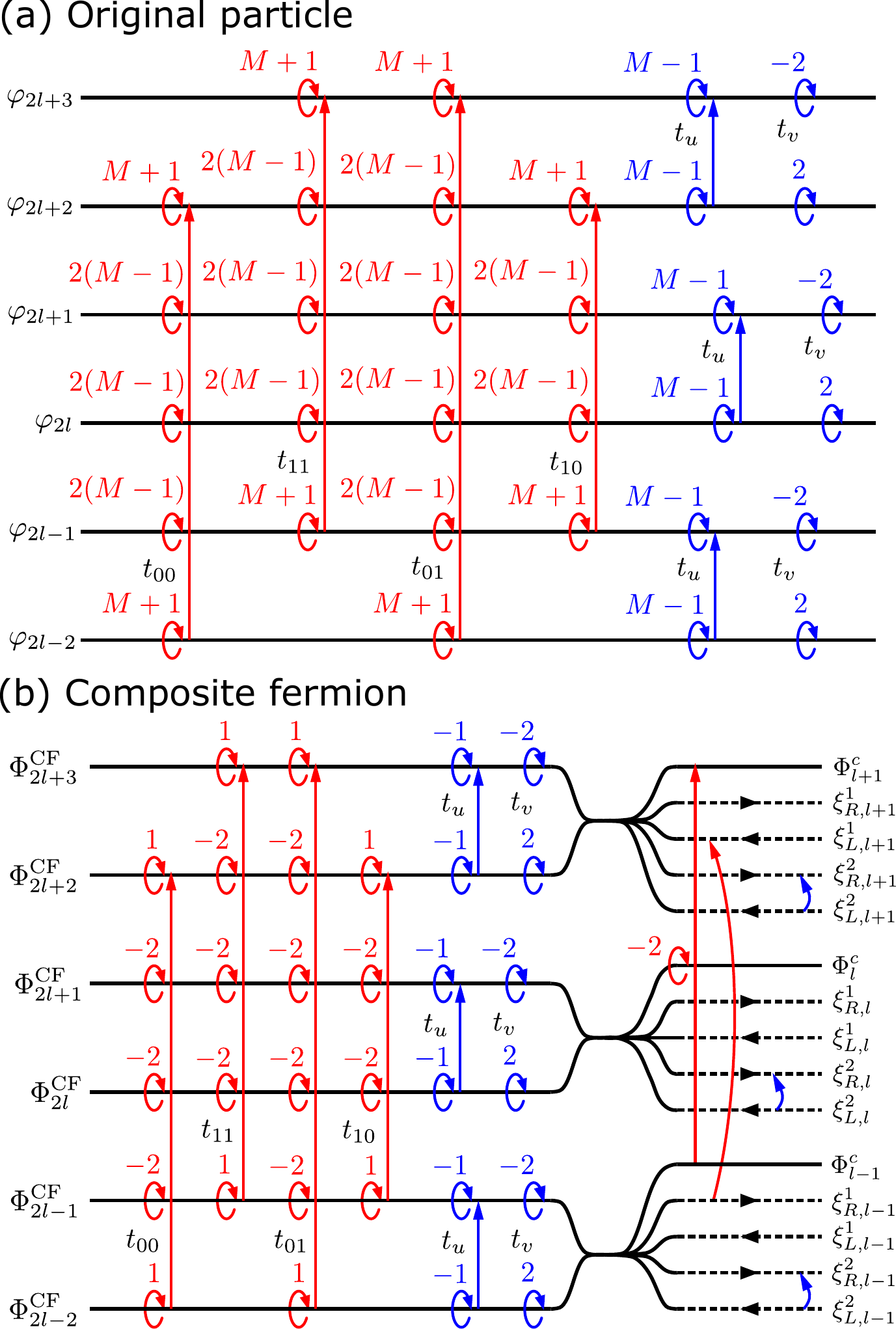}
\caption{Tunneling Hamiltonian for the Bonderson-Slingerland hierarchy state at $\nu=2/(2M-1)$ in terms of (a) the original particles and (b) composite fermions. 
The same notation as in Fig.~\ref{fig:Pfaffian} is used.}
\label{fig:BSHierarchy}
\end{figure}

We first consider the first-level Bonderson-Slingerland hierarchy states at $\nu=2/(2M-1)$ obtained by condensing bound quasielectron pairs on top of the $\nu=1/M$ Pfaffian state.
Similarly to the construction of the Haldane-Halperin hierarchy states in Sec.~\ref{sec:HaldaneHalperin}, we consider the coupled-wire Hamiltonian involving tunnelings between second-neighbor $l$'s, 
\begin{align} \label{eq:TunnelingBS}
H'_1 &= \int_x \sum_l \Biggl\{ \sum_{a=0}^1 \sum_{b=0}^1 t_{ab} \kappa_{2l+a} \kappa_{2l+b+4} \nonumber \\
&\qquad\quad \times \exp\!\left[ i \biggl(\! \varphi_{2l+a} +\sum_{c=0}^5 \Gamma^c_{ab}(M) \theta_{2l+c} -\varphi_{2l+b+4} \biggr)\!\right] \nonumber \\
&\qquad +t_u \kappa_{2l} \kappa_{2l+1} e^{i[\varphi_{2l} +(M-1) \theta_{2l} -\varphi_{2l+1} +(M-1) \theta_{2l+1}]} \nonumber \\
&\qquad +t_v e^{i(2\theta_{2l} -2\theta_{2l+1})} +\textrm{H.c.} \Biggr\}. 
\end{align}
Here, $\bfGamma_{ab}(M)$ are integer vectors 
\begin{align} \label{eq:IntegerVectorBS}
\begin{pmatrix} \bfGamma_{00} \\ \bfGamma_{11} \\ \bfGamma_{01} \\ \bfGamma_{10} \end{pmatrix} 
= \begin{pmatrix} M+1 & 2(M-1) & * & * & M+1 & 0 \\ 0 & M+1 & * & * & 2(M-1) & M+1 \\ M+1 & 2(M-1) & * & * & 2(M-1) & M+1 \\ 0 & M+1 & * & * & M+1 & 0 \end{pmatrix} \! , 
\end{align}
where $*=2(M-1)$. 
This tunneling Hamiltonian is pictorially given in Fig.~\ref{fig:BSHierarchy}~(a). 
%%%%%%%%%%%%%%%%%%%%%%%%%%%%%%%%%%%%%
%%%%%%%%%%%%%%%%%%%%%%%%%%%%%%%%%%%%%
The Bonderson-Slingerland hierarchy state is modified from its parent Pfaffian state only in the bosonic (charge) part of electron and quasiparticle operators while the neutral part remains unchanged. 
In order to see this, we successively perform the flux attachment and vortex duality transformations such that the charge part of the tunneling Hamiltonian becomes a simple boson hopping Hamiltonian to induce Bose condensation as we have done for the Haldane-Halperin hierarchy states in Sec.~\ref{sec:HaldaneHalperin}. 
To this end, we perform the $2\pi M$ flux attachment \eqref{eq:TransFluxAttach} to write the tunneling Hamiltonian \eqref{eq:TunnelingBS} in terms of the composite fermions, 
\begin{align} \label{eq:TunnelingBSCF}
H'_1 &= \int_x \sum_l \Biggl\{ \sum_{a=0}^1 \sum_{b=0}^1 t_{ab} \kappa_{2l+a} \kappa_{2l+b+4} \nonumber \\
&\qquad\quad \times \exp \!\left[i \biggl( \Phi^\CF_{2l+a} +\sum_{c=0}^5 \Gamma^c_{ab}(0) \Theta^\CF_{2l+c} -\Phi^\CF_{2l+b+4} \biggr)\!\right] \nonumber \\
&\qquad +t_u \kappa_{2l} \kappa_{2l+1} e^{i(\Phi^\CF_{2l} -\Theta^\CF_{2l} -\Phi^\CF_{2l+1} -\Theta^\CF_{2l+1})} \nonumber \\
&\qquad +t_v e^{i(2\Theta^\CF_{2l} -2\Theta^\CF_{2l+1})} +\textrm{H.c.} \Biggr\}. 
\end{align}
Using the charge and neutral bosonic fields in Eq.~\eqref{eq:FieldChargeNeutral}, this Hamiltonian can be written as 
\begin{align}
H'_1 &= \int_x \sum_l \Bigl[ t_{00} \kappa_{2l} \kappa_{2l+4} e^{i(\Phi^c_l -2\Theta^c_{l+1} -\Phi^c_{l+2} +\phi^n_{R,l} -\phi^n_{L,l+2})} \nonumber \\
&\ \ \ +t_{11} \kappa_{2l+1} \kappa_{2l+5} e^{i(\Phi^c_l -2\Theta^c_{l+1} -\Phi^c_{l+2} -\phi^n_{R,l} +\phi^n_{L,l+2})} \nonumber \\
&\ \ \ +t_{01} \kappa_{2l} \kappa_{2l+5} e^{i(\Phi^c_l -2\Theta^c_{l+1} -\Phi^c_{l+2} +\phi^n_{R,l} +\phi^n_{L,l+2})} \nonumber \\
&\ \ \ +t_{10} \kappa_{2l+1} \kappa_{2l+4} e^{i(\Phi^c_l -2\Theta^c_{l+1} -\Phi^v_{l+2} -\phi^n_{R,l} -\phi_{L,l+2})} \nonumber \\
&\ \ \ +t_u \kappa_{2l} \kappa_{2l+1} e^{i(\phi^n_{R,l} +\phi^n_{L,l})} +t_v e^{i(\phi^n_{R,l} -\phi^n_{L,l})} +\textrm{H.c.} \Bigr]. 
\end{align}
Similarly to the Pfaffian state, when forward scattering interactions and the coupling constants are appropriately tuned, we will end up with the Hamiltonian, 
\begin{align}
H'_1 &= \int_x \sum_l \Bigl[ g \cos (\Phi^c_l -2\Theta^c_{l+1} -\Phi^c_{l+2}) \xi^1_{R,l} \xi^1_{L,l+2} \nonumber \\
&\ \ \ +\frac{g_v-g_u}{2} i\xi^1_{R,l} \xi^1_{L,l} +\frac{g_v+g_u}{2} i\xi^2_{R,l} \xi^2_{L,l} \Bigr].
\end{align}
The neutral Majorana fermions $\xi^2_{r,l}$ are gapped within each $l$. 
If $g_v=g_u$ and $g$ flows to the strong-coupling limit, there will be \emph{two} unpaired Majorana fermion modes $\xi^1_{r,l}$ at the boundary.
However, this is not a desired property, and we consider a modified tunneling Hamiltonian
\begin{align}
H_1 &= \int_x \sum_l \Bigl[ g \cos (\Phi^c_l -2\Theta^c_{l+1} -\Phi^c_{l+2}) \nonumber \\
&\ \ \ \times \xi^1_{R,l} \xi^1_{L,l+1} \xi^1_{R,l+1} \xi^1_{L,l+2} + ig_u \xi^2_{R,l} \xi^2_{L,l} \Bigr], 
\end{align}
which neither changes the filling fraction nor excites other quasiparticles. 
This tunneling Hamiltonian will pair up the Majorana fermions $\xi^1_{r,l}$ from neighboring $l$'s to open a bulk gap in the neutral sector, leaving a \emph{single} unpaired Majorana fermion mode at the boundary.

Looking at the charge part, the tunneling term involves $e^{-i2\Theta^c_l}$, which is nothing but the $4k_F$ backscattering operator $e^{-i(2\theta_{2l} +2\theta_{2l+1})}$ in the original particles and thus hops a bound pair of quasielectrons with charge $-1/M$ from the dual wire $l-1/2$ to $l+1/2$. 
Thus it excites the quasiparticles with charge $-1/M$ on top of the charge-$1$ boson condensate. 
This situation is similar to the $\nu=2/(2M-1)$ hierarchy state obtained from the parent $\nu=1/M$ Laughlin state. 
Applying the vortex duality transformation to the charge part, one can see that those quasiparticles are condensed into the Laughlin $\nu=1/2$ state. 
At the final stage, the charge part of the kinetic action will produce a Chern-Simons term with the $K$ matrix in the hierarchy basis, 
\begin{align}
\bfK = \begin{pmatrix} M & -1 \\ -1 & 2 \end{pmatrix}.
\end{align}
This $K$ matrix for $M=2$ is used to describe the (charge) bosonic part of the operator content for the $\nu=2/3$ Bonderson-Slingerland hierarchy state \cite{Bonderson08}. 
The Abelian quasiparticles corresponding to this $K$ matrix with integer quasiparticle charges are combined with $\mathbf{1}$ or $\xi$ to generate Abelian quasiparticles of the Bonderson-Slingerland hierarchy, while non-Abelian quasiparticles are obtained by combining $\sigma$ and the quasiparticles corresponding to the same $K$ matrix with half-integer quasiparticle charges.

There are also other Bonderson-Slinegerland hierarchy states obtained by exciting bound pairs of quasiholes with charge $1/M$ and condensing them into the Laughlin $\nu=1/2$ state. 
These states are realized at filling fraction $\nu=2/(2M+1)$. 
The corresponding tunneling Hamiltonian is given by Eq.~\eqref{eq:TunnelingBS} with setting $*=2(M+1)$ in the integer vectors \eqref{eq:IntegerVectorBS}. 
Following the above argument, we find the tunneling Hamiltonian, 
\begin{align}
H_1 &= \int_x \sum_l \Bigl[ g \cos (\Phi^c_l +2\Theta^c_{l+1} -\Phi^c_{l+2}) \nonumber \\
&\ \ \ \times \xi^1_{R,l} \xi^1_{L,l+1} \xi^1_{R,l+1} \xi^1_{L,l+2} + ig_u \xi^2_{R,l} \xi^2_{L,l} \Bigr], 
\end{align}
which excites quasiparticles with charge $1/M$ on top of the charge-1 condensate. 
In analogy to the $\nu=2/(2M+1)$ hierarchy state on top of the Laughlin $\nu=1/M$ state, the charge part admits the Chern-Simons theory with the $K$ matrix, 
\begin{align}
\bfK = \begin{pmatrix} M & 1 \\ 1 & -2 \end{pmatrix}. 
\end{align}
This $K$ matrix again describes the bosonic part of operator content for the corresponding Bonderson-Slingerland hierarchy state, which has been analyzed for $\nu=2/5$ in Ref.~\cite{Bonderson08}. 

Higher-level hierarchy states can also be obtained by taking the above first-level states as parent states in a straightforward manner. 
On the other hand, it is at the moment unclear how to construct coupled-wire models for the hierarchy states obtained by condensation of the $\pm 1/M$ quasiparticles in the $\xi$ channel. 
The Bonderson-Slingerland hierarchy on top of the $Z_k$ Read-Rezayi states at $\nu=k/[k(M-1)+2]$ \cite{Read99,Bonderson10} can also be constructed by taking the corresponding coupled-wire model proposed by Teo and Kane \cite{Teo14}. 
In particular, $k$-quasiparticle bound states with charge $\pm 1/M$ in the trivial fusion channel for the neutral sector are excited by the $2k k_F$ backscattering operators $e^{i2(\theta_{kl} +\theta_{kl+1} +\cdots +\theta_{kl+k-1})}$ and condensed into the Laughlin $\nu=1/2$ state to yield hierarchy states modified only in the charge sector. 

There is also another proposal of hierarchy states from the Pfaffian state, known as the Levin-Halperin hierarchy states \cite{Levin09}. 
In fact, the observed quantum Hal plateau at $\nu=2+6/13$ \cite{Kumar10}, which is unlikely to be a Jain hierarchy state, may be explained by the PH conjugate of the $\nu=7/13$ Levin-Halperin state obtained from a quasielectron condensation on the Pfaffian state, in view of a strong evidence that the plateau at $\nu=2+1/2$ is in a Pfaffian state \cite{Banerjee18}.
Although it is interesting to ask how the energetics of quasiparticle excitations above the Pfaffian state leads to the Levin-Halperin hierarchy state in the coupled-wire approach, this state turned out to be Abelian and is actually obtained as a hierarchy state from the Halperin 331 state. 
The coupled-wire construction of the Haldane-Halperin hierarchy states given in Sec.~\ref{sec:HaldaneHalperin} appears to be able to account for this hierarchy state as well, but so far we could not obtain a physical coupled-wire Hamiltonian by a naive application, and a further extension will be required.

\subsection{Anti-Pfaffian state}
\label{sec:AntiPfaffian}

\begin{figure}
\includegraphics[clip,width=0.42\textwidth]{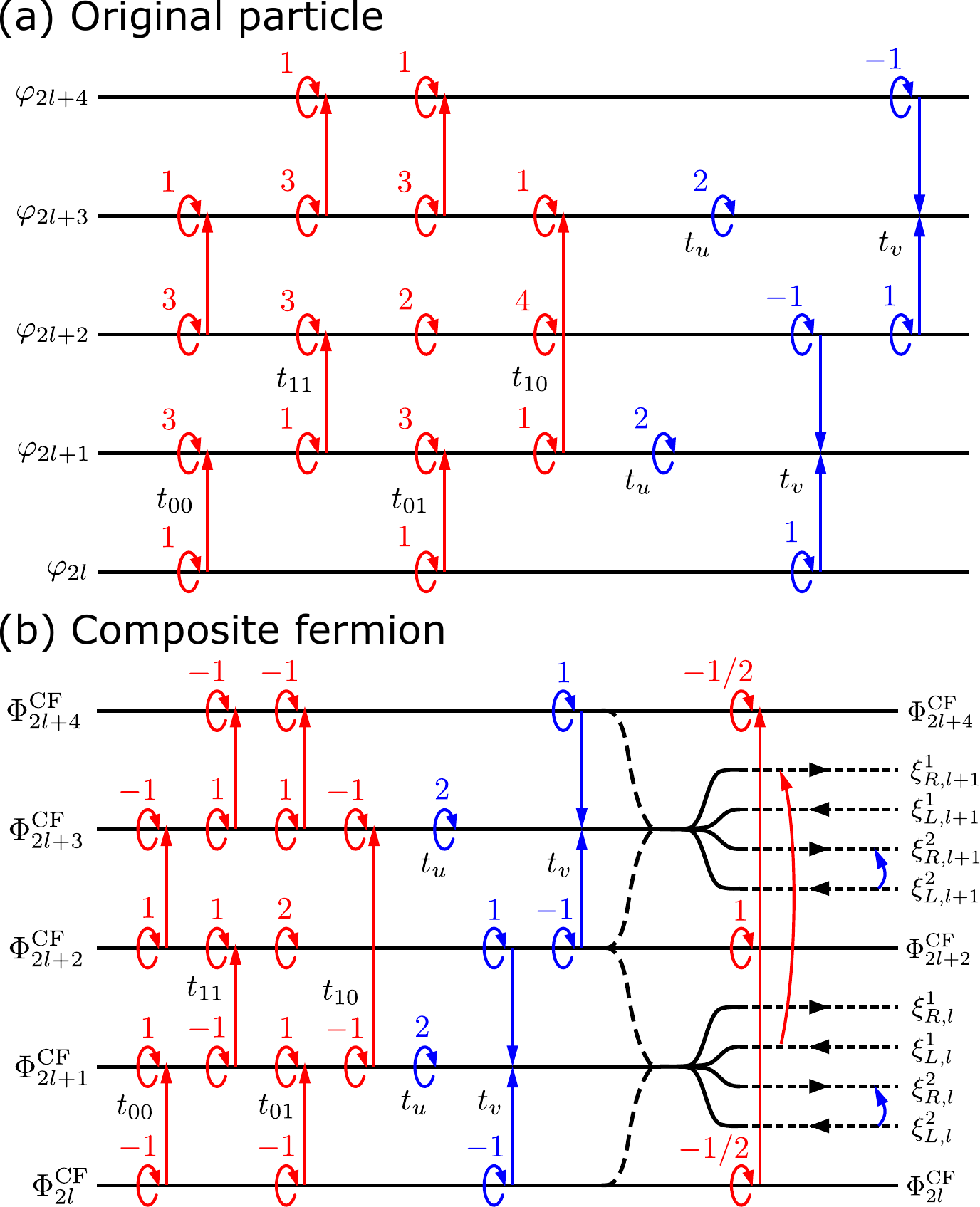}
\caption{Tunneling Hamiltonian for the anti-Pfaffian state at $\nu=1/2$ in terms of (a) the original particles and (b) composite fermions. 
A similar notation to Fig.~\ref{fig:Pfaffian} is used.}
\label{fig:AntiPfaffian}
\end{figure}

In the previous sections, we have applied the PH transformation defined in Sec.~\ref{sec:PHConjugate} to several Abelian FQH states and the CFL at $\nu=1/2$. 
In this section we apply it to the Pfaffian state at $\nu=1/2$ to obtain its PH conjugate called the anti-Pfaffian state \cite{Levin07,SSLee07}. 
We consider the tunneling Hamiltonian \eqref{eq:TunnelingPfaffian} for $M=2$ and apply the PH transformation \eqref{eq:PHTrans}. 
Rewriting the PH conjugated tunneling Hamiltonian in terms of the original bosonic fields by using Eq.~\eqref{eq:HoleFromElectron}, we obtain the tunneling Hamiltonian for the anti-Pfaffian state
\begin{align} \label{eq:TunnelingAPf}
H_1 &= \int_x \sum_l \Biggl\{ \sum_{a=0}^1 \sum_{b=0}^1 t_{ab}
 \exp \!\left[i \sum_{c=0}^4 \bigl( \gamma^c_{ab} \varphi_{2l+c} +\Gamma^c_{ab} \theta_{2l+c} \bigr)\right] \nonumber \\
&\ \ \ +t_u e^{i2\theta_{2l+1}} +t_v e^{i(\varphi_{2l} +\theta_{2l} -2\varphi_{2l+1} +\varphi_{2l+2} -\theta_{2l+2})} +\textrm{H.c.} \Biggr\}
\end{align}
with the integer vectors
\begin{align} \label{eq:IntegerVecAPf}
\begin{split}
\begin{pmatrix} \bfgamma_{00} \\ \bfgamma_{11} \\ \bfgamma_{01} \\ \bfgamma_{10} \end{pmatrix} &= \begin{pmatrix} 1 & -1 & 1 & -1 & 0 \\ 0 & 1 & -1 & 1 & -1 \\ 1 & -1 & 0 & 1 & -1 \\ 0 & 1 & 0 & -1 & 0 \end{pmatrix}, \\
\begin{pmatrix} \bfGamma_{00} \\ \bfGamma_{11} \\ \bfGamma_{01} \\ \bfGamma_{10} \end{pmatrix} &= \begin{pmatrix} 1 & 3 & 3 & 1 & 0 \\ 0 & 1 & 3 & 3 & 1 \\ 1 & 3 & 2 & 3 & 1 \\ 0 & 1 & 4 & 1 & 0 \end{pmatrix}.
\end{split}
\end{align}
Here, we have omitted Klein factors for simplicity of discussion, but we suppose that they can be supplemented properly by considering how the interaction is microscopically built from electrons. 
The tunneling Hamiltonian is pictorially given in Fig.~\ref{fig:AntiPfaffian}~(a).
%%%%%%%%%%%%%%%%%%%%%%%%%%%%%%%%%%%%%
%%%%%%%%%%%%%%%%%%%%%%%%%%%%%%%%%%%%%
Although this tunneling Hamiltonian for the anti-Pfaffian state does not resemble the tunneling Hamiltonian for the Pfaffian state in Eq.~\eqref{eq:TunnelingPfaffian}, both tunneling Hamiltonians can have exactly the same scaling dimensions, when the kinetic action is appropriately tuned to have the PH symmetry, and have the form of Eq.~\eqref{eq:ActionHFLL} with $u=\tu$. 
Thus, the Pfaffian and anti-Pfaffian states are degenerate and either is chosen when the PH symmetry is broken. 

By construction, the anti-Pfaffian state is understood as the spinless chiral $p$-wave superconductor of the \emph{composite holes}, which have been defined in Eq.~\eqref{eq:CompositeHole} in our coupled-wire language, with orbital angular momentum $\ell=1$ or Majorana Chern number $C_M=-1$ \cite{Barkeshli15}. 
We now explain how the anti-Pfaffian state can be understood as a chiral superconductor of \emph{composite fermions} with $\ell=3$ or $C_M=-3$ \cite{Barkeshli15}, by examining its edge structure in our coupled-wire model \eqref{eq:TunnelingAPf}. 
We note that the first two terms in Eq.~\eqref{eq:TunnelingAPf} with $t_{00}$ and $t_{11}$ give rise to the anti-331 state \cite{GYang13} in their strong-coupling limit, which corresponds to the composite fermion pairing with $\ell=4$ or  $C_M=-4$. 

In terms of composite fermions, the tunneling Hamiltonian \eqref{eq:TunnelingAPf} is written as [see also Fig.~\ref{fig:AntiPfaffian}~(b)]
\begin{align}
H_1 &\sim \int_x \sum_l \Biggl\{ \sum_{a=0}^1 \sum_{b=0}^1 t_{ab}
 \exp\!\left[ i \sum_{c=0}^4 \bigl( \gamma^c_{ab} \Phi^\CF_{2l+c} +\tGamma^c_{ab} \Theta^\CF_{2l+c} \bigr)\right] \nonumber \\
&\ \ \ +t_u e^{i2\Theta^\CF_{2l+1}} +t_v e^{i(\Phi^\CF_{2l} -\Theta^\CF_{2l} -2\Phi^\CF_{2l+1} +\Phi^\CF_{2l+2} +\Theta^\CF_{2l+2})} \nonumber \\
&\ \ \ +\textrm{H.c.} \Biggr\}
\end{align}
with the integer vectors $\bfgamma_{ab}$ defined in Eq.~\eqref{eq:IntegerVecAPf} and
\begin{align}
\begin{pmatrix} \tbfGamma_{00} \\ \tbfGamma_{11} \\ \tbfGamma_{01} \\ \tbfGamma_{10} \end{pmatrix} &= \begin{pmatrix} -1 & 1 & 1 & -1 & 0 \\ 0 & -1 & 1 & 1 & -1 \\ -1 & 1 & 2 & 1 & -1 \\ 0 & -1 & 0 & -1 & 0 \end{pmatrix}.
\end{align}
We introduce chiral neutral bosonic fields
\begin{align}
\begin{split}
\phi^n_{R,l} &= \frac{1}{2} (\Phi^\CF_{2l} -\Theta^\CF_{2l} -2\Phi^\CF_{2l+1} -2\Theta^\CF_{2l+1} +\Phi^\CF_{2l+2} +\Theta^\CF_{2l+2}), \\
\phi^n_{L,l} &= \frac{1}{2} (\Phi^\CF_{2l} -\Theta^\CF_{2l} -2\Phi^\CF_{2l+1} +2\Theta^\CF_{2l+1} +\Phi^\CF_{2l+2} +\Theta^\CF_{2l+2}),
\end{split}
\end{align}
which are actually the PH conjugate of the neutral fields defined for the Pfaffian state in Eq.~\eqref{eq:FieldChargeNeutral} in the sense of the PH transformation \eqref{eq:PHTrans}. 
They satisfy the commutation relations, 
\begin{align}
[\phi^n_{r,l}(x), \phi^n_{r',l'}(x')] &= i\pi r \delta_{r,r'} \delta_{l,l'} \sgn (x-x') \nonumber \\
&\ \ \ +\frac{i\pi}{4} \delta_{l,l'+1} -\frac{i\pi}{4} \delta_{l,l'-1}.
\end{align}
We may then define neutral fermion operators by Eq.~\eqref{eq:NeutralFermi} with an appropriate assignment of the Klein factors. 
When the coupling constants are fine tuned, we end up with the tunneling Hamiltonian, 
\begin{align}
H_1 &\sim \int_x \sum_l \Biggl\{ g \cos \biggl[\frac{1}{2}(\Phi^\CF_{2l} -\Theta^\CF_{2l} +2\Theta^\CF_{2l+2} \nonumber \\
&\qquad\qquad\qquad\quad -\Phi^\CF_{2l+4} -\Theta^\CF_{2l+4})\biggr] \xi^1_{L,l} \xi^1_{R,l+1} \nonumber \\
&\qquad\quad +\frac{g_v-g_u}{2} i\xi^1_{L,l} \xi^1_{R,l} +\frac{g_v+g_u}{2} i\xi^2_{L,l} \xi^2_{R,l} \Biggr\}, 
\end{align}
where we have used the Majorana fermion fields \eqref{eq:DiracToMajorana}. 
Let us make a further assumption that $g_v=g_u$. 
The Majorana fermions $\xi^2_{r,l}$ are gapped within each $l$. 
The other Majorana fermions $\xi^1_{r,l}$ are gapped in the bulk by pairing left- and right-moving modes from adjacent $l$'s while a single gapless chiral mode remains unpaired at each boundary as in the case of the Pfaffian state. 
However, its chirality is opposite to that of the Pfaffian state and thus the Majorana fermions $\xi^1_{r,l}$ contribute to Majorana Chern number $-1$ in the present case. 
When the massive Majorana fermions are integrated out, we will have a residual tunneling term consisting of even composite fermion wires, 
\begin{align}
\cos (\Phi^\CF_{2l} -\Theta^\CF_{2l} +2\Theta^\CF_{2l+2} -\Phi^\CF_{2l+4} -\Theta^\CF_{2l+4}). 
\end{align}
This tunneling term produces the Halperin $(1,1,-1)$ state, which is a state analogous to the integer quantum Hall ferromagnet or the Halperin (1,1,1) state \cite{XGWen92b,Sondhi93}. 
The latter state hosts a chiral charged fermion edge mode responsible for the Hall conductance $\sigma_{xy}=1$ and a gapless (pseudo) Goldstone mode associated with spontaneous breaking of $U(1)$ symmetry related to the conservation of the charge difference between two layers or species. 
By contrast, the Halperin $(1,1,-1)$ state of our interest hosts a neutral chiral Dirac fermion mode at the boundary and a gapless Goldstone mode from a charge-$2$ condensate in the bulk, which is to be Higgsed by the Chern-Simons gauge field. 
From these observations, we deduce that the tunneling Hamiltonian \eqref{eq:TunnelingAPf} leads to a completely gapped bulk spectrum with one neutral Dirac mode and one neutral Majorana mode propagating in the same direction at the boundary. 
Therefore, we conclude that the anti-Pfaffian state is interpreted as a chiral superconductor of the composite fermions with $\ell=3$ or $C_M=-3$. 

Similarly, one can also consider the PH conjugates of the Bonderson-Slinegrland hierarchy states at $\nu=1/3$ or $\nu=3/5$ \cite{Bonderson08}, whose electron counterparts at $\nu=2/3$ or $\nu=2/5$ are both obtained from the $\nu=1/2$ Pfaffian state and constructed in the previous section. 
One may also generalize the construction to the anti-Read-Rezayi states at $\nu=2/(2+k)$ \cite{Bishara08}, which are the PH conjugates of the $Z_k$ Read-Rezayi states at $\nu=k/(k+2)$ \cite{Read99}.

\subsection{Other composite fermion pairings}
\label{sec:OtherPairing}

We can construct a variety of Abelian and non-Abelian states at $\nu=1/M$, which will be understood as different patterns of pairing of composite fermions. 
In particular, the chiral $p$-wave pairing state of composite fermions with angular momentum $\ell=1$ or Majorana Chern number $C_M=-1$ is quite an intriguing state at $\nu=1/2$ because of the following reasons.
First, the topological order of the corresponding Pfaffian state is consistent with the PH symmetry expected in the half-filled Landau level \cite{Barkeshli15}. 
Such a PH-symmetric Pfaffian state is called PH-Pfaffian state. 
Second, the state has chiral central charge $1/2$, which is consistent with a recent measurement of thermal Hall conductance at $\nu=5/2$ \cite{Banerjee18}. 
Interestingly, Son has proposed that the $s$-wave pairing of Dirac composite fermions with an explicit PH symmetry gives rise to a non-Abelian state with the same PH-symmetric topological order \cite{Son15}. 
There are also many other theoretical attempts to explain the experimentally observed phenomena at $\nu=5/2$ \cite{Zucker16,BLian18,Mross18,ChongWang18,Simon18,Feldman18}. 

\begin{figure}
\includegraphics[clip,width=0.42\textwidth]{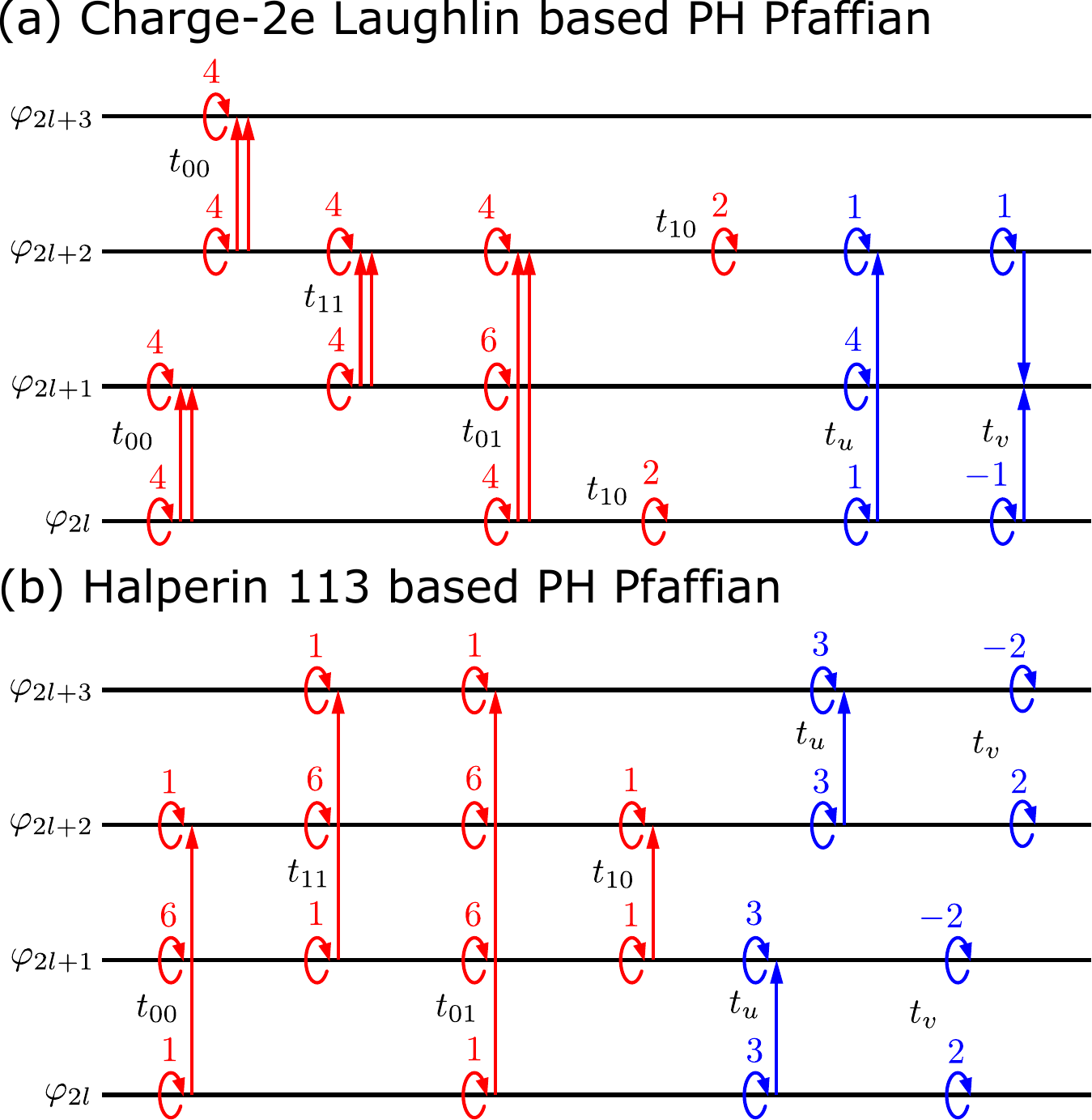}
\caption{Tunneling Hamiltonians for the $\ell=1$ paired state of the composite fermions at $\nu=1/2$, which has the same topological order as the PH Pfaffian state but with the explicitly broken PH symmetry. 
When only the coupling constants $t_{00}$ and $t_{11}$ are relevant, the models describe (a) the Laughlin $\nu=1/8$ state of charge-2 bosons or (b) the Halperin 113 state.}
\label{fig:PHPfaffian}
\end{figure}

It is thus tempting to consider a possible coupled-wire model for the PH-Pfaffian state, and here we propose two models that realize the $\ell=1$ pairing of composite fermions as shown in Fig.~\ref{fig:PHPfaffian}.  
%%%%%%%%%%%%%%%%%%%%%%%%%%%%%%%%%%%%%
%%%%%%%%%%%%%%%%%%%%%%%%%%%%%%%%%%%%%
These two models do not have a manifest PH symmetry and are related to each other by the PH transformation \eqref{eq:PHTrans}. 
The analysis similar to those in the previous subsections suggests that these models have the PH-Pfaffian state, in their phase diagram, next to the Abelian phases such as the Laughlin $\nu=1/8$ state of tightly bound charge-2 electrons and the Halperin 113 state, which correspond to the $\ell=0$ and $\ell=2$ pairing, respectively.
We note that Refs.~\cite{Mross15,Mross16a} have proposed a coupled-wire model for the T-Pfaffian state on a 2D surface of an interacting 3D topological insulator, which can also be used to construct the PH-Pfaffian state through the duality transformation while explicitly preserving the PH symmetry in a non-onsite fashion. 
In this approach, the PH-Pfaffian state is interpreted as the $s$-wave pairing of the Dirac composite fermions \cite{Son15}, while we have here proposed coupled-wire models for the PH-Pfaffian state that is rather interpreted as the $\ell=1$ pairing of the composite fermions and manifestly breaks the PH symmetry in the Hamiltonian level (see also Ref.~\cite{Kane17} for a more phenomenological construction).
We also note that the coupled-wire model in Fig.~\ref{fig:PHPfaffian}~(b) can be seen as a model of the bosonic Moore-Read state in a negative magnetic flux after applying $6\pi$-flux attachment transformation. 
The corresponding wave function has been proposed in Ref.~\cite{Jolicoeur07} and studied as a possible candidate for a PH Pfaffian state that may appear in the lowest Landau level \cite{Balram18,Mishmash18,Antonic18}.

In Ref.~\cite{Read00}, Read and Green have also discussed the spin-singlet chiral $d$-wave pairing of composite fermions. 
The corresponding Abelian state at $\nu=1/2$ will be a generalized hierarchy state obtained by condensing charge-$2/5$ quasielectrons into the Laughlin $\nu=1/4$ state on top of the $\nu=2/5$ hierarchy state. 
Thus, its coupled-wire Hamiltonian can be constructed in the way discussed in Sec.~\ref{sec:HaldaneHalperin}. 
As it is a composite-fermion pairing state with $\ell=-4$, there can be a neighboring non-Abelian state corresponding to the $\ell=-3$ pairing, which is expected to be the Blok-Wen $U(1) \times SU(2)_2$ state \cite{Blok92}, if a single Majorana fermion is gapped out in a trivial way by appropriately modulating interactions for the generalized hierarchy state. 
It is therefore interesting to seek the coupled-wire Hamiltonians for Abelian and non-Abelian states that realize the composite-fermion pairings with different angular momenta, whose quasiparticle statistics or topological order in the neutral sector follows the Kitaev's 16-fold way \cite{Kitaev06}. 
We do not pursue this direction further in this paper and leave detailed analysis for a separate paper.

\section{Conclusion and outlooks}
\label{sec:Conclusion}

In this paper we have shown, with the help of explicit nonlocal transformations of the flux attachment and vortex duality, that the coupled-wire construction admits physical interpretations of various quantum Hall states in terms of composite fermions or composite bosons on the ground of microscopic Hamiltonians.
Abelian hierarchy states are constructed as either IQH states of composite fermions (the Jain sequence) or condensates of quasiparticle excitations (the Haldane-Halperin sequence).
We also constructed the coupled-wire models for the CFL with an open Fermi surface; the constructed CFL Hamiltonian at $\nu=1/2$ is symmetric under the particular PH transformation that does not have a lattice analog.
Various pairing states of composite fermions, including the Pfaffian and anti-Pfaffian states, were also discussed. 

While some of the coupled-wire models discussed in this paper have been constructed previously, what we have actually accomplished in this paper is to develop a systematic approach to generating suitable coupled-wire Hamiltonians for various FQH states including hierarchy states, paired composite fermion states, and their PH conjugates. 
The coupled-wire construction gives us an alternative way to gain physical pictures of these FQH states, being complementary to other standard approaches such as trial wave functions and the effective Chern-Simons theory. 
Moreover, the coupled-wire construction has an advantage in its controllability of strong interactions in microscopic models, which is not easily achieved in the other approaches. 

There are several quantum Hall states whose trial wave functions are written in terms of conformal blocks of nonunitary or irrational CFT, such as the Haldane-Rezayi \cite{Haldane88}, Haffnian \cite{XGWen94a,XGWen94b}, or Gaffnian states \cite{Simon07}. 
It has been anticipated that such quantum Hall states are in fact compressible due to violation of the bulk-boundary correspondence and correspond to a phase transition point between certain FQH states \cite{Read09}. 
However, it is not clear how the structure of nonunitary or irrational CFTs appears in a microscopic Hamiltonian. 
In this regard it is an interesting open problem to construct coupled-wire Hamiltonians for those exotic quantum Hall states.
If those states are indeed located at a phase transition and if we can identify the FQH states next to the phase transition, then we can perhaps construct a microscopic model for the interface between the two FQH states, and coupling those interfaces would eventually lead to a bulk theory of the desired quantum Hall transition. 

Another question that one can naturally raise is whether the coupled-wire construction can incorporate coupling with geometry. 
In particular, when a FQH state has rotational symmetry as in a quantum Hall droplet on a sphere, the so-called shift coming from orbital spin becomes an important diagnosis for quantum Hall states \cite{XGWen92c}. 
Although it is not obvious how the shift is encoded in the coupled-wire models, it is likely that FQH states with different shifts have different realizations of coupled-wire Hamiltonian. 
For example, the PH conjugate of the Laughlin $\nu=1/3$ state and the Halperin 112 state are both realized at the same filling fraction $\nu=2/3$ and have the same quasiparticle statistics but have different shifts \cite{XGWen95}. 
The tunneling Hamiltonian for the PH conjugate of the $\nu=1/3$ Laughlin state is given in Ref.~\cite{Kane02} or in Eq.~\eqref{eq:TunnelingPHLaughlin}, but for the Halperin 112 state it is more natural to consider the tunneling Hamiltonian of the form
\begin{align}
H_1 &\sim g \int_x \sum_l \Bigl[ e^{i(\varphi_{2l} +\theta_{2l} +2\theta_{2l+1} -\varphi_{2l+2} +\theta_{2l+2} +2\theta_{2l+3})} \nonumber \\
&\ \ \ +e^{i(\varphi_{2l+1}+\theta_{2l+1} +2\theta_{2l} -\varphi_{2l+3} +\theta_{2l+3} +2\theta_{2l+2})} +\textrm{H.c.} \Bigr],
\end{align}
which is symmetric under exchanging even ($2l$) and odd ($2l+1$) layers. 
As these tunneling Hamiltonians lead to microscopically different forms of electron operators, it may result in different shifts. 
Application of this line of argument is left for future work.

\section*{Acknowledgement}

Y.F. acknowledges Yin-Chen He, Philippe Lecheminant, and Keisuke Totsuka for stimulating discussions, which inspired this work.
Y.F. would also like to thank Thierry Jolicoeur for useful discussions and the program TopMat of the PSI2 project funded by the IDEX Paris-Saclay, ANR-11-IDEX-0003-02, in which this work was finalized. 
This work was supported in part by JSPS KAKENHI Grants No.~JP17H07362 and No.~JP15K05141.

\appendix

\begin{widetext}
\section{Vortex duality to composite boson}
\label{app:VortexAction}

In this appendix, we provide the detailed derivation of the vortex theory \eqref{eq:ActionVCB} from the composite boson theory. 
Here we further add an interwire forward scattering interaction to Eq.~\eqref{eq:SLLplusA} and start with the action in terms of the original bosonic fields,
\begin{align} \label{eq:FullActionBoson}
S_0 = \int_{\tau,x} \sum_j \biggl[ \frac{i}{\pi} \partial_x \theta_j (\partial_\tau \varphi_j -A^\ext_{0,j}) +\frac{v}{2\pi} (\partial_x \varphi_j -A^\ext_{1,j})^2 +\frac{u}{2\pi} (\partial_x \theta_j)^2 +\frac{w}{8\pi} (\partial_x \varphi_j +m\partial_x \theta_j -\partial_x \varphi_{j+1} +m\partial_x \theta_{j+1})^2 \biggr].
\end{align}
As discussed in Sec.~\ref{sec:Laughlin}, the additional forward scattering term has an important effect to stabilize the Laughlin $\nu=1/m$ state. 
After the $2\pi m$ flux attachment transformation \eqref{eq:TransFluxAttach}, we obtain the composite boson action
\begin{align} \label{eq:FullActionCB}
S_0 &= \int_{\tau,x} \sum_j \biggl[ \frac{i}{\pi} \partial_x \Theta^\CB_j \Bigl( \partial_\tau \Phi^\CB_j -\frac{1}{2} (S a_{0,j-1/2}) -A_{0,j}^\ext \Bigr) +\frac{v}{2\pi} (\partial_x \Phi^\CB_j -a_{1,j} -A_{1,j}^\ext)^2 +\frac{u}{2\pi} (\partial_x \Theta^\CB_j)^2 \nonumber \\
&\ \ \ +\frac{w}{8\pi} (\Delta \partial_x \Phi^\CB_j)^2 +\frac{i}{2\pi m} a_{1,j} (\Delta a_{0,j-1/2}) \biggr]. 
\end{align}
Integrating out the Lagrange multiplier $a_0$ yields the constraint \eqref{eq:FluxAttach_a1}. 
Assuming that the action is subject to this constraint, we can rewrite the action in terms of $\Phi^\CB_j$, $\Theta^\CB_j$, and $a_{1,j}$ as 
\begin{align}
S_0 &= \int_{\tau,x} \sum_j \biggl[ \frac{i}{\pi} \partial_x \Theta^\CB_j (\partial_\tau \Phi^\CB_j -A^\ext_{0,j}) +\frac{v}{8\pi} (S \partial_x \Phi^\CB_j)^2 +\frac{v+w}{8\pi} (\Delta \partial_x \Phi^\CB_j)^2 +\frac{u}{2\pi} (\partial_x \Theta^\CB_j)^2 \nonumber \\
&\ \ \ -\frac{v}{\pi} \partial_x \Phi^\CB_j (a_{1,j} +A^\ext_{1,j}) +\frac{v}{2\pi} (a_{1,j} +A^\ext_{1,j})^2 \biggr].
\end{align}
Substituting Eqs.~\eqref{eq:CBfromVCB} and \eqref{eq:GaugeCBandVCB}, we find the action in terms of $\Phi^\VCB_{j+1/2}$, $\Theta^\VCB_{j+1/2}$, and $\alpha_{1,j+1/2}$, 
\begin{align}
S_0 &= \int_{\tau,x} \sum_j\biggl[ \frac{i}{\pi} \partial_x \Theta^\VCB_{j+1/2} \Bigl( \partial_\tau \Phi^\VCB_{j+1/2} +2iv (\Delta^{-1,T} A^\ext_{1,j}) \Bigr) +\frac{m^2 v}{8\pi} (S \partial_x \Phi^\VCB_{j-1/2})^2 +\frac{u-m^2 v}{8\pi} (\Delta \partial_x \Phi^\VCB_{j-1/2})^2 \nonumber \\
&\ \ \ +\frac{v+w}{2\pi} (\partial_x \Theta^\VCB_{j+1/2})^2 -\frac{m^2 v}{\pi} \partial_x \Phi^\VCB_{j+1/2} \Bigl( \alpha_{1,j+1/2} +\frac{i}{2m^2v} (\Delta A^\ext_{0,j}) -\frac{1}{2m} (SA^\ext_{1,j}) \Bigr) +\frac{m^2 v}{2\pi} (\alpha_{1,j+1/2})^2 +\frac{v}{2\pi} (A^\ext_{1,j})^2 \biggr],
\end{align}
where we have defined $\Delta^{-1}$ as the inverse of lattice derivative $\Delta$. 
In the matrix notation, $S_{jj'} = \delta_{j+1,j'}+\delta_{jj'}$, $\Delta_{jj'} = \delta_{j+1,j'}-\delta_{jj'}$, and $2\Delta^{-1}_{jj'}= \sgn (j-j'-1/2)$.
We note that this action is subject to the constraint \eqref{eq:FluxAttach_alpha1}. 
We can impose the constraint \eqref{eq:FluxAttach_alpha1} via a Lagrange multiplier $\alpha_{0,j}$ by adding the term, 
\begin{align}
S^\textrm{$\alpha$-constraint}_0 = \int_{\tau,x} \sum_j \biggl[ -i \frac{m}{2\pi} \Bigl( \alpha_{1,j+1/2} +\frac{1}{m} \sum_{j' \neq j} \sgn (j'-j) \partial_x \Theta^\VCB_{j'+1/2} \Bigr) (\alpha_{0,j+1} -\alpha_{0,j}) \biggr], 
\end{align}
which splits into the temporal component of the minimal coupling between the vortex and $\alpha_\mu$ and the level-$m$ Chern-Simons term in the $\alpha_2=0$ gauge. 
We subsequently shift the gauge field as 
\begin{align}
\begin{split}
\alpha_{0,j} &\to \alpha_{0,j} -iv (\Delta^{-1} S A^\ext_{1,j}), \\
\alpha_{1,j+1/2} &\to \alpha_{1,j+1/2} -\frac{i}{2m^2 v} (\Delta A^\ext_{0,j}) +\frac{1}{2m} (SA^\ext_{1,j}), 
\end{split}
\end{align}
to cancel the minimal coupling between the vortex and the electromagnetic field. 
We finally obtain 
\begin{align} \label{eq:FullActionVCB}
S_0 &= \int_{\tau,x} \sum_j \biggl[ \frac{i}{\pi} \partial_x \Theta^\VCB_{j+1/2} \Bigl( \partial_\tau \Phi^\VCB_{j+1/2} -\frac{1}{2} (S\alpha_{0,j}) \Bigr) +\frac{m^2 v}{2\pi} (\partial_x \Phi^\VCB_{j+1/2} -\alpha_{1,j+1/2})^2 +\frac{v+w}{2\pi} (\partial_x \Theta^\VCB_{j+1/2})^2 \nonumber \\
&\ \ \ +\frac{u-m^2 v}{8\pi} (\Delta \partial_x \Phi^\VCB_{j-1/2})^2 -\frac{v}{2\pi} \partial_x \Theta^\VCB_{j+1/2} (\Delta A^\ext_{1,j}) -i\frac{m}{2\pi} \alpha_{1,j+1/2} (\Delta \alpha_{0,j})-\frac{i}{4\pi} (S A^\ext_{1,j}) (\Delta \alpha_{0,j}) \nonumber \\
&\ \ \ +\frac{i}{2\pi} A^\ext_{0,j} (\Delta \alpha_{1,j-1/2}) -\frac{1}{4\pi mv} (\Delta \alpha_{0,j}) (\Delta A^\ext_{0,j}) -\frac{1}{8\pi m^2 v} (\Delta A^\ext_{0,j})^2 +\frac{v}{8\pi} (\Delta A^\ext_{1,j})^2 \biggr]. 
\end{align}
Now the vortices are decoupled from the electromagnetic field $A^\ext_\mu$, apart from the doping by the magnetic field $(\Delta A^\ext_{1,j})$. 
Instead, the vortices couple to the Chern-Simons gauge field $\alpha_\mu$. 
The external electromagnetic field couples to the $2\pi$ flux of $\alpha_\mu$ through a discrete analog of a mutual
Chern-Simons coupling $(-i/2\pi) \epsilon_{\mu \nu \lambda} A^\ext_\mu \partial_\nu \alpha_\lambda$ in the $A^\ext_2 = \alpha_2 =0$ gauge. 
Higher order derivatives like $(\Delta A^\ext_{\mu,j})^2$ will only matter to the short-distance physics. 
Omitting them yields the vortex action \eqref{eq:ActionVCB}. 

In Sec.~\ref{sec:HaldaneHalperin}, we consider the first-level hierarchy states at filling fraction $\nu=1/(q+1/2p)$ with $p$ integer. 
Setting $m=q$ in the above, we further attach $4\pi p$ flux to the vortices and define the bosonic fields corresponding to composite quasiparticles, 
\begin{align}
\Phi^\CQ_{j+1/2} = \Phi^\VCB_{j+1/2} + 2p\sum_{j' \neq j} \sgn (j'-j) \Theta^\VCB_{j'+1/2}, \ \ \ 
\Theta^\CQ_{j+1/2} = \Theta^\VCB_{j+1/2}.
\end{align}
We then apply the vortex duality, 
\begin{align}
\Phi^\VCQ_j = \sum_{j'} \sgn (j'-j+1/2) \Theta^\CQ_{j'+1/2}, \ \ \ \Theta^\VCQ_j = \frac{1}{2} (\Phi^\CQ_{j+1/2} -\Phi^\CQ_{j-1/2}). 
\end{align}
As we consider the tunneling Hamiltonian \eqref{eq:TunnelingFirstHierarchy} with $p_1=p$, we replace the interwire forward scattering interaction in Eq.~\eqref{eq:FullActionBoson} by
\begin{align}
H_0^\textrm{inter-forward} = \frac{w}{8\pi} \int_{\tau,x} \sum_j \bigl( p \partial_x \varphi_{j-1} +qp \partial_x \theta_{j-1} +2(qp-1) \partial_x \theta_j -p \partial_x \varphi_{j+1} +qp \partial_x \theta_{j+1} \bigr)^2.
\end{align}
Just repeating the above procedure and introducing a new gauge field $\beta_\mu$, the final vortex action is given by 
\begin{align}
S_0 &= \int_{\tau,x} \sum_j \biggl[ \frac{i}{\pi} \partial_x \Theta^\VCQ_j \Bigl( \partial_\tau \Phi^\VCQ_j -\frac{1}{2} (S\beta_{0,j-1/2}) \Bigr) +\frac{2q^2 p^2 v}{\pi} (\partial_x \Phi^\VCQ_j -\beta_{1,j})^2 +\frac{q^2 v+w}{2\pi} (\partial_x \Theta^\VCQ_j)^2 \nonumber \\
&\ \ \ +\frac{v(1-4q^2 p^2)}{8\pi} (\Delta \partial_x \Phi^\VCQ_j)^2 +\frac{u-q^2 v}{8\pi} \Bigl( p \partial_x \Phi^\VCQ_{j-1} +2\partial_x \Theta^\VCQ_j -p \partial_x \Phi^\VCQ_{j+1} \Bigr)^2 -\frac{q^2 v}{2\pi} \partial_x \Theta^\VCQ_j (\Delta \alpha_{1,j-1/2}) \nonumber \\
&\ \ \ +\frac{v}{4\pi} \partial_x \Phi^\VCQ_j (\Delta^T \Delta A_{1,j}^\ext) -i\frac{q}{2\pi} \alpha_{1,j+1/2} (\Delta \alpha_{0,j}) -i\frac{2p}{2\pi} \beta_{1,j} (\Delta \beta_{0,j-1/2}) -\frac{i}{4\pi} (S\alpha_{1,j-1/2}) (\Delta \beta_{0,j-1/2}) \nonumber \\
&\ \ \ +\frac{i}{4\pi} (S \alpha_{0,j}) (\Delta \beta_{1,j}) -\frac{i}{4\pi} (SA^\ext_{1,j}) (\Delta \alpha_{0,j}) +\frac{i}{2\pi} A^\ext_{0,j} (\Delta \alpha_{1,j-1/2}) -\frac{1}{4\pi qv} (\Delta \alpha_{0,j}) (\Delta A^\ext_{0,j}) -\frac{1}{32\pi q^2 p^2 v} (\Delta \alpha_{0,j})^2 \nonumber \\
&\ \ \ +\frac{q^2 v}{8\pi} (\Delta \alpha_{1,j-1/2})^2 +\frac{1}{128\pi q^2 p^2 v} (\Delta^T \Delta \alpha_{0,j})^2 +\frac{1}{32\pi q^2 p^2 v} (\Delta^T S \alpha_{0,j}) (\Delta \beta_{0,j-1/2}) -\frac{1}{8\pi q^2 v} (\Delta A^\ext_{0,j})^2 +\frac{v}{8\pi} (\Delta A^\ext_{1,j})^2 \biggr].
\end{align}
This yields the action \eqref{eq:Action332VCQ} for $q=3$ and $p=1$ by replacing $\alpha_\mu \to \alpha^1_\mu$ and $\beta_\mu \to \alpha^2_\mu$.

\section{Derivation of the hole theory}
\label{app:Hole}

\subsection{Fermion at $\nu=1/2$}
\label{app:HoleTheory}

We now turn our attention to the special case of $m=1$ in the vortex action \eqref{eq:FullActionVCB}, which is considered in Secs.~\ref{sec:PHConjugate} and \ref{sec:AntiCFL} to have the hole description. 
In this case, we first attach the $2\pi$ flux to an electron to convert it into a composite boson and then apply the vortex duality to the composite boson. 
We thus focus on the action, 
\begin{align} \label{eq:FullActionVCBm1}
S_0 &= \int_{\tau,x} \sum_j \biggl[ \frac{i}{\pi} \partial_x \Theta^\VCB_{j+1/2} \Bigl( \partial_\tau \Phi^\VCB_{j+1/2} -\frac{1}{2} (S\alpha_{0,j}) \Bigr) +\frac{v}{2\pi} (\partial_x \Phi^\VCB_{j+1/2} -\alpha_{1,j+1/2})^2 +\frac{v+w}{2\pi} (\partial_x \Theta^\VCB_{j+1/2})^2 \nonumber \\
&\ \ \ +\frac{u-v}{8\pi} (\Delta \partial_x \Phi^\VCB_{j-1/2})^2 -\frac{i}{2\pi} \alpha_{1,j+1/2} (\Delta \alpha_{0,j}) -\frac{v}{2\pi} \partial_x \Theta^\VCB_{j+1/2} (\Delta A^\ext_{1,j}) -\frac{i}{4\pi} (S A^\ext_{1,j}) (\Delta \alpha_{0,j}) \nonumber \\
&\ \ \ +\frac{i}{2\pi} A^\ext_{0,j} (\Delta \alpha_{1,j-1/2}) -\frac{1}{4\pi v} (\Delta \alpha_{0,j}) (\Delta A^\ext_{0,j}) -\frac{1}{8\pi v} (\Delta A^\ext_{0,j})^2 +\frac{v}{8\pi} (\Delta A^\ext_{1,j})^2 \biggr]. 
\end{align}
In this case, the vortices are attached to the $-2\pi$ flux thorough the level-1 Chern-Simons term and hence converted to a fermion. 
This fermion is indeed interpreted as a hole in our coupled-wire model. 
To see this, we shift the gauge field as 
\begin{align}
\alpha_{0,j} &\to \alpha_{0,j} -A^\ext_{0,j}, \\
\alpha_{1,j+1/2} &\to \alpha_{1,j+1/2} -\frac{1}{2} (SA^\ext_{1,j}).
\end{align}
Then the action \eqref{eq:FullActionVCBm1} becomes 
\begin{align} \label{eq:FullActionVCBm1a}
S_0 &= \int_{\tau,x} \sum_j \biggl[ \frac{i}{\pi} \partial_x \Theta^\VCB_{j+1/2} \Bigl( \partial_\tau \Phi^\VCB_{j+1/2} -\frac{1}{2} (S\alpha_{0,j}) +\frac{1}{2} (SA^\ext_{0,j}) \Bigr) +\frac{v}{2\pi} \Bigl( \partial_x \Phi^\VCB_{j+1/2} -\alpha_{1,j+1/2} +\frac{1}{2} (SA^\ext_{1,j}) \Bigr)^2 \nonumber \\
&\ \ \ +\frac{v+w}{2\pi} (\partial_x \Theta^\VCB_{j+1/2})^2 +\frac{u-v}{8\pi} (\Delta \partial_x \Phi^\VCB_{j-1/2})^2 -\frac{i}{2\pi} \alpha_{1,j+1/2} (\Delta \alpha_{0,j}) +\frac{i}{4\pi} (SA^\ext_{1,j}) (\Delta A^\ext_{0,j}) \nonumber \\
&\ \ \ -\frac{v}{2\pi} \partial_x \Theta^\VCB_{j+1/2} (\Delta A^\ext_{1,j}) +\frac{1}{8\pi v} (\Delta A^\ext_{0,j})^2 +\frac{v}{8\pi} (\Delta A^\ext_{1,j})^2 -\frac{1}{4\pi v} (\Delta \alpha_{0,j}) (\Delta A^\ext_{0,j}) \biggr]. 
\end{align}
Now the vortex couples to the electromagnetic field $A^\ext_\mu$ with the opposite charge to the electron. 
There is also the Chern-Simons term of the electromagnetic field, $(i/4\pi) \epsilon_{\mu \nu \lambda} A^\ext_\mu \partial_\nu A^\ext_\lambda$, which corresponds to the Hall response $+e^2/h$ from the filled lowest Landau level. 
Integrating out $\alpha_\mu$ leads us to naturally define the bosonic fields given in Eq.~\eqref{eq:HoleField}.
In terms of those bosonic fields, the action \eqref{eq:FullActionVCBm1a} is written as 
\begin{align} \label{eq:FullActionHole}
S_0 &= \int_{\tau,x} \sum_j \biggl[ \frac{i}{\pi} \partial_x \theta^\hole_{j+1/2} \Bigl( \partial_\tau \varphi^\hole_{j+1/2} +\frac{1}{2} (SA^\ext_{0,j}) +\frac{iv}{2} (\Delta A^\ext_{1,j}) \Bigr) +\frac{v}{2\pi} \Bigl( \partial_x \varphi^\hole_{j+1/2} +\frac{1}{2} (SA^\ext_{1,j}) -\frac{i}{2v} (\Delta A^\ext_{0,j}) \Bigr)^2 \nonumber \\
&\ \ \ +\frac{v+w}{2\pi} (\partial_x \theta^\hole_{j+1/2})^2 +\frac{u-v}{8\pi} (\partial_x \varphi^\hole_{j-1/2} -\partial_x \theta^\hole_{j-1/2} -\partial_x \varphi^\hole_{j+1/2} -\partial_x \theta^\hole_{j+1/2})^2 \nonumber \\
&\ \ \ +\frac{i}{4\pi} (SA^\ext_{1,j}) (\Delta A^\ext_{0,j})  +\frac{1}{8\pi v} (\Delta A^\ext_{0,j})^2 +\frac{v}{8\pi} (\Delta A^\ext_{1,j})^2 \biggr].
\end{align}
As there is no dynamical gauge field in this action, the bosonic fields $\varphi^\hole_{j+1/2}$ and $\theta^\hole_{j+1/2}$ must be local in terms of the original bosonic fields $\varphi_j$ and $\theta_j$. 
Indeed, we find 
\begin{align}
\varphi^\hole_{j+1/2} = -\frac{1}{2} (\varphi_j +\theta_j +\varphi_{j+1} -\theta_{j+1}), \ \ \ 
\theta^\hole_{j+1/2} = -\frac{1}{2} (\varphi_j +\theta_j -\varphi_{j+1} +\theta_{j+1}). 
\end{align}
By using this, we can directly obtain Eq.~\eqref{eq:FullActionHole} from the action for electrons \eqref{eq:FullActionBoson}. 
One may notice that apart from a coupling with the electromagnetic field, the action \eqref{eq:FullActionHole} maintains the same form after the replacement $(\varphi_j, \theta_j) \to (-\varphi^\hole_{j+1/2}, \theta^\hole_{j+1/2})$ with complex conjugation in Eq.~\eqref{eq:FullActionBoson} for $m=1$ when $w=u-v$. 
Thus, the action is symmetric under the PH transformation, although this PH symmetry never be realized as a true microscopic symmetry, as discussed in the main text.

This PH symmetry is equivalent to the self duality between the composite boson action \eqref{eq:FullActionCB} for $m=1$ and the vortex action \eqref{eq:FullActionVCBm1} under $(\Phi^\CB_j, \Theta^\CB_j, a_\mu) \leftrightarrow (\Phi^\VCB_{j+1/2}, -\Theta^\VCB_{j+1/2}, \alpha_\mu)$ with complex conjugation, which has been found in Ref.~\cite{Mross17a}. 
The tunneling Hamiltonian for the CFL at $\nu=1/2$, which is given in Eq.~\eqref{eq:TunnelingHFLL}, is written in terms of the composite boson and the vortex as 
\begin{align}
H_1 &= \int_x \sum_j \Bigl[ g_R \kappa_j \kappa_{j+1} e^{i(\Phi^\CB_j +2\Theta^\CB_j -\Phi^\CB_{j+1})} +g_L \kappa_j \kappa_{j+1} e^{i(\Phi^\CB_j -\Phi^\CB_{j+1} +2\Theta^\CB_{j+1})} +\textrm{H.c.} \Bigr] \nonumber \\
&= \int_x \sum_j \Bigl[ g_R \kappa_j \kappa_{j+1} e^{i(\Phi^\VCB_{j-1/2} -\Phi^\VCB_{j+1/2} -2\Theta^\VCB_{j+1/2})} +g_L \kappa_{j-1} \kappa_j e^{i(\Phi^\VCB_{j-1/2} -2\Theta^\VCB_{j-1/2} -\Phi^\VCB_{j+1/2})} +\textrm{H.c.} \Bigr].
\end{align}
This also satisfies the above self-dual property when $g_R=g_L$ and the Klein factor is appropriately chosen.

\subsection{Two-component boson at $\nu=1/2+1/2$}
\label{app:HoleTheoryBoson}

We here provide the derivation of the hole theory for the CFL of two-component bosons at $\nu=1/2+1/2$, which is focused on in Sec.~\ref{sec:CFLTwoComp}. 
We introduce two species of bosonic fields $\varphi^\sigma_j(x)$ and $\theta^\sigma_j(x)$ labeled by $\sigma=\ua,\da$.
We then assume that these bosonic fields satisfy the commutation relations, 
\begin{align}
\begin{split}
[\theta^\ua_j(x), \varphi^\ua_{j'}(x')] &= i\pi \delta_{j,j'} \Theta (x-x'), \\
[\theta^\da_j(x), \varphi^\da_{j'}(x')] &= i\pi \delta_{j,j'} (\Theta(x-x')-1), 
\end{split}
\end{align}
while the other commutators vanish.
The CFL at $\nu=1/2+1/2$ may be described by the following action,
\begin{align}
\label{eq:ActionTwoCompApp}
S_0 &= \int_{\tau,x} \sum_{j,\sigma} \biggl[ \frac{i}{\pi} \partial_x \theta^\sigma_j (\partial_\tau \varphi^\sigma_j -A^\sigma_{0,j}) +\frac{v_\sigma}{2\pi} (\partial_x \varphi^\sigma_j -A^\sigma_{1,j})^2 +\frac{u_\sigma}{2\pi} (\partial_x \theta^\sigma_j)^2 +\frac{\tu_\sigma -v_\sigma}{8\pi} \Bigl( (\Delta \partial_x \varphi^\sigma_j) -(S\partial_x \theta^{-\sigma}_j) \Bigr)^2 \biggr], \\
\label{eq:TunnelingTwoCompApp}
H_1 &= \int_x \sum_{j,\sigma} \Bigl[ g^\sigma_R e^{i(\varphi^\sigma_j +2\theta^\sigma_j +\theta^{-\sigma}_j -\varphi^\sigma_{j+1} +\theta^{-\sigma}_{j+1})} +g^\sigma_L e^{i(\varphi^\sigma_j +\theta^{-\sigma}_j -\varphi^\sigma_{j+1} +2\theta^\sigma_{j+1} +\theta^{-\sigma}_{j+1})} +\textrm{H.c.} \Bigr],
\end{align}
where the symbol $-\sigma$ stands for $\da$($\ua$) for $\sigma=\ua$($\da$). 
We have separately coupled each species of the bosonic fields to the external gauge fields $A^\ua_\mu$ and $A^\da_\mu$ and worked on the $A^\ua_2 = A^\da_2 =0$ gauge. 
Setting $v_\sigma \equiv v$, $u_\sigma = \tu_\sigma \equiv u$, and $g^\sigma_R = g^\sigma_L \equiv g$, we find the action given by Eqs.~\eqref{eq:ActionTwoComp} and \eqref{eq:TunnelingTwoComp}. 
The tunneling terms involving $e^{i\varphi^\ua_j \pm i\theta^\da_j}$ or $e^{i\varphi^\da_j \pm i\theta^\ua_j}$ look unusual for bosonic systems at first sight, but such tunnelings are indeed possible when the boson numbers of two species are separately conserved and can be realized as certain correlated hoppings in the lattice systems \cite{Fuji16}. 

We first show that the action describes a CFL with two Fermi surfaces by the $2\pi$ flux attachment to both species of boson. 
To see this, we introduce the bosonic fields corresponding to the composite fermions, 
\begin{align}
\Phi^{\CF,\sigma}_j = \varphi^\sigma +\sum_{j' \neq j} \sgn (j'-j) (\theta^\sigma_{j'} +\theta^{-\sigma}_{j'}), \ \ \ \Theta^{\CF,\sigma}_j = \theta^\sigma_j,
\end{align}
which satisfy the commutation relations, 
\begin{align} \label{eq:MutualCFComm}
\begin{split}
[\Phi^{\CF,\ua}_j(x), \Phi^{\CF,\ua}_{j'}(x')] &= -i\pi \sgn (j-j'), \\
[\Theta^{\CF,\ua}_j(x), \Phi^{\CF,\ua}_{j'}(x')] &= i\pi \delta_{j,j'} \Theta (x-x'), \\
[\Phi^{\CF,\da}_j(x), \Phi^{\CF,\da}_{j'}(x')] &= i\pi \sgn (j-j'), \\
[\Theta^{\CF,\da}_j(x), \Phi^{\CF,\da}_{j'}(x')] &= i\pi \delta_{j,j'} (\Theta (x-x')-1),
\end{split}
\end{align}
while the other commutators vanish. 
This transformation makes the kinetic terms nonlocal while preserves the tunneling terms in a local form. 
The nonlocality of the kinetic terms is cured by introducing an auxiliary field $a_{1,j}$ with the constraint, 
\begin{align}
a_{1,j} = \sum_{j' \neq j} \sgn (j'-j) (\partial_x \Theta^{\CF,\ua}_{j'} +\partial_x \Theta^{\CF,\da}_{j'}). 
\end{align}
We implement this constraint by a Lagrange multiplier $a_{0,j+1/2}$ as we have routinely done. 
We then find 
\begin{align}
S_0 &= \int_{\tau,x} \sum_{j,\sigma} \biggl[ \frac{i}{\pi} \partial_x \Theta^{\CF,\sigma}_j \Bigl( \partial_\tau \Phi^{\CF, \sigma}_j -\frac{1}{2} (Sa_{0,j-1/2}) -A^\sigma_{0,j} \Bigr) +\frac{v_\sigma}{2\pi} (\partial_x \Phi^{\CF,\sigma}_j -a_{1,j} -A^\sigma_{1,j})^2 +\frac{u_\sigma}{2\pi} (\partial_x \Theta^{\CF,\sigma}_j)^2 \nonumber \\
&\ \ \ +\frac{\tu_\sigma -v_\sigma}{8\pi} \Bigl( (\Delta \partial_x \Phi^{\CF,\sigma}_j) +(S \partial_x \Theta^{\CF,\sigma}_j) \Bigr)^2 +\frac{i}{2\pi} a_{1,j} (\Delta a_{0,j-1/2}) \biggr], \\
H_1 &= \int_{\tau,x} \sum_{j,\sigma} \Bigl[ g^\sigma_R e^{i(\Phi^{\CF,\sigma}_j +\Theta^{\CF,\sigma}_j -\Phi^{\CF,\sigma}_{j+1} -\Theta^{\CF,\sigma}_{j+1})} +g^\sigma_L e^{i(\Phi^{\CF,\sigma}_j -\Theta^{\CF,\sigma}_j -\Phi^{\CF,\sigma}_{j+1} +\Theta^{\CF,\sigma}_{j+1})} +\textrm{H.c.} \Bigr]. 
\end{align}
From the commutation relations \eqref{eq:MutualCFComm}, we can regard that the operators $e^{i\Phi^{\CF,\sigma}_j \pm i\Theta^{\CF,\sigma}_j}$ are fermionic operators anticommuting between the same species. 
While these operators commute between different species, we can multiply the Klein factors $\kappa_\sigma$ obeying $\{ \kappa_\sigma, \kappa_{\sigma'} \} = 2\delta_{\sigma, \sigma'}$ for each species to define fully anticommuting fermionic operators $\psi^{\CF,\sigma}_{R/L,j} \propto \kappa_\sigma e^{i\Phi^{\CF,\sigma}_j \pm i\Theta^{\CF,\sigma}_j}$. 
Such Klein factors do not appear explicitly in the action since the fermionic operators for each species always appear in a bilinear form due to the separate charge conservation.
Thus we find that the action describes the CFL with two Fermi surfaces of the composite fermions, each of which carries different spins. 

We then wish to obtain the hole description of the action given by Eqs.~\eqref{eq:ActionTwoCompApp} and \eqref{eq:TunnelingTwoCompApp}. 
To proceed, we first apply the mutual $2\pi$ flux attachment to the bosonic action \cite{Senthil13}.
We hence define new bosonic fields corresponding to the mutual composite bosons,
\begin{align}
\Phi^{\CB,\sigma}_j = \varphi^\sigma_j +\sum_{j' \neq j} \sgn (j'-j) \theta^{-\sigma}_{j'}, \ \ \ \Theta^{\CB,\sigma}_j = \theta^\sigma_j, 
\end{align}
which satisfy the commutation relations,
\begin{align}
\begin{split}
[\Theta^{\CB,\ua}_j(x), \Phi^{\CB,\ua}_{j'}(x')] &= i\pi \delta_{j,j'} \Theta (x-x'), \\
[\Theta^{\CB,\da}_j(x), \Phi^{\CB,\da}_{j'}(x')] &= i\pi \delta_{j,j'} (\Theta (x-x') -1), 
\end{split}
\end{align}
while the other commutators vanish. 
We now introduce Lagrange multipliers $b^\sigma_{0,j+1/2}$ in order to implement the constraints, 
\begin{align}
b^\sigma_{1,j} &= \sum_{j' \neq j} \sgn (j'-j) \partial_x \Theta^{\CB,-\sigma}_{j'},
\end{align}
We then find the action in terms of the mutual composite bosons, 
\begin{align}
S_0 &= \int_{\tau,x} \sum_{j,\sigma} \biggl[ \frac{i}{\pi} \partial_x \Theta^{\CB,\sigma}_j \Bigl( \partial_\tau \Phi^{\CB,\sigma}_j -\frac{1}{2} (Sb^\sigma_{0,j-1/2}) -A^\sigma_{0,j} \Bigr) +\frac{v_\sigma}{2\pi} (\partial_x \Phi^{\CB,\sigma}_j -b^\sigma_{1,j} -A^\sigma_{1,j})^2 +\frac{u_\sigma}{2\pi} (\partial_x \Theta^{\CB,\sigma}_j)^2 \nonumber \\
&\ \ \ +\frac{\tu_\sigma -v_\sigma}{8\pi} (\Delta \partial_x \Phi^{\CB,\sigma}_j)^2 +\frac{i}{2\pi} b^\sigma_{1,j} (\Delta b^{-\sigma}_{0,j-1/2}) \biggr], \\
H_1 &= \int_x \sum_{j,\sigma} \Bigl[ g^\sigma_R e^{i(\Phi^{\CB,\sigma}_j +2\Theta^{\CB,\sigma}_j -\Phi^{\CB,\sigma}_{j+1})} +g^\sigma_L e^{i(\Phi^{\CB,\sigma}_j -\Phi^{\CB,\sigma}_{j+1} +2\Theta^{\CB,\sigma}_{j+1})} +\textrm{H.c.} \Bigr]. 
\end{align}
Here the action has a discrete analog of the mutual Chern-Simons term $(i/2\pi) \epsilon_{\mu \nu \lambda} b^\ua_\mu \partial_\nu b^\da_\lambda$ in the $b^\sigma_2=0$ gauge.
We then apply the vortex duality for each species of the mutual composite bosons. 
The corresponding transformation is given by
\begin{align}
\Phi^{\VCB,\sigma}_{j+1/2} = \sum_{j'} \sgn (j'-j-1/2) \Theta^{\CB,-\sigma}_{j'}, \ \ \ \Theta^{\VCB,\sigma}_{j+1/2} = \frac{1}{2} (\Phi^{\CB,-\sigma}_{j+1} -\Phi^{\CB,-\sigma}_j). 
\end{align}
We remark that the label $\sigma$ for the vortices is changed from that for the mutual composite bosons. 
These bosonic fields satisfy the commutation relations, 
\begin{align}
\begin{split}
[\Theta^{\VCB,\ua}_{j+1/2}(x), \Phi^{\VCB,\ua}_{j'+1/2}(x')] &= i\pi \delta_{j,j'} \Theta(x-x'), \\
[\Theta^{\VCB,\da}_{j+1/2}(x), \Phi^{\VCB,\da}_{j'+1/2}(x')] &= i\pi \delta_{j,j'} (\Theta (x-x')-1). 
\end{split}
\end{align}
The gauge fields $\beta^\sigma_{1,j+1/2}$ coupled to the vortices are subject to the constraints written in terms of the vortex or composite boson fields,
\begin{align} \label{eq:ConstraintBetaTwoComp}
\beta^\sigma_{1,j+1/2} = -\sum_{j' \neq j} \sgn (j'-j) \partial_x \Theta^{\VCB,-\sigma}_{j'+1/2} 
= \frac{1}{2} (\partial_x \Phi^{\CB,\sigma}_{j+1} +\partial_x \Phi^{\CB,\sigma}_j). 
\end{align}
The constraints for the mutual Chern-Simons gauge fields $b^\sigma_{1,j}$ are also written in terms of the vortex fields by 
\begin{align}
b^\sigma_{1,j} = \frac{1}{2} (\partial_x \Phi^{\VCB,\sigma}_{j+1/2} +\partial_x \Phi^{\VCB,\sigma}_{j-1/2}). 
\end{align}
Substituting these expressions, introducing Lagrange multipliers $\beta^\sigma_{0,j}$ to implement the constraint \eqref{eq:ConstraintBetaTwoComp}, and subsequently shifting the gauge fields as 
\begin{align}
\begin{split}
\beta^\sigma_{0,j} &\to \beta^\sigma_{0,j} -iv_{-\sigma} (\Delta^{-1} S A^{-\sigma}_{1,j}), \\
\beta^\sigma_{1,j+1/2} &\to \beta^\sigma_{1,j+1/2} -\frac{i}{2v_\sigma} (\Delta A^{-\sigma}_{0,j}) +\frac{1}{2} (S A^\sigma_{1,j}), 
\end{split}
\end{align}
we obtain the vortex action, 
\begin{align}
S_0 &= \int_{\tau,x} \sum_{j,\sigma} \biggl[ \frac{i}{\pi} \partial_x \Theta^{\VCB,\sigma}_{j+1/2} \Bigl( \partial_\tau \Phi^{\VCB,\sigma}_{j+1/2} -\frac{1}{2}(S \beta^\sigma_{0,j}) \Bigr) +\frac{v_\sigma}{2\pi} (\partial_x \Phi^{\VCB,\sigma}_{j+1/2} -\beta^\sigma_{1,j+1/2})^2 +\frac{\tu_{-\sigma}}{2\pi} (\partial_x \Theta^{\VCB,\sigma}_{j+1/2})^2 \nonumber \\
&\ \ \ +\frac{u_{-\sigma} -v_\sigma}{8\pi} (\Delta \partial_x \Phi^{\VCB,\sigma}_{j-1/2})^2 -\frac{v_{-\sigma}}{2\pi} \partial_x \Theta^{\VCB,\sigma}_{j+1/2} (\Delta A^{-\sigma}_{1,j}) -\frac{i}{2\pi} \beta^\sigma_{1,j+1/2} (\Delta \beta^{-\sigma}_{0,j}) -\frac{i}{4\pi} (SA^\sigma_{1,j}) (\Delta \beta^{-\sigma}_{0,j}) \nonumber \\
&\ \ \ +\frac{i}{2\pi} A^\sigma_{0,j} (\Delta \beta^{-\sigma}_{1,j-1/2}) -\frac{1}{8\pi v_{-\sigma}} (\Delta A^\sigma_{0,j})^2 +\frac{v_\sigma}{8\pi} (\Delta A^\sigma_{1,j})^2 -\frac{1}{4\pi v_{-\sigma}} (\Delta \beta^\sigma_{0,j}) (\Delta A^\sigma_{0,j}) \biggr], \\
H_1 &= \int_x \sum_{j,\sigma} \Bigl[ g^\sigma_R e^{i(\Phi^{\VCB,-\sigma}_{j-1/2} -\Phi^{\VCB,-\sigma}_{j+1/2} -2\Theta^{\VCB,-\sigma}_{j+1/2})} +g^\sigma_L e^{i(\Phi^{\VCB,-\sigma}_{j-1/2} -2\Theta^{\VCB,-\sigma}_{j-1/2} -\Phi^{\VCB,-\sigma}_{j+1/2})} +\textrm{H.c.} \Bigr]. 
\end{align}
Now the vortices are coupled to the gauge fields $\beta^\sigma_\mu$ with a discrete analog of the mutual Chern-Simons term $-(i/2\pi) \epsilon_{\mu \nu \lambda} \beta^\ua_\mu \partial_\nu \beta^\da_\lambda$ with the opposite sign to that for the mutual composite bosons. 
Under the duality transformation $(\Phi^{\CB,\sigma}_j, \Theta^{\CB,\sigma}_j, b^\sigma_\mu) \leftrightarrow (\Phi^{\VCB,\sigma}_{j+1/2}, -\Theta^{\VCB,\sigma}_{j+1/2}, \beta^\sigma_\mu$) with complex conjugation, the theory is self dual when $u_\sigma = \tu_{-\sigma}$ and $g^\sigma_R = g^{-\sigma}_L$. 
There is also another duality transformation involving the interchange of two species $\ua \leftrightarrow \da$, under which the theory is self dual when $v_\ua = v_\da$, $u_\sigma = \tu_\sigma$, and $g^\sigma_R = g^\sigma_L$. 
For these two self-duality conditions to be satisfied at the same time, we must have $v_\sigma \equiv v$, $u_\sigma = \tu_\sigma \equiv u$, and $g^\sigma_R = g^\sigma_L \equiv g$.

We finally integrate out the mutual Chern-Simons gauge fields to obtain the hole description of the CFL action. 
We first shift the gauge fields as 
\begin{align}
\beta^\sigma_{0,j} \to \beta^\sigma_{0,j} -A^\sigma_{0,j}, \ \ \ \beta^\sigma_{1,j+1/2} \to \beta^\sigma_{1,j+1/2} -\frac{1}{2} (SA^\sigma_{1,j}). 
\end{align}
This makes the vortices couple to the external gauge fields and generates a discrete analog of the mutual Chern-Simons term $(i/2\pi) \epsilon_{\mu \nu \lambda} A^\ua_\mu \partial_\nu A^\da_\lambda$, which gives the Hall response of the bosonic IQH state. 
Integrating out $\beta^\sigma_0$ yields the constraints,
\begin{align}
\beta^\sigma_{1,j+1/2} = -\sum_{j' \neq j} \sgn (j'-j) \partial_x \Theta^{\VCB,-\sigma}_{j'+1/2} +\frac{i}{2v_\sigma} (\Delta A^{-\sigma}_{0,j}).
\end{align}
This leads us to define the bosonic fields, 
\begin{align}
\begin{split}
\varphi^{\hole,\sigma}_{j+1/2} &= \Phi^{\VCB,\sigma}_{j+1/2} +\sum_{j' \neq j} \sgn (j'-j) \Theta^{\VCB,-\sigma}_{j'+1/2}, \\
\theta^{\hole,\sigma}_{j+1/2} &= \Theta^{\VCB,\sigma}_{j+1/2}.
\end{split}
\end{align}
We finally obtain the CFL action in terms of the holes, 
\begin{align}
S_0 &= \int_{\tau,x} \sum_{j,\sigma} \biggl[ \frac{i}{\pi} \partial_x \theta^{\hole,\sigma}_{j+1/2} \Bigl( \partial_\tau \varphi^{\hole,\sigma}_{j+1/2} +\frac{1}{2} (SA^\sigma_{0,j}) +\frac{iv_{-\sigma}}{2} (\Delta A^{-\sigma}_{1,j}) \Bigr) +\frac{v_\sigma}{2\pi} \Bigl( \partial_x \varphi^{\hole,\sigma}_{j+1/2} +\frac{1}{2} (SA^\sigma_{1,j}) -\frac{i}{2v_\sigma} (\Delta A^{-\sigma}_{0,j}) \Bigr)^2 \nonumber \\
&\ \ \ +\frac{\tu_{-\sigma}}{2\pi} (\partial_x \theta^{\hole,\sigma}_{j+1/2})^2 +\frac{u_{-\sigma} -v_\sigma}{8\pi} \Bigl( (\Delta \partial_x \varphi^{\hole,\sigma}_{j-1/2}) +(S\partial_x \theta^{\hole,-\sigma}_{j-1/2}) \Bigr)^2 +\frac{i}{4\pi} (SA^\sigma_{1,j}) (\Delta A^{-\sigma}_{0,j}) \nonumber \\
&\ \ \ +\frac{1}{8\pi v_{-\sigma}} (\Delta A^\sigma_{0,j})^2 +\frac{v_\sigma}{8\pi} (\Delta A^\sigma_{1,j})^2 \biggr], \\
H_1 &= \int_x \sum_{j,\sigma} \Bigl[ g^\sigma_R e^{i(\varphi^{\hole,-\sigma}_{j-1/2} -\theta^{\hole,-\sigma}_{j-1/2} -\varphi^{\hole,-\sigma}_{j+1/2} -2\theta^{\hole,\sigma}_{j+1/2} -\theta^{\hole,-\sigma}_{j+1/2})} +g^\sigma_L e^{i(\varphi^{\hole,-\sigma}_{j-1/2} -2\theta^{\hole,\sigma}_{j-1/2} -\theta^{\hole,-\sigma}_{j-1/2} -\varphi^{\hole,-\sigma}_{j+1/2} -\theta^{\hole,-\sigma}_{j+1/2})} +\textrm{H.c.} \Bigr]. 
\end{align}
This hole theory is related to the original action by a local transformation, 
\begin{align}
\varphi^{\hole,\sigma}_{j+1/2} = -\frac{1}{2} (\varphi^\sigma_j +\theta^{-\sigma}_j +\varphi^\sigma_{j+1} -\theta^{-\sigma}_{j+1}), \ \ \ 
\theta^{\hole,\sigma}_{j+1/2} = -\frac{1}{2} (\varphi^{-\sigma}_j +\theta^\sigma_j -\varphi^{-\sigma}_{j+1} +\theta^\sigma_{j+1}). 
\end{align}
The PH transformation is now defined by 
\begin{align}
\varphi^\sigma_j  \to -\varphi^{\hole,\sigma}_{j+1/2}, \ \ \ \theta^\sigma_j \to \theta^{\hole,\sigma}_{j+1/2}
\end{align}
with complex conjugation, although this transformation is not well-defined in the pure 2D lattice system as in the fermionic case.
The CFL action has the PH symmetry of this sense when $u_\sigma = \tu_{-\sigma}$ and $g^\sigma_R = g^{-\sigma}_L$, which is indeed the self duality condition discussed above. 
This transformation can be conveniently expressed in terms of the bosonic fields $\phi^\sigma_j = \varphi^\sigma +\theta^{-\sigma}_j$ and $\tilde{\phi}^\sigma_j = \varphi^\sigma_j -\theta^{-\sigma}_j$. 
The pairs of the bosonic fields $\{ \phi^\sigma_j \}$ and $\{ \tilde{\phi}^\sigma_j \}$ have the opposite chirality to each other and actually describe counter-propagating edge modes of the bosonic IQH state \cite{Fuji16}. 
In terms of these fields, the PH transformation is given by $\phi^\sigma_j \to \tilde{\phi}^\sigma_{j+1}$ and $\tilde{\phi}^\sigma_j \to \phi^\sigma_j$ with complex conjugation.
This is equivalent to the antiunitary PH symmetry considered for the $N_f=2$ QED${}_3$ in Ref.~\cite{Mross16a} when $\{ \phi^\sigma_j \}$ and $\{ \tilde{\phi}^\sigma_j \}$ are separated by half the lattice spacing. 
\end{widetext}

\section{Some details about Pfaffian states}
\label{app:Pfaffian}

\subsection{Review of Teo-Kane's construction}
\label{app:PfaffianTeoKane}

For the paper to be self-contained, we here briefly review the Teo-Kane's construction of the Pfaffian state at $\nu=1/M$ \cite{Teo14}. 
The corresponding tunneling Hamiltonian is given in Eq.~\eqref{eq:TunnelingPfaffian}. 
By grouping two adjacent wires, Teo and Kane introduced the chiral bosonic fields corresponding to the charge ($c$) and neutral ($n$) sectors, 
\begin{align} \label{eq:ChargeNeutralField}
\begin{split}
\tphi^c_{R,l} &= \frac{1}{2} [ \varphi_{2l} +\varphi_{2l+1} +(M+1) \theta_{2l} +(3M-1) \theta_{2l+1} ], \\
\tphi^c_{L,l} &= \frac{1}{2} [ \varphi_{2l} +\varphi_{2l+1} -(3M-1) \theta_{2l} -(M+1) \theta_{2l+1}], \\
\tphi^n_{R,l} &= \frac{1}{2} [ \varphi_{2l} -\varphi_{2l+1} +(M+1) \theta_{2l} +(M-3) \theta_{2l+1} ], \\
\tphi^n_{L,j} &= \frac{1}{2} [ \varphi_{2l} -\varphi_{2l+1} +(M-3) \theta_{2l} +(M+1) \theta_{2l+1} ], 
\end{split}
\end{align}
which satisfy the commutation relations, 
\begin{align} \label{eq:CommChargeNeutralTK}
\begin{split}
[\tphi^c_{r,l}(x), \tphi^c_{r',l'}(x')] &= i\pi rM \delta_{r,r'} \delta_{l,l'} \sgn (x-x') \\
&\ \ \  +i\pi M \delta_{l,l'} \epsilon_{r,r'}, \\
[\tphi^n_{r,l}(x), \tphi^n_{r,l'}(x')] &= i\pi r \delta_{r,r'} \delta_{l,l'} \sgn (x-x') \\
&\ \ \ +i\pi \delta_{l,l'} \epsilon_{r,r'}, \\
[\tphi^c_{r,l}(x), \tphi^n_{r',l'}(x')] &= -\frac{i\pi}{2} (M-1) \delta_{l,l'}. 
\end{split}
\end{align}
In terms of these bosonic fields, the tunneling Hamiltonian \eqref{eq:TunnelingPfaffian} can be written as 
\begin{align} \label{eq:TunnelingPfaffianTK}
H_1 &= \int_x \sum_l \Bigl[ t_{00} \kappa_{2n} \kappa_{2n+1} e^{i(\tphi^c_{R,l} -\tphi^c_{L,l+1} +\tphi^n_{R,l} -\tphi^n_{L,l+1})} \nonumber \\
&\ \ \ +t_{11} \kappa_{2n+1} \kappa_{2n+3} e^{i(\tphi^c_{R,l} -\tphi^c_{L,l+1} -\tphi^n_{R,l} +\tphi^n_{L,l+1})} \nonumber \\
&\ \ \ +t_{01} \kappa_{2n} \kappa_{2n+3} e^{i(\tphi^c_{R,l} -\tphi^c_{L,l+1} +\tphi^n_{R,l} +\tphi^n_{L,l+1})} \nonumber \\
&\ \ \ +t_{10} \kappa_{2n+1} \kappa_{2n+2} e^{i(\tphi^c_{R,l} -\tphi^c_{L,l+1} -\tphi^n_{R,l} -\tphi^n_{L,l+1})} \nonumber \\
&\ \ \ +t_u \kappa_{2n} \kappa_{2n+1} e^{i(\tphi^n_{R,l} +\tphi^n_{L,l})} +t_v e^{i(\tphi^n_{R,l} -\tphi^n_{L,l})} +\textrm{H.c.} \Bigr].
\end{align}
The commutation relation \eqref{eq:CommChargeNeutralTK} suggests that vertex operators of the neutral bosonic fields, 
\begin{align} \label{eq:NeutralFermiTK}
\tpsi^n_{r,l}(x) = \frac{\eta_l}{\sqrt{2\pi \alpha}} e^{i\tphi^n_{r,l}(x)}, 
\end{align}
can be regarded as fermionic operators when appropriate forward scattering interactions are incorporated into the kinetic action such that these vertex operators have conformal weight $1/2$. 
In order to make sure that these operators anticommute between different $l$'s, we choose the Klein factor to be $\eta_l = \kappa_{2l}$ for the fermionic case (even $M$), while we introduce new Majorana operators obeying $\{ \eta_l, \eta_{l'} \} = 2\delta_{ll'}$ for the bosonic case (odd $M$). 
The neutral sector can be split into two Ising CFTs and the fermionic operator is then written in terms of two Majorana fermions as, 
\begin{align} \label{eq:NeutralMajoranaTK}
\tpsi^n_{r,l} = \frac{1}{\sqrt{2}} (\txi^1_{r,l} +i\txi^2_{r,l}). 
\end{align}
As we will see below, the Ising CFT associated with $\txi^2_{r,l}$ is gapped by the interaction within each $l$. 
The operator hopping unit charge between adjacent $l$'s is given by $e^{i\tphi^c_{r,l}} \txi^1_{r,l}$, which is identified with the electron operator of the Pfaffian state. 
Quasiparticles are excited in pair by the bare electron $2k_F$ backscattering operators, 
\begin{align}
\begin{split}
e^{i2\theta_{2l}} &= e^{i(\tphi^c_{R,l} -\tphi^c_{L,l})/2M +i(\tphi^n_{R,l} -\tphi^n_{L,l})/2}, \\
e^{i2\theta_{2l+1}} &= e^{i(\tphi^c_{R,l} -\tphi^c_{L,l})/2M -i(\tphi^n_{R,l} -\tphi^n_{L,l})/2}.
\end{split}
\end{align}
From these, one can read off the smallest quasiparticle charge $\pm 1/2M$. 
However, the neutral sector of the quasiparticles operators still requires integration of the massive Majorana fermion $\txi^2_{r,l}$. 
Indeed, Teo and Kane discussed that the neutral sector can be identified with the spin field of the Ising CFT. 
It does not have a bosonic (vertex) representation since it has two fusion channels.
We now wish to express the tunneling Hamiltonian \eqref{eq:TunnelingPfaffianTK} in terms of the bosonic charge modes and these neutral Majorana modes. 
Since we need some care about the Klein factors, we below separately treat the fermionic and bosonic cases. 

\subsection{Klein factor}
\label{app:KleinPfaffian}

For the bosonic (odd $M$) case, we have assigned that $\kappa_j$ to be just a constant: $\kappa_j=1$. 
Introducing the new Klein factors $\eta_l$, the tunneling Hamiltonian \eqref{eq:TunnelingPfaffianTK} is written in terms of the fermionic operators \eqref{eq:NeutralFermiTK} as 
\begin{align} \label{eq:TunnelingPfaffianTKB}
H_1 &= 2\pi \alpha \int_x \sum_l \Bigl[ -\eta_l \eta_{l+1} e^{i(\tphi^c_{R,l} -\tphi^c_{L,l+1})} \nonumber \\
&\ \ \ \times \bigl( t_{00} e^{-i\pi (M-1)/2} \tpsi^n_{R,l} \tpsi^{n\dagger}_{L,l+1} +t_{01} \tpsi^n_{R,l} \tpsi^n_{L,l+1} \nonumber \\
&\ \ \ +t_{10} \tpsi^{n\dagger}_{R,l} \tpsi^{n\dagger}_{L,l+1} +t_{11} e^{i\pi (M-1)/2} \tpsi^{n\dagger}_{R,l} \tpsi^n_{L,l+1} \bigr) \nonumber \\
&\ \ \ +it_u \tpsi^n_{R,l} \tpsi^n_{L,l} -it_v \tpsi^n_{R,l} \tpsi^{n\dagger}_{L,l} +\textrm{H.c.} \Bigr]. 
\end{align}
We now choose the coupling constants to be $t_{00} = -e^{i\pi (M-1)/2} g$, $t_{11} = -e^{-i\pi (M-1)/2} g$, $t_{01} = t_{10} = -g$, $t_u = -g_u$, and $g_v = -g_v$ with $g$, $g_u$, and $g_v$ being real. 
Then the tunneling Hamiltonian \eqref{eq:TunnelingPfaffianTKB} takes a simple form in terms of the Majorana fermions \eqref{eq:NeutralMajoranaTK}, 
\begin{align}
H_1 &= 2\pi \alpha \int_x \sum_l \Bigl[ 4g \eta_l \eta_{l+1} \cos (\tphi^c_{R,l} -\tphi^c_{L,l+1}) \txi^1_{R,l} \txi^1_{L,l+1} \nonumber \\
&\ \ \ +\frac{g_v -g_u}{2} i \txi^1_{R,l} \txi^1_{L,l} +\frac{g_v +g_u}{2} i \txi^2_{R,l} \txi^2_{L,l} \Bigr].
\end{align}
When $g_v=g_u$, the neutral Majorana fermions $\txi^2_{r,l}$ are gapped within each $l$, while the residual charge and neutral modes are paired up between neighboring $l$'s to open a gap. 
Thus, the charged boson $\tphi^c_{r,l}$ and the neutral Majorana fermion $\txi^1_{r,l}$ are left at the boundaries. 
The vertex operators of the charged bosonic fields $\eta_l e^{i\tphi^c_{r,l}}$ carry charge 1 and follow the fermionic statistics. 
They are combined with the Majorana fermions $\txi^1_{r,l}$ to form the electron operator with the bosonic statistics. 
This observation is consistent with what is expected for the Pfaffian state. 

For the fermionic (even $M$) case, $\kappa_j$ are Majorana operators. 
In terms of the fermionic fields \eqref{eq:NeutralFermiTK}, the tunneling Hamiltonian \eqref{eq:TunnelingPfaffianTK} is written as 
\begin{align}
H_1 &= 2\pi \alpha \int_x \sum_l \Bigl[ e^{i(\tphi^c_{R,l} -\tphi^c_{L,l+1})} \nonumber \\
&\ \ \ \times \bigl( t_{00} e^{-i\pi (M-1)/2} \tpsi^n_{R,l} \tpsi^{n\dagger}_{L.l+1} \nonumber \\
&\ \ \ +t_{11} e^{i\pi (M-1)/2} \kappa_{2l} \kappa_{2l+1} \kappa_{2l+2} \kappa_{2l+3} \tpsi^{n\dagger}_{R,l} \tpsi^n_{L,l+1} \nonumber \\
&\ \ \ -t_{01} \kappa_{2l+2} \kappa_{2l+3} \tpsi^n_{R,l} \tpsi^n_{L,l+1} -t_{10} \kappa_{2l} \kappa_{2l+1} \tpsi^{n\dagger}_{R,l} \tpsi^{n\dagger}_{L,l+1} \bigr) \nonumber \\
&\ \ \ +ig_u \kappa_{2l} \kappa_{2l+1} \tpsi^n_{R,l} \tpsi^n_{L,l} -ig_v \tpsi^n_{R,l} \tpsi^{n\dagger}_{L,l} +\textrm{H.c.} \Bigr].
\end{align}
In this Hamiltonian, the Klein factors appear only in the bilinear form $\kappa_{2l} \kappa_{2l+1}$. 
Since they are commuting with each other, we can simultaneously diagonalize them and replace them by their eigenvalues, say, $\kappa_{2l} \kappa_{2l+1} =i$. 
Choosing the coupling constants to be $t_{00} = e^{i\pi (M-1)/2} g$, $t_{11} = e^{i\pi (M-1)/2} g$, $t_{01} = t_{10} = ig$, $t_u = ig_u$, and $t_v = -g_v$, we find the tunneling Hamiltonian in terms of the Majorana fermions \eqref{eq:NeutralMajoranaTK}, 
\begin{align}
H_1 &= 2\pi \alpha \int_x \sum_l \Bigl[ 4g \cos (\tphi^c_{R,l} -\tphi^c_{L,l+1}) \txi^1_{R,l} \txi^1_{L,l+1} \nonumber \\
&\ \ \ +\frac{g_v-g_u}{2} i \txi^1_{R,l} \txi^1_{L,l} +\frac{g_v+g_u}{2} i \txi^2_{R,l} \txi^2_{L,l} \Bigr].
\end{align}
Now the vertex operators $e^{i\tphi^c_{r,l}}$ carry charge 1 and follow the bosonic statistics. 
They are combined with the neutral Majorana fermions $\txi^1_{r,l}$ to form the electron operators with the fermionic statistics. 

The above treatment of the Klein factors is essentially the same for the composite fermion formulation of the Pfaffian state discussed in Sec.~\ref{sec:GeneralPfaffian}. 
With the same choice of the coupling constants and the Klein factors, we can find the tunneling Hamiltonian for the bosonic case, 
\begin{align}
H_1 &= 2\pi \alpha \int_x \sum_l \Bigl[ 4g \eta_l \eta_{l+1} \cos (\Phi^c_l -\Phi^c_{l+1}) \xi^1_{R,l} \xi^1_{L,l+1} \nonumber \\
&\ \ \ +\frac{g_v -g_u}{2} i \xi^1_{R,l} \xi^1_{L,l} +\frac{g_v +g_u}{2} i \xi^2_{R,l} \xi^2_{L,l} \Bigr], 
\end{align}
while for the fermionic case, 
\begin{align}
H_1 &= 2\pi \alpha \int_x \sum_l \Bigl[ 4g \cos (\Phi^c_l -\Phi^c_{l+1}) \xi^1_{R,l} \xi^1_{L,l+1} \nonumber \\
&\ \ \ +\frac{g_v -g_u}{2} i \xi^1_{R,l} \xi^1_{L,l} +\frac{g_v +g_u}{2} i \xi^2_{R,l} \xi^2_{L,l} \Bigr]. 
\end{align}
By setting $g \to g/ 8\pi \alpha$ and $g_{u,v} \to g_{u,v}/ 2\pi \alpha$, we obtain the tunneling Hamiltonian \eqref{eq:TunnelingPfaffianCF3}. 

\bibliography{QHHierarchyCW}

\end{document}